\documentclass[a4paper,12pt]{article}
\usepackage[vmargin=1.18in,hmargin=1.1in]{geometry}
\usepackage[english]{babel}
\usepackage[utf8x]{inputenc}
\usepackage[T1]{fontenc}
\usepackage{microtype}
\usepackage{parskip}
\usepackage[dvipsnames]{xcolor}
\usepackage{graphicx}
\usepackage{subcaption}
\usepackage{longtable}
\usepackage{multirow}
\usepackage{listings}
\usepackage{makecell}
\usepackage{array}
\usepackage{float}
\usepackage{wrapfig}
\usepackage{dsfont}
\usepackage{rotating}
\usepackage{booktabs}
\usepackage{enumerate}
\usepackage{extarrows}
\usepackage{tikz}
\usetikzlibrary{chains}
\usetikzlibrary{arrows}
\usepackage{pgf}
\usepackage[normalem]{ulem}
\usepackage{placeins}
\usepackage{needspace}

\usepackage{amsmath}
\usepackage{amssymb}
\usepackage{amsthm}
\usepackage{bm}
\usepackage[linesnumbered,ruled]{algorithm2e}
\usepackage{algorithmicx}
\usepackage{mathtools}
\usepackage{mathrsfs}
\usepackage{stmaryrd}

\usepackage[round,authoryear]{natbib}
\usepackage[
  colorlinks,
  citecolor=blue,
  linkcolor=red,
  anchorcolor=red,
  urlcolor=blue
]{hyperref}
\mathtoolsset{showonlyrefs}
\usepackage{authblk}
\usepackage{pifont}
\newcommand{\Letter}{\ding{41}}
\usepackage{todonotes}

\theoremstyle{plain}

\newtheorem{definition}{Definition}
\newtheorem{theorem}{Theorem}
\newtheorem{lemma}{Lemma}

\newtheorem{example}{Example}
\newtheorem{corollary}{Corollary}
\newtheorem{remark}{Remark}
\newtheorem{assumption}{Assumption}
\newtheorem{proposition}{Proposition}

\newcommand{\NN}{\mathbb{N}}

\newcommand{\PP}{\mathbb{P}}

\newcommand{\var}{\mathrm{Var}}
\newcommand{\cov}{\mathrm{Cov}}
\newcommand{\indep}{\perp\!\!\!\perp}

\newcommand{\indc}[1]{\mathds{1}\left\{#1\right\}}

\newcommand{\argmin}[1]{\underset{#1}{\arg\!\min}}

\usepackage{xspace}

\def\@#1\@{\begin{align}#1\end{align}}
\def\$#1\${\begin{align*}#1\end{align*}}

\definecolor{myblue}{rgb}{.8, .8, 1}
\definecolor{mathblue}{rgb}{0.2472, 0.24, 0.6} 
\definecolor{mathred}{rgb}{0.6, 0.24, 0.442893}
\definecolor{mathyellow}{rgb}{0.6, 0.547014, 0.24}

\newcommand{\cA}{{\mathcal{A}}}
\newcommand{\cB}{{\mathcal{B}}}

\newcommand{\cD}{{\mathcal{D}}}

\newcommand{\cF}{{\mathcal{F}}}
\newcommand{\cG}{{\mathcal{G}}}
\newcommand{\cH}{{\mathcal{H}}}

\newcommand{\cM}{{\mathcal{M}}}
\newcommand{\cN}{{\mathcal{N}}}

\newcommand{\E}{\mathbb{E}}
\newcommand{\R}{\mathbb{R}}

\newcommand{\diag}{\textnormal{diag}}

\renewcommand{\widetilde}{\tilde}
\DeclareMathOperator{\esssup}{ess\,sup}
\DeclareMathOperator{\Tr}{Tr}

\newcommand{\ras}{\xrightarrow[]{a.s.}}
\newcommand{\iid}{\textnormal{i.i.d.}}

\newcommand{\convd}{\xrightarrow{d}}
\newcommand{\convp}{\xrightarrow{p}}
\newcommand{\convdp}{\xrightarrow{d,p}}

\newcommand{\V}{\mathrm{Var}}
\newcommand{\independent}{{\perp \! \! \! \perp}}
\newcommand{\indicator}{\mathds{1}}

\newcommand{\iidsim}{\stackrel{\mathrm{i.i.d.}}{\sim}}
\newcommand{\expit}{\mathrm{expit}}

\title{Conditional independence testing through \\ marginal permutation}
\author{Jiahao Ai and Ziang Niu\textsuperscript{\Letter}}
\date{}

\begin{document}
\maketitle
\begingroup
\renewcommand{\thefootnote}{\Letter}
\footnotetext{Corresponding author, \texttt{ziangniu@wharton.upenn.edu}}
\endgroup

\begin{abstract}
Marginal permutation is exact for testing marginal independence but generally invalid for conditional independence because it breaks the dependence between the exposure ($\bm X$) and covariates ($\bm Z$). In this paper, we show that empirical studentization can nevertheless make this deliberately misspecified resampling scheme asymptotically valid for important conditional independence tests based on scores. The resulting procedures are novel in the sense that they are valid in finite samples when the exposure is marginally independent of the outcome ($\bm Y$) and covariates, and asymptotically valid under the broader conditional independence null, $\bm X\indep \bm Y\mid \bm Z$, subject to method-specific conditions. For generalized linear model scores, separately studentizing the observed and permuted scores yields validity under a correctly specified null mean despite \emph{arbitrary} conditional variance misspecification. Moreover, the permutation calibration also applies to tests based on generalized and projected covariance measures. Their validity accommodates machine learning nuisance estimates under a condition on the stability of refitting after permutation, for which we provide sufficient conditions for common regression learners. Permutation calibration preserves first-order local power while better capturing the finite-sample discreteness and skewness induced by sparse observations. Simulations and single-cell CRISPR screen analysis show substantial calibration gains with our methods when effective sample size is small.
\end{abstract}


\section{Introduction}\label{sec:intro}

\subsection{Independence testing under uncertainty about confounding}\label{subsec:motivation}

Many scientific questions reduce to testing whether a variable of interest is associated with an outcome. A recurring complication is uncertainty about confounding: factors that influence the outcome may or may not also be associated with the variable of interest, and their role can differ across comparisons. Single-cell genomics provides a prominent example.

\begin{example}[Potential confounding of perturbation effects in CRISPR screens]\label{ex:crispr-confounding}
\normalfont
Perturbation effects are assessed through associations between detected guide indicators and gene expression. Batch (including biological replicate) and sequencing depth can affect both guide detection and measured expression~\citep{barry2021sceptre,barry2024robust}. In particular, \citet{barry2024robust} show that shared variation across replicates can create an apparent association even for a non-targeting guide, although technical factors need not confound every perturbation--gene pair.
\end{example}

Similar uncertainty arises in other large-scale association studies.
In pooled microbiome studies, study origin can confound associations between a tested taxon's abundance and disease to varying degrees across taxa~\citep{gibbons2018correcting}. In genetic association scans, ancestry can likewise confound genotype--phenotype associations to varying degrees across markers~\citep{price2006principal}.

To formalize this problem, let $\bm X$ denote the variable of interest, $\bm Y$ the outcome, and $\bm Z$ the adjustment covariates. When $\bm X$ is independent of $\bm Z$, the absence of an association with $\bm Y$ conditional on $\bm Z$ is equivalent to the marginal independence (MI) null \mbox{$H_0^{\mathrm{MI}}:\bm X\indep(\bm Y,\bm Z)$}. More generally, shared dependence on $\bm Z$ can induce marginal associations under the conditional independence (CI) null \mbox{$H_0^{\mathrm{CI}}:\bm X\indep\bm Y\mid\bm Z$}. Whether the additional independence required for MI holds may be unknown and can vary across tested variables.

Marginal permutation shuffles $\bm X$ while holding $(\bm Y,\bm Z)$ fixed. Under MI, this transformation preserves the null distribution and can provide finite-sample calibration without accurate estimation of covariate effects or large-sample approximations~\citep{lehmannRomano2022}. Under CI, however, $\bm X$ may depend on $\bm Z$; shuffling then disrupts this relationship, and ordinary permutation calibration can fail. By contrast with the distribution-free MI guarantee, any test that controls finite-sample level over the entire CI null has power at most its nominal level against any alternative~\citep{shah2020hardness}. This motivates the following question:

\begin{center}
\emph{Can marginal permutation tests retain finite-sample validity under MI \\ while achieving (asymptotic) validity under a suitably structured CI null?}
\end{center}
In this paper, we set out to answer this question in the affirmative by establishing a novel general validity guarantee: for appropriately constructed and empirically studentized generalized scores, marginal permutation is asymptotically valid under CI---despite disrupting the $\bm X$--$\bm Z$ relationship---while retaining finite-sample validity under the stronger MI null.

\subsection{Contributions of this work}

Concretely, we make three main contributions.

\begin{itemize}
  \item \textbf{A validity framework for marginally permuted score statistics.}
  We formulate a framework based on generalized scores in which marginal permutation recomputes the weight for the exposure while holding the residual for the outcome fixed. Under the stronger null $H_0^{\mathrm{MI}}$, implementations that are equivariant under permutation are valid in finite samples without accuracy requirements on nuisance estimates. We establish a general theorem giving oracle moment and nuisance approximation conditions under which empirical studentization yields standard normal limits for observed and conditional permutation distributions, even though permutation destroys the association between $\bm X$ and $\bm Z$. This framework accommodates generalized linear model (GLM) scores, generalized covariance measures (GCM, \citet{shah2020hardness}), and projected covariance measures (PCM, \citet{lundborg2022projected}).
  \item \textbf{Permuted GLM score tests robust to variance misspecification.}
  By separately studentizing the observed GLM score and each marginal permutation using their respective empirical score variances, we obtain a new score test that is asymptotically valid under a correct null conditional mean model and the stated regularity conditions, without requiring CI null. Moreover, our procedure allows arbitrary misspecification of the working conditional variance. Under marginal independence, our proposed test is valid in finite samples even when the working likelihood is completely misspecified.
  \item \textbf{Doubly robust tests calibrated by permutation.}
  We develop versions of the GCM and PCM calibrated by permutation with nuisance functions estimated from the data. Under conditional independence, our theory establishes asymptotic validity, subject to the double robustness assumptions of the original GCM and PCM procedures and an additional condition on the stability of refitting after permutation. We give sufficient conditions for this stability for ordinary least squares, linear ridge regression, kernel ridge regression, and, separately, the lasso. Under marginal independence, both procedures are valid in finite samples without any accuracy requirements on nuisance estimates.
\end{itemize}

\begin{figure}[!ht]
  \centering
  \includegraphics[width=\textwidth]{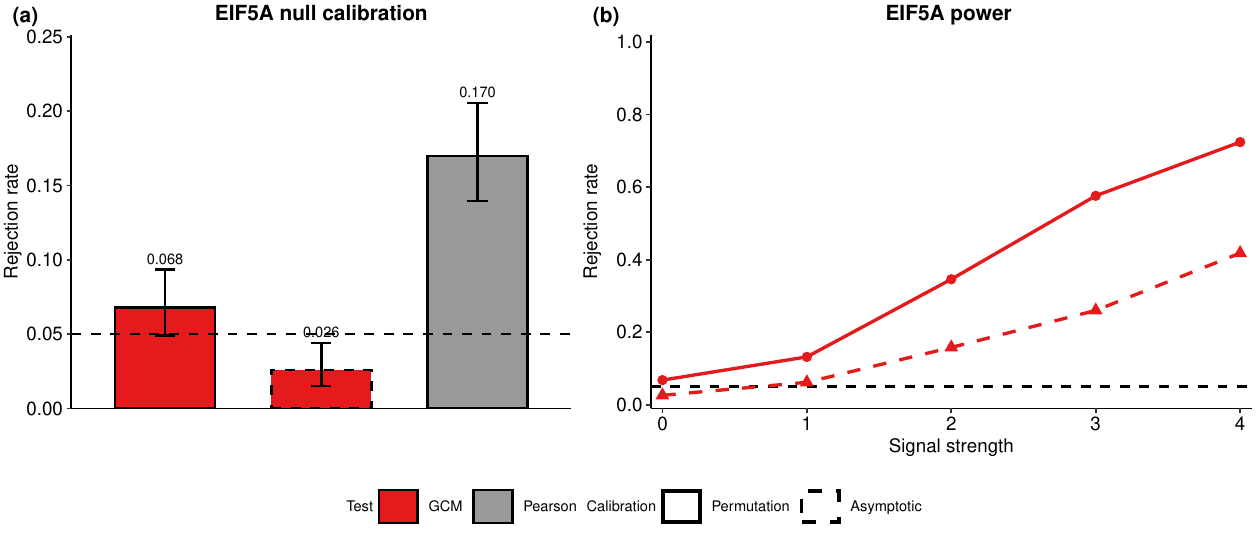}
  \caption{Semisynthetic CCLE calibration and power for \texttt{EIF5A}, a mutation target with high sparsity, at $s=10$. Panel~(a) shows null rejection with 95\% Monte~Carlo intervals; panel~(b) shows rejection over the paired signal strength grid. Pearson test appears only as an unconditional null baseline in panel~(a), and dashed black lines mark $0.05$.}
  \label{fig:ccle_eif5a_intro}
\end{figure}

For the permuted GLM, GCM, and PCM testing procedures, we further show that permutation calibration preserves first-order local power in the local alternative regimes. Simulations, a semisynthetic experiment, and a single-cell CRISPR screen analysis demonstrate substantial calibration gains when the effective sample size is small. Figure~\ref{fig:ccle_eif5a_intro} previews this behavior in the semisynthetic Cancer Cell Line Encyclopedia (CCLE) analysis and demonstrates the advantage of our proposed test.

\subsection{Related work}

Our work lies at the intersection of several strands of the literature, yet differs substantially from each in its resampling mechanism and validity guarantees.

\paragraph{Robust permutation with studentization} Studentized permutation tests are precedents for our result giving both validity guarantees. This literature uses studentization to combine finite-sample exactness under a strong null invariant under permutation with asymptotic validity for a weaker hypothesis~\citep{janssen1997studentized,chung2013exact,diciccio2017robust,wu2021randomization}. \citet{toulis2026approximate} provides a recent general treatment of approximate randomization tests, including asymptotic and nonasymptotic guarantees when unobserved invariant variables are replaced by estimated proxies such as regression residuals. \citet{barry2024robust} empirically investigated this robustness phenomenon in single-cell CRISPR screens. \citet{barry2025permuted} established the corresponding validity theory for a score test with information studentization in a specialized negative binomial model motivated by bulk RNA-seq differential expression analyses. Our contribution is a general theory for establishing the asymptotic validity of marginal permutation tests based on empirically studentized generalized scores under explicit oracle conditions (Theorem~\ref{thm:product-score-validity}). We instantiate this theory for GLM, GCM, and PCM statistics, leading to a new class of testing procedures (Sections~\ref{sec:glm-warmup} and~\ref{sec:dr}). Section~\ref{subsec:comparison-info-studentization} compares our approach with the score test using information studentization proposed by \citet{barry2025permuted}.

\paragraph{Conditional resampling and permutation methods for CI testing} Conditional resampling methods instead seek to preserve the exposure--covariate relationship. The conditional randomization tests (CRTs, \citep{candes,liu2022fast}) are exact with a known conditional exposure law, but guarantees with an estimated exposure law depend on the model or hold asymptotically~\citep{niu2023reconciling,li2023maxway}. Permutation methods instead change how values are shuffled across observations to account for $\bm Z$. Some methods make certain shuffles more likely using an estimated model of how the shuffled variable depends on the covariates, or shuffle values only among observations with similar $\bm Z$~\citep{berrett2020conditional,kim2022local}. Other methods use an assumed increasing or decreasing relationship with $\bm Z$ to choose which pairs to swap~\citep{hore2025isotonicity}. By contrast, our marginal permutations are unrestricted and deliberately break that relationship: validity under the CI null rests on the structure of the generalized scores rather than on a simulated conditional null, so no fully specified generative model for $\bm X\mid\bm Z$ is required, although the GCM and PCM still estimate conditional means or projections. Unlike the standard CRT exactness guarantee, our finite-sample MI guarantee does not require knowledge of the conditional exposure law, $\bm X\mid\bm Z$.

\paragraph{Robust GLM inference} Robust GLM inference provides a close point of comparison. Quasi-likelihood and sandwich procedures seek validity under variance or likelihood misspecification~\citep{wedderburn1974quasi,white1982maximum}. Empirical studentization of the observed score considered in this paper has important precursors in robust score tests based on empirical covariance estimates~\citep{breslow1990tests,boos1992generalized}. These earlier tests use regular parametric or quasi-likelihood nuisance estimation and asymptotic calibration. Recent sign-flip procedures provide related robustness results using resampling~\citep{hemerik2020robust,desantis2025inference}. Our GLM tests permit general higher-order moment misspecification while preserving the desirable dual validity: asymptotic validity under the specified conditional mean null together with finite-sample validity under the MI null, a novel property not shared by the earlier robust score tests.

\subsection{Organization and notation}

The remainder of the paper is organized as follows. Section~\ref{sec:method} introduces the generalized score framework, its finite-sample validity under MI, and a theorem based on approximation by oracle scores for asymptotic validity. Sections~\ref{sec:glm-warmup} and~\ref{sec:dr} develop the GLM score test robust to variance misspecification and the GCM and PCM tests with permutation calibration, respectively. Sections~\ref{sec:simulation} and~\ref{sec:realdata} evaluate these methods through simulations and genomic applications. The supplementary materials provide implementation details, additional experiments, regularity conditions, proofs, and analyses of local power.

\paragraph{Notation} We observe $n$ independent and identically distributed copies $(X_i,Y_i,Z_i)_{i=1}^n$ of a generic random triple $(\bm X,\bm Y,\bm Z)$, where $\bm X,\bm Y\in\R$ and $\bm Z\in\R^p$. Let $P_X$ and $P_Z$ denote the marginal laws of $\bm X$ and $\bm Z$, respectively, and $P_{XZ}$ their joint law. Bold type distinguishes generic random variables from unbold indexed sample entries and data objects. The data vectors are $X=(X_1,\ldots,X_n)^\top$ and $Y=(Y_1,\ldots,Y_n)^\top$, and the data matrix $Z\in\R^{n\times p}$ has rows $Z_i^\top$. We use $\cG$ to denote the symmetric group of $\{1,\ldots,n\}$ (or of the inference data for the PCM); a generic element is $\pi\in\cG$, and $\Pi$ (resp.\ $\Pi_b$) denotes a uniform random draw from $\cG$ (resp.\ the $b$-th Monte~Carlo draw), independent of the data. We use $\convd$ to denote convergence in distribution, $\convp$ for convergence in probability, and $\ras$ for almost sure convergence. Let $\cF_n$ be a sigma-field containing the observed sample, $T_n$, and any algorithmic randomness held fixed across permutations. We write $\E_\Pi$ for expectation over $\Pi$ alone, holding the observed sample and any algorithmic randomness fixed across permutations. For conditional convergence in distribution given $\cF_n$, we write
\[
W_n\mid\cF_n\convdp W
\quad\Longleftrightarrow\quad
\PP(W_n\le t\mid\cF_n)\convp\PP(W\le t),
\]
where the convergence on the right holds at every continuity point $t$ of the distribution function of $W$. For a level $\alpha\in(0,1)$, $\mathbb{Q}_{1-\alpha}(\,\cdot\mid\cF_n)$ denotes the conditional $(1-\alpha)$-quantile given the observed sample, and $\varphi_n^\alpha\in\{0,1\}$ the rejection indicator.

\section{Permutation for generalized score statistics}
\label{sec:method}\label{subsec:gen-score}

Our general testing framework is based on \emph{generalized score statistics}, encompassing a broad class of test statistics for conditional independence. Concretely, given independent observations $(X_i,Y_i,Z_i), i=1,\ldots,n$, each statistic is built from a \emph{weight} $\hat\nu_i$ (a transformation of $X_i$, possibly using a nuisance estimator learned from the data) and a \emph{response residual} $\hat\epsilon_i$ (an adjusted residual of $Y_i$ held fixed across permutations). The product $\hat R_i\equiv\hat\nu_i\,\hat\epsilon_i$ measures the signed alignment between the exposure weight and the outcome residual: positive values indicate agreement in sign, while negative values indicate opposite signs. This motivates the test statistic:
\begin{equation}\label{eq:gen-score}
  T_n\equiv\frac{\frac{1}{\sqrt n}\sum_{i=1}^n \hat R_i}{\sqrt{\widehat V_n}},\quad\text{where}\quad \widehat V_n\;\equiv\;\frac1n\sum_{i=1}^n \hat R_i^2-\Bigl(\frac1n\sum_{i=1}^n \hat R_i\Bigr)^2.
\end{equation}
We use the empirical variance operator $\mathcal{S}_n:\mathbb{R}^n\to\mathbb{R}$, defined by $\mathcal{S}_n(a)=n^{-1}\sum_{i=1}^na_i^2-(n^{-1}\sum_{i=1}^na_i)^2$. Thus $\mathcal{S}_n(\hat R)=\widehat V_n$, where
$\hat R\equiv (\hat R_1,\ldots,\hat R_n)$. The choice of $(\hat\nu_i,\hat\epsilon_i,\widehat V_n)$ specifies the test.

\subsection{Permutation calibration for generalized score test statistic} 

Permutation calibration holds the fitted response residuals $\hat\epsilon_i$ fixed and recomputes the weights $\hat\nu_i$ after shuffling $X$. In particular, for $\pi\in\cG$, let $\hat\nu_i^{(\pi)}$ denote the weight obtained by replacing $X_i$ with $X_{\pi(i)}$ and recomputing the fit involving $X$ using the procedure for the given method to obtain
\begin{align}\label{eq:gen-perm-score}
  \hat R_i^{(\pi)}&\equiv\hat\nu_i^{(\pi)}\,\hat\epsilon_i,\quad\widehat V_n^{(\pi)}\equiv\mathcal S_n\bigl(\hat R^{(\pi)}\bigr),\quad T_n^{(\pi)}\equiv\frac{n^{-1/2}\sum_{i=1}^n \hat R_i^{(\pi)}}{\sqrt{\widehat V_n^{(\pi)}}}.
\end{align}
With $\Pi$ a uniform random permutation independent of the data and $\mathbb{Q}_{1-\alpha}(T_n^{(\Pi)}\mid \cF_n)$ the conditional $(1-\alpha)$-quantile of $T_n^{(\Pi)}$ given the sample, the test with permutation calibration at level $\alpha$ rejects when
\begin{align}\label{eq:gen-perm-test}
  \varphi_n^\alpha\equiv\indc{T_n> \mathbb{Q}_{1-\alpha}(T_n^{(\Pi)}\mid \cF_n)}=1.
\end{align}
Algorithm~\ref{alg:gen} gives the generic implementation.

\begin{algorithm}[!ht]
\caption{Generalized score test calibrated by permutation}
\label{alg:gen}
\KwIn{Data $(X_i,Y_i,Z_i)_{i=1}^n$, level $\alpha\in(0,1)$, number of permutations $B\in\NN$.} Fit any nuisance estimators required by $\hat\nu_i$ and $\hat\epsilon_i$; compute $\hat R_i=\hat\nu_i\,\hat\epsilon_i$ and the observed statistic $T_n$ as in~\eqref{eq:gen-score}\;
\For{$b=1,\ldots,B$}{
  Draw a permutation $\Pi_b$ of $\{1,\ldots,n\}$\;
  Refit $X_{\Pi_b(i)}\mid Z_i$ with permuted sample $(X_{\Pi_b(i)},Z_i)_{i=1}^n$ to obtain $\hat\nu_i^{(\Pi_b)}$\;
  Form $\hat R_i^{(\Pi_b)}=\hat \nu_i^{(\Pi_b)}\hat \epsilon_i$ and the permuted statistic $T_{n}^{(\Pi_b)}$ analogously to $T_n$\;
}
Compute the (right-sided) $p$-value
$p=\frac{1}{B+1}\bigl(1+\sum_{b=1}^B\indc{T_{n}^{(\Pi_b)}\ge T_n}\bigr)$\;
Reject the null hypothesis at level $\alpha$ when $p\le\alpha$.
\end{algorithm}

\subsection{Finite-sample validity under $H_0^{\mathrm{MI}}$}

The first validity result for the permutation calibration holds at any sample size and \emph{regardless of the choice} of weight, response residual, or nuisance estimator.

\begin{proposition}[Finite-sample validity under a null invariant under permutation]
\label{prop:mi-exact}\label{subsec:mi}\label{thm:gcm-finite}\label{thm:pcm-finite}
Let $T_n$ be any measurable statistic and $\cG$ be any subgroup of permutations under whose action the joint distribution of the data is invariant. Then the test~\eqref{eq:gen-perm-test} calibrated by permutation is valid in finite samples: for every $n$ and $\alpha\in(0,1)$,
\begin{align*}
  \PP\bigl(\varphi_n^\alpha=1\bigr)\le \alpha,\qquad\PP(p\le \alpha)\le \alpha.
\end{align*}
In particular, this applies under $H_0^{\mathrm{MI}}: \bm X\indep(\bm Y,\bm Z)$ with $\cG$ the permutation group on $\{1,\ldots,n\}$.
\end{proposition}

A proof of Proposition~\ref{prop:mi-exact} can be found in~\citet{hemerik2018}. As long as the joint distribution of the data is invariant under the permutation group $\cG$ used to calibrate the test, Proposition~\ref{prop:mi-exact} shows that the permuted statistics $T_n^{(\Pi_b)}$ are exchangeable with $T_n$, and the resulting Monte~Carlo $p$-value is super-uniform. Such guarantee does not rely on \emph{any} conditions on accuracy of nuisance estimates, although inaccurate fits may reduce power.

\subsection{Asymptotic validity via oracle approximation}
\label{subsec:shared-limit}

Marginal permutation can change both the nuisance targets and the variance of the score. The following theorem gives a general criterion under which empirical studentization nevertheless yields a common standard normal limit for the observed and permuted statistics, ensuring asymptotically valid permutation calibration. These conditions are verified below under the conditional mean null for the GLM score and under conditional independence for the GCM and PCM.

\begin{theorem}[Asymptotic validity of permuted product scores]
\label{thm:product-score-validity}
Let $T_n$ and $T_n^{(\Pi)}$ be the empirically studentized statistics defined in~\eqref{eq:gen-score} and~\eqref{eq:gen-perm-score}, respectively. Suppose there exist an auxiliary sigma-field $\cH_n\subseteq\cF_n$, oracle scores $U_{n1},\ldots,U_{nn}$, and a real array $(C_{n,ik})_{1\le i,k\le n}$, measurable with respect to $\cF_n$ and satisfying $\sum_{k=1}^nC_{n,ik}=0$ for every $i=1,\ldots,n$, so that the following conditions hold.
\begin{enumerate}[(i)]
\item \textbf{Observed oracle regularity.}
Conditional on $\cH_n$, the scores $U_{n1},\ldots,U_{nn}$ are i.i.d.\ with $\E[U_{n1}\mid\cH_n]=0$ and $
\sigma_{s,n}^2\equiv\E[U_{n1}^2\mid\cH_n]$, where $\PP(\sigma_{s,n}^2>0)\rightarrow1$, and
\[
\frac{\E[U_{n1}^4\mid\cH_n]}{n\sigma_{s,n}^4}=o_\PP(1).
\]
\item \textbf{Permutation oracle regularity.} Define $\sigma_{p,n}^2\equiv\frac1{n^2}\sum_{i=1}^n\sum_{k=1}^nC_{n,ik}^2$ and suppose $\PP(\sigma_{p,n}^2>0)\rightarrow1$. Then, defining $C_{n,\cdot k}\equiv\frac1n\sum_{i=1}^nC_{n,ik}$, we have
\begin{align}\label{eq:product-score-array-conditions}
  \frac{\sum_{k=1}^nC_{n,\cdot k}^2}{(n-1)\sigma_{p,n}^2}=o_\PP(1),\quad\frac{\max_{1\le i,k\le n}C_{n,ik}^2}{n\sigma_{p,n}^2}=o_\PP(1).
\end{align}
\item \textbf{Approximation of fitted scores by oracle scores.}
Writing $\Delta_{s,ni}\equiv\hat R_i-U_{ni}$ and $\Delta_{p,ni}\equiv\hat R_i^{(\Pi)}-C_{n,i\Pi(i)}$, suppose, for $r\in\{s,p\}$,
\begin{equation}\label{eq:product-score-nuisance-transfer}
  \frac{1}{\sqrt n\,\sigma_{r,n}}\sum_{i=1}^n\Delta_{r,ni}=o_\PP(1),\qquad \frac{1}{n\sigma_{r,n}^2}\sum_{i=1}^n\Delta_{r,ni}^2=o_\PP(1).
\end{equation}
For $r=p$, these $o_\PP(1)$ statements are under the joint law of the data and the independent permutation $\Pi$.
\end{enumerate}
Then, $T_n\convd N(0,1)$ and $T_n^{(\Pi)}\mid\cF_n\convdp N(0,1)$. Consequently, $\mathbb Q_{1-\alpha}(T_n^{(\Pi)}\mid\cF_n)\convp z_{1-\alpha}$ and $\PP\!\{T_n>\mathbb Q_{1-\alpha}(T_n^{(\Pi)}\mid\cF_n)\}
\rightarrow\alpha$, for every fixed $\alpha\in(0,1)$.
\end{theorem}

Theorem~\ref{thm:product-score-validity} builds on the studentized permutation theory of \citet{chung2013exact} and \citet{diciccio2017robust}. The proof is given in Appendix~\ref{app:proof-product-score}. We now present the following discussion of the conditions in Theorem~\ref{thm:product-score-validity}.

\begin{remark}[Conditions in Theorem~\ref{thm:product-score-validity}]
  Theorem~\ref{thm:product-score-validity} gives conditions under which oracle scores approximate marginally permuted product scores with estimated nuisance functions. The two array conditions in part~(ii) ensure that column centering has a negligible effect on the permutation variance and that no individual array entry dominates. For both the observed ($r=s$) and permuted ($r=p$) statistics, the first bound in part~(iii) makes the difference between the fitted and oracle score numerators negligible relative to the oracle standard deviation $\sigma_{r,n}$. The second makes the corresponding difference in empirical variances negligible relative to the oracle variance $\sigma_{r,n}^2$. The final calibration step follows the strategy of establishing a common limit used in studentized permutation and randomization tests~\citep{janssen1997studentized,chung2013exact,diciccio2017robust,wu2021randomization} and related resampling procedures~\citep{hemerik2020robust,desantis2025inference,niu2023reconciling}.
\end{remark}

\begin{remark}[Auxiliary sigma-field $\cH_n$]
  The auxiliary sigma-field $\cH_n$ accommodates constructions using sample splitting: it may be taken to be trivial for the GLM and GCM applications below, while for the PCM it contains information from the training fold and any randomness used to fit the training models. The conditions in Theorem~\ref{thm:product-score-validity} provide a common route to validity for the GLM, GCM, and PCM procedures developed below.
\end{remark}

\section{GLM score test calibrated by permutation}
\label{sec:glm-warmup}\label{subsec:glm}

We first instantiate Theorem~\ref{thm:product-score-validity} using a working generalized linear model from an exponential family. Let $\vartheta$ denote a dispersion or shape parameter. For reference measure $\lambda$, let the working family be the regular one-parameter natural exponential family
\begin{align*}
  p_{\mathrm{nat}}(y;\omega,\vartheta)
  &\equiv \frac{h_\vartheta(y)e^{\omega y}}{\mathcal Z_\vartheta(\omega)}=h_\vartheta(y)\exp\{\omega y-A_\vartheta(\omega)\},
\end{align*}
where $\mathcal Z_\vartheta(\omega)\equiv\int e^{\omega u}h_\vartheta(u)\,d\lambda(u)$ and $A_\vartheta(\omega)\equiv\log\mathcal Z_\vartheta(\omega)$ is the log-partition function. Here $\omega$ ranges over an open natural parameter space on which $0<\mathcal Z_\vartheta(\omega)<\infty$, and we assume $A_\vartheta''(\omega)>0$. Let $\cM$ be the range of $A_\vartheta'$ and define the working density $p(y;\mu,\vartheta)\equiv p_{\mathrm{nat}}(y;(A_\vartheta')^{-1}(\mu),\vartheta)$, where $(A_\vartheta')^{-1}$ is the inverse of $A_\vartheta'$. Let $g:\cM\to\R$ be a strictly monotone differentiable link. For $\gamma\in\R$ and $\beta\in\R^p$, the working model is
\begin{equation}\label{eq:glm-family-identities}
  \bm Y\mid \bm X=x,\bm Z=z\sim p\!\left(\,\cdot\,;g^{-1}(x\gamma+z^\top\beta),\vartheta\right).
\end{equation}
The variance function induced by the working exponential family is $V_\vartheta(\mu)\equiv A_\vartheta''((A_\vartheta')^{-1}(\mu))$. The scalar coefficient $\gamma$ is the parameter of interest, while $\beta$ collects the nuisance coefficients on $\bm Z$. Under the correctly specified working model, testing conditional independence corresponds to testing $H_0^{\mathrm{CI}}:\gamma=0$~\citep{candes}. Differentiating the exponential family density parameterized by its mean gives, for $b\in\R^p$,
\begin{align*}
  s_i(b;\vartheta)\equiv \frac{Y_i-\mu_i(b)}{V_\vartheta(\mu_i(b))\,g'(\mu_i(b))}\quad\text{where}\quad \mu_i(b)\equiv g^{-1}(Z_i^\top b).
\end{align*}
The null estimator $\hat\beta$ is a measurable root of the nuisance score equation
\begin{align}\label{eq:glm-score-eq}
  U_n(b;\hat\vartheta)\equiv\frac1n\sum_{i=1}^n Z_i s_i(b;\hat\vartheta),\quad
  U_n(\hat\beta;\hat\vartheta)=0\quad\text{whenever a root exists}.
\end{align}
Define $\hat\mu_i\equiv \mu_i(\hat\beta), \hat\epsilon_i\equiv s_i(\hat\beta;\hat\vartheta)$ and $\hat W_i\equiv 1/(V_{\hat\vartheta}(\hat\mu_i)\,g'(\hat\mu_i)^2)$ and $\hat W=\diag(\hat W_1,\ldots,\hat W_n)$. The quantity $\hat W_i$ is the weight from the working Fisher information for the individual linear predictor.

\subsection{Permuted score test with working GLM likelihood}

To display the GLM fit in the generalized score template~\eqref{eq:gen-score}, we consider the residuals of the treatment against the nuisance design using the same working information weights: for $i=1,\ldots,n$,
\begin{equation}\label{eq:glm-nu}
  \hat\theta\equiv (Z^\top\hat W Z)^{-1} Z^\top\hat W X\quad\text{and}\quad\hat\nu_i^{\mathrm{GLM}}\equiv X_i-Z_i^\top\hat\theta.
\end{equation}
Define $\hat R^{\mathrm{GLM}}\equiv (\hat R_1^{\mathrm{GLM}},\ldots,\hat R_n^{\mathrm{GLM}})$, where $\hat R_i^{\mathrm{GLM}}\equiv \hat\nu_i^{\mathrm{GLM}}\hat\epsilon_i$. The GLM score statistic is therefore exactly of the form
\eqref{eq:gen-score}:
\begin{align}\label{eq:glm-Tn}
  T_n^{\mathrm{GLM}}\equiv\frac{\frac1{\sqrt n}\sum_{i=1}^n\hat R_i^{\mathrm{GLM}}}{\sqrt{\widehat V_n^{\mathrm{GLM}}}}\quad\text{where}\quad \widehat V_n^{\mathrm{GLM}}\equiv\mathcal{S}_n(\hat R^{\mathrm{GLM}}).
\end{align}
$(T_n^{\mathrm{GLM}},\widehat{V}_n^{\mathrm{GLM}})$ is the GLM score analogue of $(T_n,\widehat{V}_n)$ in~\eqref{eq:gen-score}. Algorithm~\ref{alg:glm} presents the resulting GLM score test calibrated by permutation.

\begin{algorithm}[!ht]
\caption{The GLM score permutation test}
\label{alg:glm}
\KwIn{Data $(X_i,Y_i,Z_i)_{i=1}^n$ with $Z_i\in\R^p$; level $\alpha\in(0,1)$; number of permutations $B\in\NN$; working density $p$, link $g$, and estimated dispersion or shape parameter $\hat\vartheta$ (only on data $(Y,Z)$).} Fit the null model by solving~\eqref{eq:glm-score-eq}; compute $\hat\mu_i$, $\hat\epsilon_i$ and $\hat W_i$ from the null fit\;
Compute $\hat\nu_i^{\mathrm{GLM}}$ from~\eqref{eq:glm-nu}, then compute the observed statistic $T_n^{\mathrm{GLM}}$ from~\eqref{eq:glm-Tn}\;
\For{$b=1,\ldots,B$}{
  Draw $\Pi_b$ independently and uniformly from $\cG$ (with replacement)\;
  Form $X^{(\pi)}\equiv (X_{\pi(1)},\ldots,X_{\pi(n)})^\top$ and $\hat\theta^{(\pi)}\equiv (Z^\top\hat W Z)^{-1}Z^\top\hat W X^{(\pi)}$\;
  Form $\hat R^{\mathrm{GLM},(\pi)}\equiv (\hat R_1^{\mathrm{GLM},(\pi)},\ldots,\hat R_n^{\mathrm{GLM},(\pi)})$, where $\hat R_i^{\mathrm{GLM},(\pi)}\equiv \hat\nu_i^{\mathrm{GLM},(\pi)}\hat\epsilon_i$ and 
  \begin{align}\label{eq:glm-perm-nu}
    \hat\nu_i^{\mathrm{GLM},(\pi)}\equiv X_{\pi(i)}-Z_i^\top\hat\theta^{(\pi)};
  \end{align}\\
  Compute $\widehat V_n^{\mathrm{GLM},(\Pi_b)}$ and
  $T_n^{\mathrm{GLM},(\Pi_b)}$ by 
  \begin{align}\label{eq:glm-emp-perm}
    T_n^{\mathrm{GLM},(\pi)}\equiv\frac{\frac1{\sqrt n}\sum_{i=1}^n\hat R_i^{\mathrm{GLM},(\pi)}}{\sqrt{\widehat V_n^{\mathrm{GLM},(\pi)}}}\quad\text{where}\quad \widehat V_n^{\mathrm{GLM},(\pi)}\equiv\mathcal{S}_n(\hat R^{\mathrm{GLM},(\pi)});
  \end{align}
}
Compute the $p$-value $p\equiv\frac{1}{B+1}\bigl(1+\sum_{b=1}^{B}\indc{T_{n}^{\mathrm{GLM},(\Pi_b)}\ge T_n^{\mathrm{GLM}}}\bigr)$\;
Reject $H_0^{\mathrm{CM}}$ at level~$\alpha$ when $p\le\alpha$.
\end{algorithm}

\begin{remark}[Computational shortcut in Algorithm~\ref{alg:glm}]
  Algorithm~\ref{alg:glm} treats the realized null fit $(\hat\vartheta,\hat\beta)$, response residuals $\hat\epsilon_i$, and working weights $\hat W_i$ as fixed across permutations. The matrix $(Z^\top\hat WZ)^{-1}Z^\top\hat W$ can therefore be cached, so each permutation requires only the matrix--vector update in~\eqref{eq:glm-perm-nu}, followed by the empirical variance and ratio in~\eqref{eq:glm-emp-perm}. We discuss the computational and theoretical consequences when $X$ is used to obtain estimate $\hat\vartheta$ in Remark~\ref{rmk:glm-dispersion-x}.
\end{remark}

\begin{remark}[Connection to empirical score tests]\label{rmk:glm-breslow}
The empirical studentizer in~\eqref{eq:glm-Tn} follows the empirical score
approach of \citet{breslow1990tests}; see also \citet{boos1992generalized}.
The fitted nuisance-score equation gives $\sum_i Z_i\hat\epsilon_i=0$,
and hence $\sum_i\hat R_i^{\mathrm{GLM}}=\sum_iX_i\hat\epsilon_i$.
Thus residualizing $X$ leaves the total likelihood score for $\gamma$
unchanged, while supplying contributions adjusted for nuisance estimation
to its variance estimator. This variance is estimated empirically rather
than through the working Fisher information.
Our procedure calibrates the statistic by marginal permutation, recomputing the exposure residuals and their empirical variance after each
shuffle.
\end{remark}

\subsection{Validity and robustness under the conditional mean null}
\label{subsec:glm-validity}

We assume that, for some $\beta$,
\begin{align}\label{eq:glm-null-mean}
  H_0^{\mathrm{CM}}:\quad \E[\bm Y\mid \bm X,\bm Z]=\mu_\beta(\bm Z)\equiv g^{-1}(\bm Z^\top\beta).
\end{align}
The true conditional variance $\var(\bm Y\mid\bm X,\bm Z)$ may depend on $\bm X$ and differ from the working variance $V_{\vartheta_\star}(\mu_\beta(\bm Z))$, where $\vartheta_\star$ is the deterministic limit of $\hat\vartheta$ in Assumption~\ref{assumption:glm-reg}. Write $V_Y(\bm Z)\equiv\var(\bm Y\mid\bm Z)$. Under~\eqref{eq:glm-null-mean}, set $\mu_i\equiv\mu_i(\beta), v_i\equiv V_{\vartheta_\star}(\mu_i),g_i'\equiv g'(\mu_i)$. Moreover, $\xi_i\equiv\frac{Y_i-\mu_i}{v_i g_i'}, W_i\equiv \frac{1}{v_i(g_i')^2}, S_i\equiv \var(\xi_i\mid Z_i) =V_Y(Z_i)/(v_i^2(g_i')^2)$. Thus $S_i/W_i=V_Y(Z_i)/v_i$ measures variance misspecification given $Z_i$. Let $M\equiv\E[W_iZ_iZ_i^\top]$, $r\equiv\E[W_iZ_iX_i]$ and $\theta\equiv M^{-1}r$. Then define
\begin{align}
  \delta_i&\equiv X_i-\theta^\top Z_i,
  &R_i&\equiv\sqrt{W_i}\,\delta_i,
  &\tau_s^2&\equiv\E[\delta_i^2\xi_i^2],
  &\tau_p^2&\equiv\E[\xi_i^2]=\E[S_i].
  \label{eq:glm-raw-score-variances}
\end{align}
Here $R_i$ is the population counterpart of the weighted residual in~\eqref{eq:glm-nu}, while $\tau_s^2$ and $\tau_p^2$ are the asymptotic variances of the observed and permuted raw scores, respectively. We first derive the asymptotic sampling and permutation distributions of the empirically studentized statistic $T_n^{\mathrm{GLM}}$ used by Algorithm~\ref{alg:glm}. To do so, we need the following conditions on the GLM nuisance fits.

\begin{assumption}[Conditions on GLM nuisance estimates]\label{assumption:glm-reg}
  The observations $(X_i,Y_i,Z_i)$ are i.i.d., satisfying~\eqref{eq:glm-null-mean}, and the nuisance dimension $p$ is fixed as $n\to\infty$. Moreover, suppose the following conditions hold: (i) $\hat\vartheta\to\vartheta_\star$ almost surely for a deterministic value $\vartheta_\star$; (ii) $\hat\beta$ is a root of~\eqref{eq:glm-score-eq} and $\hat\beta\to\beta$ almost surely; and (iii) $\E[X_i]=0$, $\E[X_i^2]=1$, and $\E[X_i^4]<\infty$.
\end{assumption}

\begin{proposition}[Sampling distribution of the empirically studentized GLM score]\label{prop:glm-unperm-emp}
Suppose Assumptions~\ref{assumption:glm-reg}, \ref{assumption:reg-outcome}, and \ref{assumption:reg-glm} hold, and suppose in addition that $\tau_s^2>0$. Then $T_n^{\mathrm{GLM}}\convd N(0,1)$.
\end{proposition}

\begin{theorem}[Validity of empirically studentized GLM score test under the conditional mean null]\label{thm:glm-perm-emp}
Under the assumptions of Proposition~\ref{prop:glm-unperm-emp}, conditions~(i)--(iii) of Theorem~\ref{thm:product-score-validity} hold with the following oracle choices:
\begin{align*}
  U_{ni}\equiv(X_i-Z_i^\top\theta)\xi_i,\qquad C_{n,ik}\equiv(X_k-\bar X)\xi_i,\qquad \bar X\equiv\frac1n\sum_{k=1}^nX_k.
\end{align*}
Consequently, Theorem~\ref{thm:product-score-validity} gives $T_n^{\mathrm{GLM}}\convd N(0,1)$ and $T_n^{\mathrm{GLM},(\Pi)}\mid\cF_n\convdp N(0,1)$. Then, $\mathbb{Q}_{1-\alpha}\bigl(T_n^{\mathrm{GLM},(\Pi)}\mid\cF_n\bigr)\convp z_{1-\alpha}$, and $\PP\!\bigl\{T_n^{\mathrm{GLM}}\ge\mathbb{Q}_{1-\alpha}\bigl(T_n^{\mathrm{GLM},(\Pi)}\mid\cF_n\bigr)\bigr\}\rightarrow\alpha$, for every fixed $\alpha\in(0,1)$.
\end{theorem}

Theorem~\ref{thm:glm-perm-emp} establishes asymptotic validity of the GLM score permutation test under the conditional mean null, even when the working variance function is misspecified. The proof is given in Appendix~\ref{app:proof-glm-perm-emp}.

\begin{remark}[Validity under arbitrary misspecification of the variance function]
  The empirically studentized permutation score does not require the working
  family to describe the full conditional law of $\bm Y\mid \bm X,\bm Z$. Subject to the
  conditions of Proposition~\ref{prop:glm-unperm-emp} and
  Theorem~\ref{thm:glm-perm-emp}, it is enough that the null conditional mean
  in~\eqref{eq:glm-null-mean} be correctly specified. The true conditional variance $\var(\bm Y\mid\bm X,\bm Z)$ may otherwise differ arbitrarily from the working variance $V_\vartheta(\mu_\beta(\bm Z))$. This robustness is useful for count outcomes,
  for which overdispersion or heteroskedasticity can make a fully parametric
  variance model fragile. Related sign-flip tests were
  developed by~\citet{hemerik2020robust} and extended to arbitrary
  misspecification of the variance function under a correct mean model
  by~\citet{desantis2025inference}. Our construction based on marginal permutation
  additionally inherits finite-sample validity under marginal
  independence established in Proposition~\ref{prop:mi-exact}, regardless of
  the quality of the nuisance fits. Under the
  conditional mean null, Theorem~\ref{thm:glm-perm-emp} establishes
  asymptotic validity. In the simulations with sparse exposures in
  Section~\ref{sec:nb-score-simulation}, where the effective sample size is
  small, empirically studentized permutation remains close to the nominal
  Type~I error level, whereas both sign-flip procedures reject more than
  twice as often as intended (Figure~\ref{fig:nb-score-ci}).
\end{remark}

\begin{remark}[Dependence of the dispersion estimate on $X$]\label{rmk:glm-dispersion-x}
  For the dispersion estimator, Theorem~\ref{thm:glm-perm-emp} requires
  $\hat\vartheta\to\vartheta_\star$ but does not require $\hat\vartheta$ to be
  measurable with respect to $(Y,Z)$ alone. Thus $\hat\vartheta$ may depend on
  the full observed sample, including $X$. Algorithm~\ref{alg:glm} holds the
  realized value of $\hat\vartheta$ fixed across permutations. This cached implementation remains valid in finite samples under the MI null whenever the
  dispersion estimator is invariant to relabeling the $X$ coordinates, in
  particular whenever it is a function of $(Y,Z)$ only. In general, if $X$
  enters dispersion estimation in a way that is not invariant to relabeling, retaining the finite-sample MI guarantee requires reestimating
  $\hat\vartheta$ from each permuted sample $(X^{(\pi)},Y,Z)$ using the same
  estimation rule. The corresponding null fit, response residuals, and
  working weights must then be recomputed before forming the permuted score.
\end{remark}

\begin{remark}[No first-order loss of local power]
  The robustness gained through permutation calibration does not entail a
  first-order loss of local power under correct model specification.
  Specifically, Appendix~\ref{subsec:glm-power} shows that, under $n^{-1/2}$-local
  alternatives in the GLM exposure coefficient $\gamma$, our test calibrated by permutation
  and the usual parametric score test with Gaussian calibration have the same limiting power curve. Thus, in the regular parametric regime, permutation calibration preserves the first-order local power of the classical score test.
\end{remark}

\subsection{A less robust route: studentization using working Fisher information}\label{subsec:comparison-info-studentization}

The empirical studentizer $\widehat V_n^{\mathrm{GLM}}$ does not generally coincide with the variance estimate obtained from the working Fisher information. This distinction is what permits robustness against arbitrary misspecification of the variance function. For comparison with the empirical studentizer used by Algorithm~\ref{alg:glm}, define the studentizer using working Fisher information and its associated statistic by
\begin{align}\label{eq:glm-tilde-Tn}
  \tilde T_n^{\mathrm{GLM}}\equiv\frac{\frac1{\sqrt n}\sum_{i=1}^n\hat\nu_i^{\mathrm{GLM}}\hat\epsilon_i}{\sqrt{\tilde V_n^{\mathrm{GLM}}}}\quad\text{and}\quad\tilde V_n^{\mathrm{GLM}}\equiv\frac1n\sum_{i=1}^n\hat W_i(\hat\nu_i^{\mathrm{GLM}})^2.
\end{align}
It is not hard to check that $n\tilde V_n^{\mathrm{GLM}}= X^\top\hat W X - X^\top\hat W Z ( Z^\top\hat W Z)^{-1} Z^\top\hat W X$, which is the working efficient Fisher information for $\gamma$. For $\pi\in\cG$, the permuted companion is
\begin{align}\label{eq:glm-tilde-perm}
  \quad\tilde T_n^{\mathrm{GLM},(\pi)}\equiv\frac{\frac1{\sqrt n}\sum_{i=1}^n\hat\nu_i^{\mathrm{GLM},(\pi)}\hat\epsilon_i}{\sqrt{\tilde V_n^{\mathrm{GLM},(\pi)}}}\quad\text{and}\quad \tilde V_n^{\mathrm{GLM},(\pi)}\equiv\frac1n\sum_{i=1}^n\hat W_i\bigl(\hat\nu_i^{\mathrm{GLM},(\pi)}\bigr)^2.
\end{align}
Now we present the sampling and permutation distributions of the GLM score with information studentization.

\begin{proposition}[Sampling distribution of the GLM score with information studentization]
\label{prop:glm-unperm}
Suppose Assumptions~\ref{assumption:glm-reg}, \ref{assumption:reg-outcome}, and \ref{assumption:reg-glm} hold. Then
\begin{align*}
  \tilde T_n^{\mathrm{GLM}}\convd N(0,\sigma_s^2)\quad\text{and}\quad\sigma_s^2\equiv\frac{\tau_s^2}{\E[R_i^2]}.
\end{align*}
If $\var(Y_i\mid X_i,Z_i)=v_i=V_{\vartheta_\star}(\mu_i)$ almost surely, then $\sigma_s^2=1$.
\end{proposition}

\begin{theorem}[Permutation distribution of the GLM score with information studentization]\label{thm:glm-perm}
Under the assumptions of Proposition~\ref{prop:glm-unperm}, let $\cF_n=\sigma((X_i,Y_i,Z_i)_{i=1}^n)$ and let $\Pi$ be an independent uniform permutation. Then
\begin{align*}
  \tilde T_n^{\mathrm{GLM},(\Pi)}\mid\cF_n\convdp N(0,\sigma_p^2)\quad\text{and}\quad\sigma_p^2\equiv\frac{\tau_p^2}{\E[W_i]}.
\end{align*}
Consequently, $\mathbb{Q}_{1-\alpha}\bigl(\tilde T_n^{\mathrm{GLM},(\Pi)}\mid\cF_n\bigr)\convp \sigma_p z_{1-\alpha}$. If $\var(Y_i\mid X_i,Z_i)=cv_i=cV_{\vartheta_\star}(\mu_i)$ almost surely for a constant $c>0$, then the test $\indicator\{\tilde T_n^{\mathrm{GLM}}\geq \mathbb{Q}_{1-\alpha}(\tilde T_n^{\mathrm{GLM}, (\Pi)}\mid \cF_n)\}$ has asymptotic level $\alpha$ under $H_0^{\mathrm{CM}}$.
\end{theorem}

Theorem~\ref{thm:glm-perm} extends the results in \citet{barry2025permuted} to general GLMs. Its proof is given in Appendix~\ref{app:proof-glm-perm}. We now discuss the robustness of the GLM score test with information studentization under variance misspecification.

\begin{remark}[Limited variance robustness of the GLM score with information studentization]
  For the statistic $\tilde T_n^{\mathrm{GLM}}$ with information studentization, the sampling scale $\sigma_s^2=\tau_s^2/\E[R_i^2]$ depends on the weighted residual of $\bm X$ after projection on $\bm Z$, whereas the permutation scale $\sigma_p^2=\tau_p^2/\E[W_i]$ depends only on the marginal second moment of the working response residual and the average working weight. Both scales equal $c$ when $\var(Y_i\mid X_i,Z_i)=cv_i$ almost surely for a constant $c>0$. Correct variance specification gives $c=1$ but they may differ otherwise. By contrast, the empirical studentizers estimate the relevant variance of the raw score separately: $\widehat V_n^{\mathrm{GLM}}$ estimates $\tau_s^2$ for the observed score, while $\widehat V_n^{\mathrm{GLM},(\Pi)}$ estimates the corresponding permutation variance. Proposition~\ref{prop:glm-unperm-emp} and Theorem~\ref{thm:glm-perm-emp} therefore yield standard normal sampling and permutation limits without requiring this multiplicative variance condition.
\end{remark}

\begin{figure}[!ht]
  \centering
  \includegraphics[width=\textwidth]{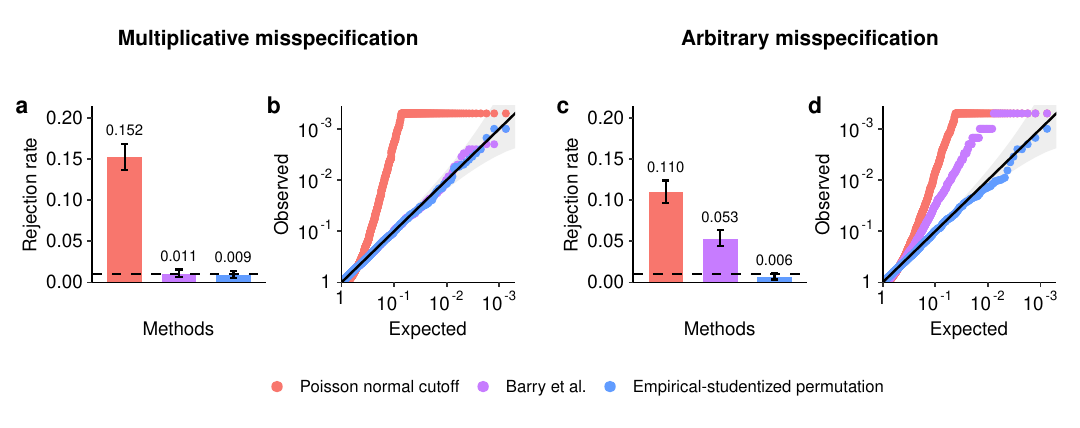}
  \caption{Variance robustness under the CI null at $\alpha=0.01$. Panels (a)--(b) use multiplicative variance misspecification, under which both permutation methods are asymptotically valid. Panels (c)--(d) go beyond the multiplicative model, under which permutation with information studentization fails but empirically studentized permutation retains asymptotic validity. The black dashed line marks the nominal level $0.01$.}
  \label{fig:glm-variance-robustness}
\end{figure}

We compare two CI designs with a binary covariate taking values $0$ and $1$.
Both use the same covariate and exposure distributions, the same conditional
outcome mean, and the same correctly specified Poisson mean model. In the
first design, the conditional outcome variance is five times the mean for
either covariate value. In the second, it is five times the mean when the
covariate is $0$ and equals the mean when it is $1$. The variance multiplier
is therefore constant in the first design and depends on the covariate in
the second. Figure~\ref{fig:glm-variance-robustness} reports simulations with $n=500$,
$2{,}000$ replicates, and $B=2{,}000$ and results align with the theoretical predictions. Appendix~\ref{app:glm-variance-simulation} gives the complete data-generating procedures.

\section{Doubly robust tests calibrated by permutation}\label{sec:dr}

The generalized score statistic~\eqref{eq:gen-score} provides a common template
for several existing doubly robust CI tests. Section~\ref{subsec:gcm}
and~\ref{subsec:pcm} develop permuted versions of the GCM of
\citet{shah2020hardness} and the PCM of~\citet{lundborg2022projected}, respectively. Throughout this section, $\cF_n$ denotes the $\sigma$-field generated by the full observed sample and any algorithmic randomness held fixed across permutations.

\subsection{GCM test calibrated by permutation}
\label{subsec:gcm}

Let $\mu_x(z)\equiv\E[\bm X\mid \bm Z=z]$ and $\mu_y(z)\equiv\E[\bm Y\mid \bm Z=z]$ and the oracle residuals $\nu_i^{\mathrm{GCM}}\equiv X_i-\mu_x(Z_i)$ and $\epsilon_i^{\mathrm{GCM}}\equiv Y_i-\mu_y(Z_i)$. The GCM test begins with flexible regression learners $\cA_x$ and $\cA_y$ to learn the conditional expectations $\hat\mu_x(z)\approx\E[\bm X\mid \bm Z=z]$ and $\hat\mu_y(z)\approx\E[\bm Y\mid \bm Z=z]$, respectively. Under $H_0^{\mathrm{CI}}$, the fitted residuals
\begin{align*}
  \widehat\nu_i^{\mathrm{GCM}}\equiv X_i-\hat\mu_x(Z_i)\quad\text{and}\quad\widehat\epsilon_i^{\mathrm{GCM}}\equiv Y_i-\hat\mu_y(Z_i)
\end{align*}
are approximately uncorrelated. The GCM test statistic is constructed based on the fitted residuals and the variance of their product: $\widehat V_n^{\mathrm{GCM}}\equiv \mathcal{S}_n(\hat R^{\mathrm{GCM}})$, where $\hat R^{\mathrm{GCM}}\equiv(\hat R_1^{\mathrm{GCM}},\ldots,\hat R_n^{\mathrm{GCM}})$ and $\hat R_i^{\mathrm{GCM}}\equiv \widehat\nu_i^{\mathrm{GCM}}\widehat\epsilon_i^{\mathrm{GCM}}$:
\begin{align}\label{eq:gcm-Tn}
  T_n^{\mathrm{GCM}}\equiv\frac{\frac{1}{\sqrt n}\sum_{i=1}^n \hat R_i^{\mathrm{GCM}}}{\sqrt{\widehat V_n^{\mathrm{GCM}}}}.
\end{align}
For the GCM, permutation changes the weight in two ways: it relabels $X_i$ as $X_{\pi(i)}$ and refits the same learner $\cA_x$ to $(X_{\pi(i)},Z_i)_{i=1}^n$. The fit $\hat\mu_y$ of $Y$ on $Z$, and hence $\widehat\epsilon_i^{\mathrm{GCM}}$, remain fixed. The permuted weight and permuted residual are defined as
\begin{align*}
  \widehat\nu_i^{\mathrm{GCM},(\pi)} \equiv X_{\pi(i)}-\hat\mu_x^{(\pi)}(Z_i)\quad\text{and}\quad\hat R_{i}^{\mathrm{GCM},(\pi)}\equiv \widehat\nu_i^{\mathrm{GCM},(\pi)}\widehat\epsilon_i^{\mathrm{GCM}}.
\end{align*}
Then, $T_n^{\mathrm{GCM},(\pi)}$ is constructed with $T_n^{(\pi)}$~\eqref{eq:gen-perm-score} with these pieces. Algorithm~\ref{alg:gcm} summarises the procedure.

\begin{algorithm}[!ht]
\caption{The GCM test calibrated by permutation}\label{alg:gcm}
\KwIn{Data $(X_i,Y_i,Z_i)_{i=1}^n$; level $\alpha\in(0,1)$; number of permutations $B\in\NN$; regression learners $\cA_x,\cA_y$.}
Apply $\cA_x$ to $(X_i,Z_i)_{i=1}^n$ and $\cA_y$ to $(Y_i,Z_i)_{i=1}^n$ to compute $T_n^{\mathrm{GCM}}$ from~\eqref{eq:gcm-Tn}\;
\For{$b=1,\ldots,B$}{
  Draw a permutation $\Pi_b$ of $\{1,\ldots,n\}$\;
  Apply $\cA_x$ to $(X_{\Pi_b(i)},Z_i)_{i=1}^n$ to obtain $\hat\mu_x^{(\Pi_b)}$ and form $\hat R_{i}^{\mathrm{GCM},(\Pi_b)}$\;
  Compute the permuted statistic $T_{n}^{\mathrm{GCM},(\Pi_b)}$ from $(\hat R_{i}^{\mathrm{GCM},(\Pi_b)})_{i=1}^n$ as in~\eqref{eq:gcm-Tn}\;
}
Compute the right-tailed $p$-value $p\equiv\frac{1}{B+1}\bigl(1+\sum_{b=1}^{B}\indc{T_{n}^{\mathrm{GCM},(\Pi_b)}\ge T_n^{\mathrm{GCM}}}\bigr)$\;
Reject $H_0^{\mathrm{CI}}$ at level~$\alpha$ when $p\le\alpha$.
\end{algorithm}

\begin{remark}[Permutation target of the refit $\hat\mu_x^{(\Pi_b)}$]
  The only difference from \citet{shah2020hardness}'s original GCM procedure is that the asymptotic cutoff is replaced by permutation calibration in Algorithm~\ref{alg:gcm}. The permutation refit $\hat \mu_x^{(\Pi_b)}$ targets the constant function $\E[\bm X]$. Conditional on the observed sample, its finite-sample permutation analogue is $\sum_{i=1}^n X_i/n$.
\end{remark}

\paragraph{Validity under CI null}
For the permuted GCM test, define the empirical mean squared errors for the outcome and exposure regressions $E_{n,y}\equiv\frac1n\sum_{i=1}^n\bigl(\mu_y(Z_i)-\hat\mu_y(Z_i)\bigr)^2$ and $E_{n,x}\equiv\frac1n\sum_{i=1}^n\bigl(\mu_x(Z_i)-\hat\mu_x(Z_i)\bigr)^2$, and the error after refitting to permuted data:
\begin{align*}
  \tilde E_{n,x}\equiv\frac1n\sum_{i=1}^n\bigl(\hat\mu_x^{(\Pi)}(Z_i)-\bar X\bigr)^2.
\end{align*}

\begin{assumption}[GCM nuisance consistency]\label{assumption:unperm}
Under the GCM conditional independence null, assume $E_{n,y}=o_\PP(1)$, $E_{n,x}=o_\PP(1)$, and $nE_{n,y}E_{n,x}=o_\PP(1)$.
\end{assumption}

\begin{assumption}[GCM permutation stability]\label{assumption:perm}
Under the GCM conditional independence null, assume $\tilde E_{n,x}=o_\PP(1)$ and $nE_{n,y}\tilde E_{n,x}=o_\PP(1)$.
\end{assumption}

\begin{theorem}[Asymptotic validity of the permuted GCM test]\label{thm:gcm-valid}
Fix $\alpha\in(0,1)$ and consider the GCM test of Algorithm~\ref{alg:gcm}, under $H_0^{\mathrm{CI}}$, Assumptions~\ref{assumption:unperm} and~\ref{assumption:perm}, the shared outcome condition Assumption~\ref{assumption:reg-outcome}, and Assumption~\ref{assumption:reg-gcm}. Take
\begin{align*}
  U_{ni}\equiv\{X_i-\mu_x(Z_i)\}\{Y_i-\mu_y(Z_i)\},\quad C_{n,ik}\equiv(X_k-\bar X)\{Y_i-\mu_y(Z_i)\}.
\end{align*}
Let $T_n=T_n^{\mathrm{GCM}}$, and let $T_n^{(\Pi)}$ be its permuted version under an independent uniform permutation, computed as in Algorithm~\ref{alg:gcm}.
The stated assumptions verify conditions~(i)--(iii)
of Theorem~\ref{thm:product-score-validity} for the displayed oracle products and array. Thus
\begin{align*}
  T_n\convd N(0,1)\quad\text{and}\quad T_n^{(\Pi)}\mid\cF_n\convdp N(0,1).
\end{align*}
Consequently, $\PP(\varphi_n^\alpha=1)\to\alpha$ under the CI null.
\end{theorem}

The proof is given in Appendix~\ref{app:proof-gcm-valid}. \citet[Theorem~6(i)]{shah2020hardness} establish asymptotic normality of the original GCM under a condition on the product of nuisance error rates and conditions on nuisance errors weighted by variance. Assumption~\ref{assumption:unperm} retains the same requirement on the product of error rates; its consistency conditions and our bounded conditional moment assumptions imply the original conditions on weighted errors. Theorem~\ref{thm:gcm-valid} additionally establishes conditional permutation normality under Assumption~\ref{assumption:perm} and the stated auxiliary regularity, thereby justifying permutation calibration under CI. This high-level condition is verified in a range of estimators including ordinary least squares, linear ridge regression, kernel ridge regression, and the lasso under their respective conditions, in Appendix~\ref{app:proof-perm-stab}.

\subsection{PCM test calibrated by permutation}
\label{subsec:pcm}

The projection covariance measure (PCM) of \citet{lundborg2022projected} was originally proposed to test whether a variable $\bm X$ is explanatory for an outcome $\bm Y$ given covariates $\bm Z$. We formulate this problem as the conditional independence null $H_{0}^{\mathrm{CI}}: \bm Y\indep \bm X\mid \bm Z$. PCM starts by partitioning the full sample $(X_i,Y_i,Z_i)_{i=1}^{n}$ into two folds $\cD_1\cup\cD_2$ of equal size $n/2$. The procedure then has two stages.

\paragraph{Training with $\cD_2$} On the training fold $\cD_2$, fit $\hat m_p(x,z)\approx\E[\bm Y\mid \bm X=x, \bm Z=z]$ by any regression learner $\mathcal{A}_p$. Conditional on $\cD_2$, define the ordinary conditional projection $q(z)\equiv\E[\hat m_p(\bm X,z)\mid \bm Z=z,\cH_n]$. Regressing $\hat m_p(X_i,Z_i)$ on $Z_i$ over $\cD_2$ gives $\hat m_q$ with any regression learner $\mathcal{A}_q$, which is intended to estimate $q$. Together the two fits on the training fold define the learned direction $\hat f(x,z)\equiv\hat m_p(x,z)-\hat m_q(z)$, which approximates the part of $\E[\bm Y\mid \bm X,\bm Z]$ that depends on $\bm X$ after adjusting for $\bm Z$.

\paragraph{Testing with $\cD_1$} Treating $\hat m_p$ and $\hat m_q$ as fixed, the inference fold $\cD_1$ produces two further regressions. First, regressing $Y$ on $Z$ gives $\hat\mu_y(z)\approx\mu_y(z)\equiv\E[\bm Y\mid \bm Z=z]$ with learner $\mathcal{A}_y$. Second, regressing $\hat f(X_i,Z_i)$ on $Z_i$ gives $\hat\mu_f(z)\approx\mu_f(z)\equiv\E[\hat f(\bm X,\bm Z)\mid \bm Z=z,\cH_n]$ with learner $\mathcal{A}_f$. Equivalently, $\mu_f(z)=q(z)-\hat m_q(z)$. Conditional on the training fold, define the oracle weight and response residual by $\nu_i^{\mathrm{PCM}}\equiv\hat f(X_i,Z_i)-\mu_f(Z_i)$ and $\epsilon_i^{\mathrm{PCM}}\equiv Y_i-\mu_y(Z_i)$, respectively. Their fitted counterparts on $\cD_1$ are defined as
\begin{align}
  \widehat\nu_i^{\mathrm{PCM}}\equiv\hat f(X_i,Z_i)-\hat\mu_f(Z_i) \quad\text{and}\quad \widehat\epsilon_i^{\mathrm{PCM}}\equiv Y_i-\hat\mu_y(Z_i).
\end{align}
The PCM instantiates the template~\eqref{eq:gen-score} with these fitted quantities and $\widehat V_n^{\mathrm{PCM}}\equiv\mathcal{S}_{n/2}(\hat R^{\mathrm{PCM}})$, where $\hat R^{\mathrm{PCM}}\equiv(\hat R_i^{\mathrm{PCM}})_{i\in\cD_1}$ and $\hat R_i^{\mathrm{PCM}}\equiv\widehat\nu_i^{\mathrm{PCM}}\widehat\epsilon_i^{\mathrm{PCM}}$. The PCM test statistic is then defined as
\begin{equation}\label{eq:pcm-Tn}
  T_n^{\mathrm{PCM}}\equiv\frac{\frac{1}{\sqrt{n/2}}\sum_{i\in\cD_1}\hat R_i^{\mathrm{PCM}}}{\sqrt{\widehat V_n^{\mathrm{PCM}}}}.
\end{equation}

Let $\cG$ denote the symmetric group of $\cD_1$, and let $\Pi\in\cG$ be a uniform random permutation independent of the data. Estimates $\hat m_p,\hat m_q$ depend only on fold $\cD_2$, which is unaffected by permutation in $\cG$, so $\widehat\epsilon_i^{\mathrm{PCM}}$ is fixed across permutations. Only the projection learner $\hat\mu_f$ is refit: recompute the learned direction at each permuted point, $\hat f(X_{\pi(i)},Z_i)=\hat m_p(X_{\pi(i)},Z_i)-\hat m_q(Z_i)$, for $i\in\cD_1$. Then obtain $\hat\mu_f^{(\pi)}$ by regressing $\hat f(X_{\pi(i)},Z_i)$ on $Z_i$ over $\cD_1$ using the same learner $\mathcal{A}_f$. The permuted fitted weight is
\begin{align*}
  \widehat\nu_i^{\mathrm{PCM},(\pi)}
  &\equiv\hat f(X_{\pi(i)},Z_i)-\hat\mu_f^{(\pi)}(Z_i),
  &
  \hat R_i^{\mathrm{PCM},(\pi)}
  &\equiv\widehat\nu_i^{\mathrm{PCM},(\pi)}\widehat\epsilon_i^{\mathrm{PCM}}.
\end{align*} 
The statistic $T_{n}^{\mathrm{PCM},(\pi)}$ is the template's $T_n^{(\pi)}$~\eqref{eq:gen-perm-score} formed from $\widehat\nu_i^{\mathrm{PCM},(\pi)}\,\widehat\epsilon_i^{\mathrm{PCM}}$ on fold $\cD_1$. Algorithm~\ref{alg:vpcm} summarises the procedure.

\begin{algorithm}[!ht]
\caption{The PCM test calibrated by permutation}
\label{alg:vpcm}
\KwIn{Data $(X_i,Y_i, Z_i)_{i=1}^{n}$; level
$\alpha\in(0,1)$; number of permutations $B\in\NN$; regression learners $\mathcal{A}_p,\mathcal{A}_q,\mathcal{A}_y,\mathcal{A}_f$ for $\hat m_p,\hat m_q,\hat\mu_y,\hat\mu_f$.}
Partition the sample into $\cD_1\cup\cD_2$ with $|\cD_1|=|\cD_2|=n/2$\;
On $\cD_2$: regress $Y$ on $(X,Z)$ to obtain $\hat m_p$, and $\hat m_p(X_i,Z_i)$ on $Z_i$ to obtain $\hat m_q$\;
On $\cD_1$: compute $\hat f(x,z)=\hat m_p(x,z)-\hat m_q(z)$; regress $Y$ on $Z$ to obtain $\hat\mu_y$, and regress $\hat f(X_i,Z_i)$ on $Z_i$ to obtain $\hat\mu_f$; compute $T_n^{\mathrm{PCM}}$ from~\eqref{eq:pcm-Tn}\;
\For{$b=1,\ldots,B$}{
  Draw an independent uniform permutation $\Pi_b$ of $\cD_1$\; 
  Apply learner $\mathcal{A}_f$ to refit the projection regression of $\hat f(X_{\Pi_b(i)},Z_i)$ on $Z_i$ to obtain $\hat\mu_f^{(\Pi_b)}$, with $\hat f(X_{\Pi_b(i)},Z_i)=\hat m_p(X_{\Pi_b(i)},Z_i)-\hat m_q(Z_i)$ for $i\in\cD_1$\;
  Compute the permuted statistic $T_n^{\mathrm{PCM},(\Pi_b)}$ from $(\hat R_i^{\mathrm{PCM},(\Pi_b)})_{i\in\cD_1}$ as in~\eqref{eq:pcm-Tn}\;
}
Compute the $p$-value $p\equiv\frac{1}{B+1}\bigl(1+\sum_{b=1}^{B}\indc{T_{n}^{\mathrm{PCM},(\Pi_b)}\ge T_{n}^{\mathrm{PCM}}}\bigr)$\;
Reject $H_{0}^{\mathrm{CI}}$ at level~$\alpha$ when $p\le\alpha$.
\end{algorithm}

\begin{remark}[Permutation target of the projection learner]\label{rmk:perm-target}
  Let $\tilde{\bm X}$ be an independent draw from the marginal law of $\bm X$, independent of $(\bm Z,\bm Y)$. For the permuted PCM test (Algorithm~\ref{alg:vpcm}), the permutation refit targets $b^{\mathrm{perm}}(z)$, where
  \[
  b^{\mathrm{perm}}(z)
  \equiv \bar q(z)-\hat m_q(z)
  =\E[\hat f(\tilde{\bm X},z)\mid\cH_n],
  \qquad
  \bar q(z)\equiv\E[\hat m_p(\tilde{\bm X},z)\mid\cH_n].
  \]
  Thus the population permutation target is obtained by evaluating the learned direction under the product law $P_X\otimes P_Z$.
\end{remark}

\paragraph{Validity} Let $n_\star=|\cD_1|=n/2$.
For the PCM, if the fits from the training stage $\hat m_p$ and $\hat m_q$ are randomized, their realized algorithmic randomness is treated as part of the training information and held fixed in all subsequent conditional arguments.  Let $\cH_n$ be the sigma-field generated by the training fold $\cD_2$ and any randomness used in fitting during training held fixed thereafter. Conditional on this training information, write
\[
\mu_f(z)\equiv\E[\hat f(\bm X,\bm Z)\mid \bm Z=z,\cH_n],
\qquad
\sigma_{f,n}^2
\equiv\E\!\left[\{\hat f(\bm X,\bm Z)-\mu_f(\bm Z)\}^2\mid\cH_n\right].
\]
This is the same variance of the learned direction used to normalize the error in the projection regression in the original PCM analysis~\citep[Eq.~(12)]{lundborg2022projected}.  Importantly, $\sigma_{f,n}^2$ is allowed to converge to zero under the null.  On the inference fold, define
\[
E_{n,y}^{\mathrm{PCM}}
\equiv\frac1{n_\star}\sum_{i\in\cD_1}\{\mu_y(Z_i)-\hat\mu_y(Z_i)\}^2,
\qquad
E_{n,f}^{\mathrm{PCM}}
\equiv\frac1{n_\star}\sum_{i\in\cD_1}\{\mu_f(Z_i)-\hat\mu_f(Z_i)\}^2.
\]
Recall from Remark~\ref{rmk:perm-target} that, if $\tilde{\bm X}$ is an independent draw from the marginal law of $\bm X$, independent of $\bm Z$, then $b^{\mathrm{perm}}(z)=\E[\hat f(\tilde{\bm X},z)\mid\cH_n]$. The permutation analogue of the variance of the learned direction is therefore the quantity under the product law $\sigma_{f,n,\mathrm{perm}}^2\equiv\E[\{\hat f(\tilde{\bm X},\bm Z)-b^{\mathrm{perm}}(\bm Z)\}^2\mid\cH_n]$, where, conditional on $\cH_n$, $\tilde{\bm X}\sim P_X$ and $\bm Z\sim P_Z$ are independent.  Thus $\sigma_{f,n}^2$ and $\sigma_{f,n,\mathrm{perm}}^2$ measure the same learned direction under $P_{XZ}$ and $P_X\otimes P_Z$, respectively.

For the conditional permutation argument we also retain the exact finite-sample permutation target.  For $i,k\in\cD_1$, let $B_{ik}\equiv\hat f(X_k,Z_i)$ and $B_{i\cdot}\equiv\frac1{n_\star}\sum_{k\in\cD_1}B_{ik}$. Then $B_{i\cdot}=\E_\Pi[\hat f(X_{\Pi(i)},Z_i)\mid\cF_n]$ exactly.  Accordingly, define the error after refitting the projection regression by
\[
\tilde E_{n,f}^{\mathrm{PCM}}
\equiv\frac1{n_\star}\sum_{i\in\cD_1}
\{\hat\mu_f^{(\Pi)}(Z_i)-B_{i\cdot}\}^2.
\]

\begin{assumption}[PCM nuisance conditions]\label{assumption:pcm-unperm}
Under the PCM conditional independence null, conditional on $\cH_n$, assume
\begin{equation}\label{eq:pcm-nuisance-normalized}
E_{n,y}^{\mathrm{PCM}}=o_\PP(1),\qquad
\frac{E_{n,f}^{\mathrm{PCM}}}{\sigma_{f,n}^2}=o_\PP(1),\qquad
\frac{n_\star E_{n,y}^{\mathrm{PCM}}E_{n,f}^{\mathrm{PCM}}}{\sigma_{f,n}^2}=o_\PP(1).
\end{equation}
The ratios are considered on $\{\sigma_{f,n}^2>0\}$, whose probability tends to one by Assumption~\ref{assumption:reg-pcm}.
\end{assumption}

\begin{assumption}[Stability of PCM refitting after permutation]\label{assumption:pcm-perm}
Under the PCM conditional independence null, assume
\begin{equation}\label{eq:pcm-perm-stability-normalized}
\frac{\tilde E_{n,f}^{\mathrm{PCM}}}{\sigma_{f,n,\mathrm{perm}}^2}=o_\PP(1),\qquad
\frac{n_\star E_{n,y}^{\mathrm{PCM}}\tilde E_{n,f}^{\mathrm{PCM}}}{\sigma_{f,n,\mathrm{perm}}^2}=o_\PP(1).
\end{equation}
The ratios are considered on $\{\sigma_{f,n,\mathrm{perm}}^2>0\}$, whose probability tends to one by Assumption~\ref{assumption:reg-pcm}.
\end{assumption}

\begin{theorem}[Asymptotic validity of the permuted PCM test]\label{thm:pcm-valid}
Fix $\alpha\in(0,1)$ and consider the PCM test of Algorithm~\ref{alg:vpcm}, under $H_0^{\mathrm{CI}}$, Assumptions~\ref{assumption:pcm-unperm}, \ref{assumption:pcm-perm}, the shared outcome condition Assumption~\ref{assumption:reg-outcome}, and Assumption~\ref{assumption:reg-pcm}. Conditional on $\cH_n$, take $U_{ni}\equiv\{\hat f(X_i,Z_i)-\mu_f(Z_i)\}\{Y_i-\mu_y(Z_i)\}$ and
\begin{align*}
  C_{n,ik}\equiv\left\{\hat f(X_k,Z_i)-\frac1{n_\star}\sum_{\ell\in\cD_1}\hat f(X_\ell,Z_i)\right\}\{Y_i-\mu_y(Z_i)\},
\end{align*}
for $i,k\in\cD_1$, where $n_\star=|\cD_1|$ and
$\mu_f(z)=\E[\hat f(\bm X,\bm Z)\mid \bm Z=z,\cH_n]$.
Let $T_n=T_n^{\mathrm{PCM}}$, and let $T_n^{(\Pi)}$ be its permuted version under an independent uniform permutation, computed as in Algorithm~\ref{alg:vpcm}.
The stated assumptions verify conditions~(i)--(iii)
of Theorem~\ref{thm:product-score-validity} for the displayed oracle products and array, with sample size $n_\star$. Thus
\begin{align*}
  T_n\convd N(0,1)\quad\text{and}\quad T_n^{(\Pi)}\mid\cF_n\convdp N(0,1).
\end{align*}
Consequently, $\PP(\varphi_n^\alpha=1)\to\alpha$ under the CI null.
\end{theorem}

The proof is given in Appendix~\ref{app:proof-pcm-valid}. Assumption~\ref{assumption:pcm-unperm} retains the conditions on normalized nuisance errors and their products from~\citet[Assumption~3(a)]{lundborg2022projected}, allowing $\sigma_{f,n}^2$ to vanish. The proof verifies the observed Lyapunov condition and the conditions on the permutation array of Theorem~\ref{thm:product-score-validity} directly. Assumption~\ref{assumption:pcm-perm} and the bound under the product law in Assumption~\ref{assumption:reg-pcm}(iii) control the errors in the fitted scores after permutation, while allowing $\sigma_{f,n,\mathrm{perm}}^2$ to vanish. Moreover, the permutation stability assumption is instantiated for a range of estimators including ordinary least squares, linear ridge regression, kernel ridge regression, and the lasso under their respective conditions, in Appendix~\ref{app:proof-perm-stab}.

\subsection{Further discussion}\label{subsec:dr-discussion}

We present several further remarks on the permuted GCM and PCM tests.

\begin{remark}[Finite-sample validity under strong null]
  Under the strong null $H_0^{\mathrm{MI}}:\bm X\indep(\bm Y,\bm Z)$, Proposition~\ref{prop:mi-exact} gives finite-sample validity for both permutation tests without assumptions on nuisance consistency, products of nuisance error rates, or stability of refitting after permutation. For PCM, this conclusion holds conditional on the training fold and its fitting randomness, with permutations restricted to the inference fold.
\end{remark}

\begin{remark}[No first-order loss of local power]
  The additional robustness supplied by permutation calibration does not entail a first-order loss of local power. For the GCM, Theorem~\ref{thm:gcm-power} shows that, under the $n^{-1/2}$-local alternatives studied in Appendix~\ref{subsec:gcm-power}, permutation and Gaussian calibration yield the same limiting power curve. Likewise, Theorem~\ref{thm:pcm-power} shows that, under the local linear alternatives studied in Appendix~\ref{subsec:pcm-power}, permutation calibration preserves the original PCM's power curve determined by a mixture of two Gaussian limiting distributions~\citep{lundborg2022projected}. Thus, in both cases, permutation calibration retains the first-order local power of the corresponding original procedure. The simulations in Section~\ref{sec:sim-additive} provide complementary finite-sample evidence: permutation calibration improves null calibration while preserving the sensitivity of GCM and PCM to their respective alternatives.
\end{remark}

\begin{remark}[Choosing the permutation direction]\label{rmk:perm-direction}
Under $H_0^{\mathrm{CI}}$, $\bm X\indep\bm Z$ implies $\bm X\indep(\bm Y,\bm Z)$, while $\bm Y\indep\bm Z$ implies $\bm Y\indep(\bm X,\bm Z)$. Thus the permutation direction can be chosen in advance according to which variable is more plausibly independent of $\bm Z$: permute $\bm X$ in the former case, or apply the procedure with $\bm X$ and $\bm Y$ interchanged in the latter. Proposition~\ref{prop:mi-exact} gives finite-sample validity whenever the corresponding MI null holds.
\end{remark}

\begin{remark}[Computational cost of permutation refitting]
  Permutation refitting is the main computational expense for Algorithms~\ref{alg:gcm}-\ref{alg:vpcm}: each draw ordinarily requires refitting the regression for permuted GCM and PCM tests. For refit learners that are affine in their response, refitting can be accelerated. Appendix~\ref{sec:practical} develops this shortcut and an adaptive scheme that draws only as many permutations as the decision requires leveraging the sequential testing framework of~\citet{besag1991sequential,fischer-barry-ramdas-EJS}.
\end{remark}

\section{Simulations}\label{sec:simulation}

We present two simulation studies. Section~\ref{sec:nb-score-simulation}
examines score tests calibrated by permutation for negative binomial outcomes,
while Section~\ref{sec:sim-additive} considers doubly robust tests calibrated by permutation with flexible nuisance regression.

\subsection{CI tests for negative binomial outcomes}
\label{sec:nb-score-simulation}

This simulation is motivated by pooled single-cell CRISPR screens, where negative binomial regression is commonly used to test whether a CRISPR perturbation indicator $\bm X$ is associated with the expression of a target gene $\bm Y$ after adjustment for observed technical covariates $\bm Z$~\citep{barry2021sceptre,barry2024robust,niu2024-spa}. The corresponding null hypothesis can be expressed as conditional independence, $\bm X\indep \bm Y\mid \bm Z$.

\begin{figure}[!ht]
  \centering
  \includegraphics[width=\textwidth]{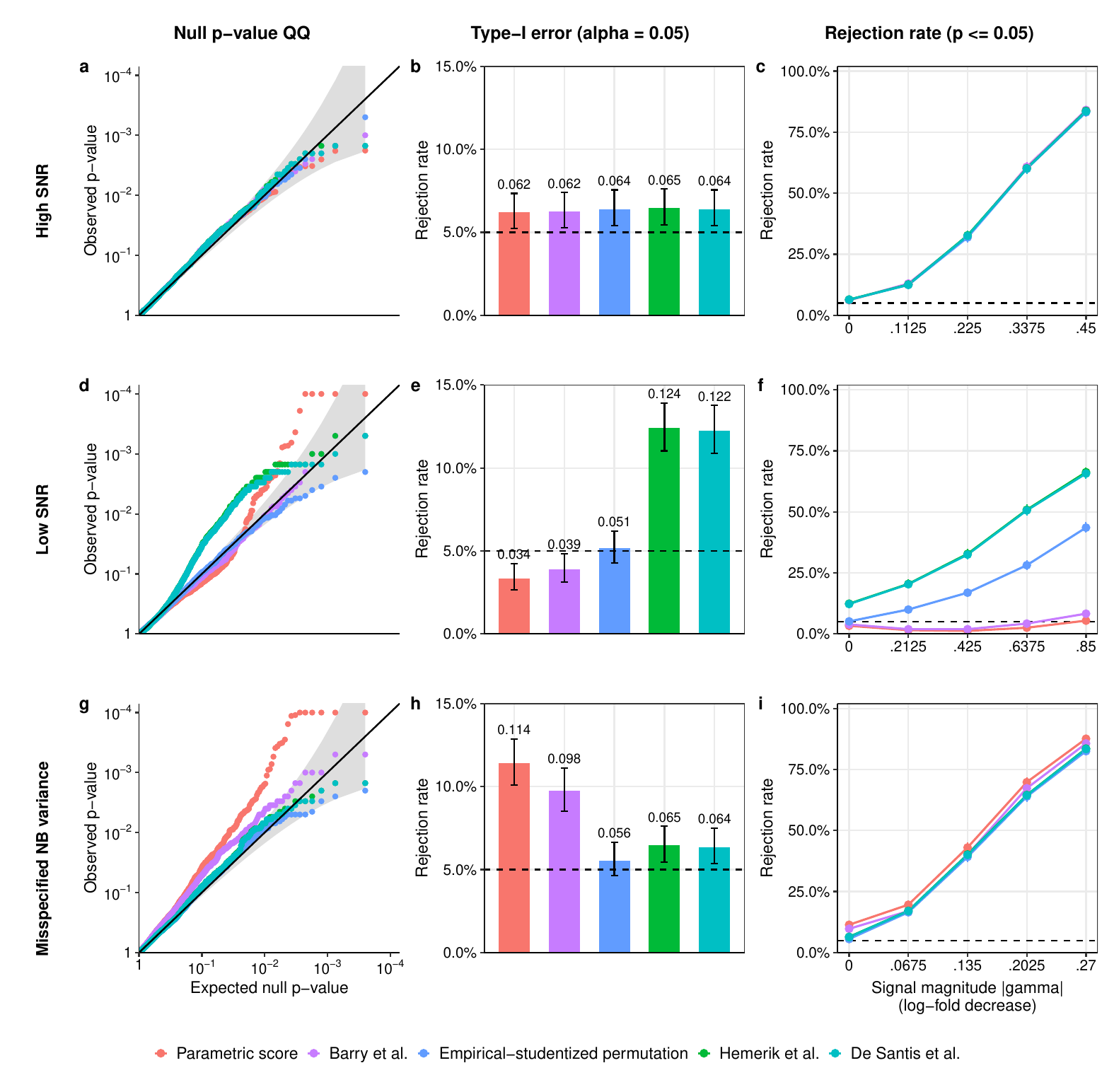}
  \caption{Score calibration under conditional independence. Rows show high SNR, low SNR, and misspecified negative binomial variance; columns show null $p$-value QQ plots, Type~I error, and rejection rate at the nominal cutoff. Signal zero is the null. Black dashed lines mark $0.05$.}
  \label{fig:nb-score-ci}
\end{figure}

\paragraph{Data-generating process}
Suppose $\bm Z\sim N(0,1),\pi(\bm Z)=\expit(a+\delta \bm Z)$ and $\mu(\bm X,\bm Z)=10\exp(0.5\bm Z+\gamma \bm X)$, then we consider the following data-generating process:
\begin{align*}
  \bm X\mid \bm Z\sim\operatorname{Ber}(\pi(\bm Z)),\qquad\bm Y\mid \bm X,\bm Z\sim\operatorname{NB}\{\mu(\bm X,\bm Z),\theta(\bm Z)\}.
\end{align*}
The size function $\theta(\cdot)$ varies by design, and the outcome variance is
$\V(\bm Y\mid \bm X,\bm Z)=\mu+\mu^2/\theta(\bm Z)$. Setting $\delta=0$
gives the marginal independence, whereas the CI design uses $\delta=0.5$. The intercept $a$ controls the prevalence of $\bm X$. Putting $p(a)\equiv \E[\expit(a+0.5\bm Z)]$, we specific the parameters $(a,\theta(\cdot))$ in the following three settings: (1) \textbf{High SNR}, with $a=p^{-1}(0.5)$ and $\theta(\bm Z)=0.5$; (2) \textbf{Low SNR}, with $a=p^{-1}(0.025)$ and $\theta(\bm Z)=0.5$; (3) \textbf{Variance misspecification}, with $a=p^{-1}(0.14)$ and $\theta(\bm Z)=46\exp(-2.1\bm Z)$. The latter two settings are used to isolate two finite-sample challenges: sparse perturbation assignments and misspecification of the working negative binomial variance model, two common features of single-cell CRISPR screens.

Each setting uses $n=400$ independent observations in each of $2{,}000$ Monte~Carlo replicates and $1{,}999$
resampling draws. Appendix~\ref{app:nb-score-mi} reports the signal grids for
$\gamma$ and the remaining Monte~Carlo settings.

\paragraph{Methods compared}
All five procedures fit the working model by maximum likelihood using the R package \texttt{MASS}~\citep{venables2002mass}. In the designs with high and low SNR, the null mean and size are estimated by negative binomial regression of $Y$ on $Z$ using the package's \texttt{glm.nb} function. We compare the standard parametric score test, permutation with information studentization~\citep{barry2025permuted}, empirically studentized permutation (ours), the robust sign-flipping test~\citep{hemerik2020robust}, and its standardized version~\citep{desantis2025inference}.

\paragraph{Results interpretation}

Figure~\ref{fig:nb-score-ci} reports null calibration and rejection rates under CI. The corresponding MI results appear in Appendix~\ref{app:nb-score-mi}. Under the CI null with high SNR, all five methods behave similarly and control Type~I error reasonably well, with indistinguishable power. Exposure sparsity (i.e., the degree of sparsity in $\bm X$) sharply separates the methods in the design with low SNR. Our method (empirically studentized permutation) remains near nominal, whereas both sign-flip procedures reject more than twice as often as intended. Such failure is mainly due to resampling mechanism: sign flipping creates symmetric resampling distribution even when the sampling distribution is highly skewed. Finally, negative binomial dispersion that depends on the covariates produces the complementary failure pattern: the parametric test and the test with information studentization reject too often because studentization using working Fisher information no longer aligns the sampling and permutation scales, whereas empirical studentization remains close to nominal.

Overall, our method shows the highest power among methods which control Type~I error.

\subsection{CI tests with complex nuisance structures}\label{sec:sim-additive}

This simulation compares GCM and PCM procedures with permutation or asymptotic calibration in continuous regression, with dCRT~\citep{liu2022fast,niu2023reconciling} included as a conditional resampling benchmark.

\paragraph{Data-generating process}
The data-generating process separates marginal independence from conditional independence and include alternatives chosen to emphasize the complementary sensitivities of the GCM and PCM statistics. Let $s(z)=\sin(2\pi z)$, and let $\bm Z_1,\varepsilon\sim N(0,1)$ be independent. We consider three designs: (1) \textbf{Marginal independence}: $\bm\xi\sim N(0,1)$, $\bm X=\bm \xi$, and $\bm Y=s(\bm Z_1)+\gamma \bm X^2+\bm\varepsilon$; (2) \textbf{Additive Gaussian}: $\bm \xi\sim N(0,1)$, $\bm X=s(\bm Z_1)+\bm\xi$, and $\bm Y=s(\bm Z_1)+\gamma \bm X^2+\bm \varepsilon$; and (3) \textbf{Heteroscedastic}: $\bm \xi+1\sim\operatorname{Exp}(1)$, $\bm X=s(\bm Z_1)(1-\bm \xi)$, and $\bm Y=s(\bm Z_1)+\gamma \bm X+\bm \varepsilon$. The latter two satisfy CI with $\bm X$ dependent on $\bm Z$. The sample size, signal grids, and replication count are reported in Appendix~\ref{app:sim-additive-parameters}.

\paragraph{Methods compared} We consider PCM and GCM procedures with permutation or asymptotic calibration. These four procedures use random forest nuisance regression and the adaptive permutation procedure in Algorithm~\ref{alg:adaptive}, with $B_{\max}=499$ draws and adaptive threshold $h=15$. We additionally consider in-sample dCRT~\citep{niu2023reconciling}. The resampling law is set to be $N\{\hat\mu_x(\bm Z),\hat\sigma_x^2\}$, where $\hat \mu_x(\cdot)$ is obtained from random forest prediction, and $\hat\sigma_x^2$ is the empirical variance, $\sum_{i=1}^n (X_i - \hat\mu_x)^2/(n-1)$. We consider $499$ resamples here. The GCM and dCRT procedures use two-sided $p$-values, with dCRT using the plus-one rule applied to the absolute statistic, whereas the PCM procedures use right-sided $p$-values. We use $0.05$ as the nominal level.

\begin{figure}[!ht]
	\centering
	\includegraphics[width=\textwidth]{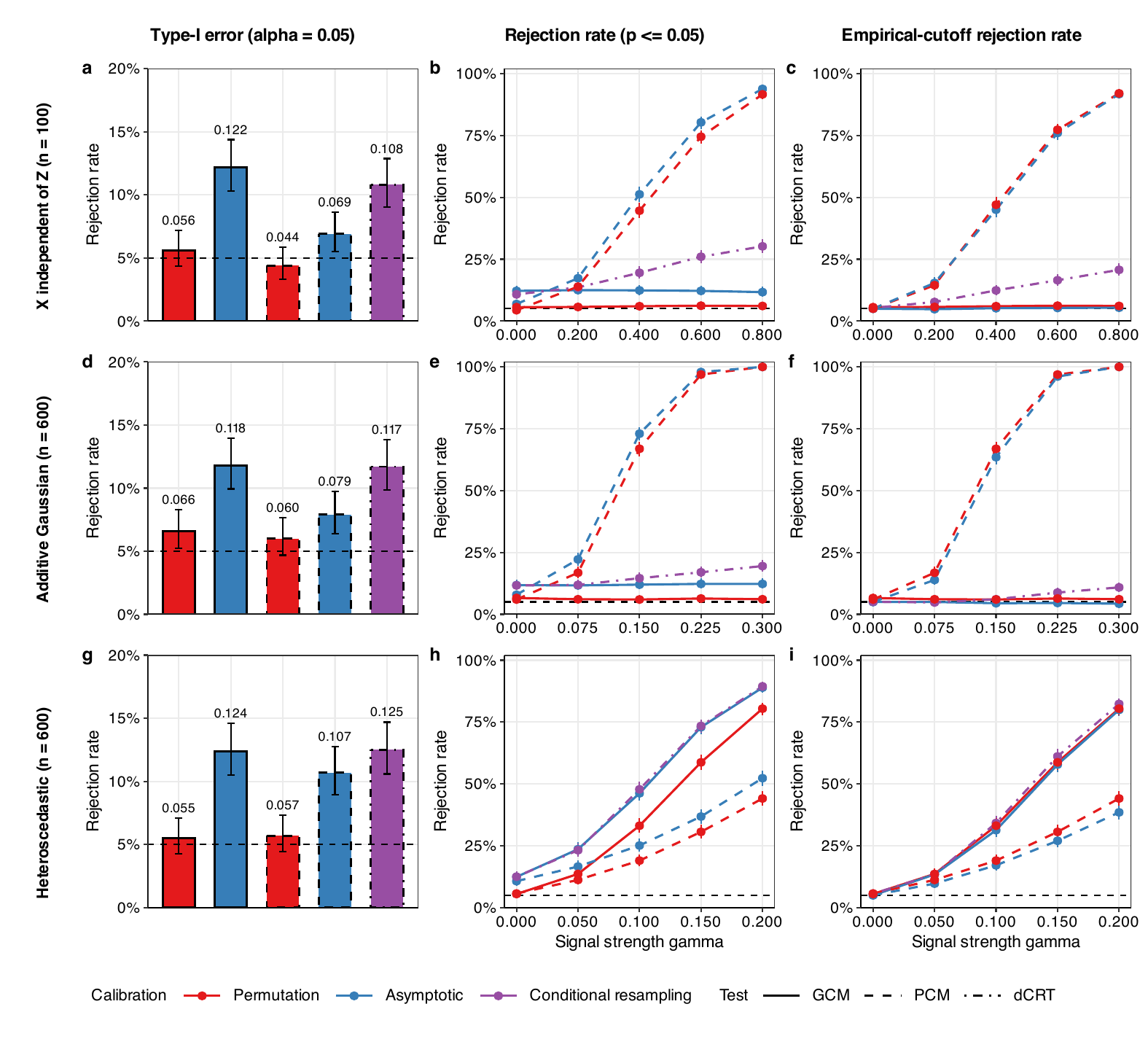}
	\caption{GCM and PCM using random forests with marginal permutation or asymptotic calibration, together with in-sample dCRT using conditional resampling calibration. Colors denote calibration and line or border types denote the test. Black dashed lines mark the nominal level $0.05$.}
	\label{fig:pcm_additive}
\end{figure}

\paragraph{Results interpretation}
Figure~\ref{fig:pcm_additive} shows that permutation calibration brings both GCM and PCM substantially closer to nominal Type~I error in all three designs, although permutation PCM remains mildly liberal under heteroscedasticity. In-sample dCRT is materially liberal at three nulls, with rejection rates $0.108$, $0.117$, and $0.125$. The quadratic alternatives in the first two rows favor PCM. Under the linear Heteroscedastic alternative, GCM is more powerful than PCM. After empirical size adjustment, the tests calibrated by permutation have power broadly similar to their asymptotic counterparts. These adjusted curves are descriptive rather than comparisons with exact correction for size, because the adaptive permutation $p$-values for GCM and PCM are censored at $0.05$.

Overall, permutation calibration improves null calibration while preserving the complementary sensitivity of PCM and GCM.

\section{Permutation benefit under low effective sample size}\label{sec:realdata}

Asymptotic calibration is least reliable when the effective sample size is low, which is precisely the regime where marginal permutation is most useful. We demonstrate this benefit in two complementary genomic settings: a semisynthetic Type~I error study on Cancer Cell Line Encyclopedia (CCLE) mutation data~\citep{barretina2012ccle}, and an FDR and power comparison on Perturb-seq data~\citep{frangieh2021perturbseq}.

\subsection{A revisit on CCLE experiment: a semisynthetic analysis}
\label{sec:ccle_synthY}

The Cancer Cell Line Encyclopedia pairs genomic characterization of human cancer cell lines with their pharmacological response profiles, and a primary use of the resource is to screen which mutations carry information about drug sensitivity beyond the remaining genomic background~\citep{barretina2012ccle}. Such a screening question is precisely a conditional independence hypothesis, which has been studied in~\citet{kook2024comets} using the asymptotic GCM and PCM tests. The mutation panel is nevertheless extremely sparse: a somatic alteration in any given gene is present in only a small fraction of the cell lines, so each binary indicator is zero for nearly every line. Among our screened targets the mutation frequencies range from $0.002$ to $0.055$, that is, from roughly one to twenty-six nonzero indicators out of $474$. The asymptotic calibration adopted in~\citet{kook2024comets} does not account for this sparsity, yet it is exactly the regime in which a normal reference is least trustworthy.

\paragraph{Semisynthetic design} The experiment uses the $n=474$ cell lines and the $p=466$ screened binary mutation indicators. For a set $\mathcal{M}$ of target mutations we test separately, for each $j\in\mathcal{M}$, whether $\bm X = \bm X_j$ is associated with a synthetic outcome $\bm Y$ given the remaining mutations $\bm Z=\bm X_{-j}$. The targets are selected by prevalence of mutation: we divide the $466$ mutations into three groups at the empirical frequency terciles and sample $10$ targets from each. We report the $|\mathcal{M}|=20$ targets from the lowest- and highest-frequency groups, corresponding to high and low sparsity, respectively. For the Type~I error check, $\bm Y$ is drawn from a sparse nonlinear function of $\bm Z$ alone, built from $s=10$ active conditioning variables and their pairwise interactions. Hence $\bm Y\independent \bm X_j\mid \bm Z$ holds by construction, while association between $\bm X_j$ and $\bm Z$ preserves correlation. For the power check, we keep this null response and add a conditional effect carried by the mutation itself. The alternative is constructed for a representative target per group, namely \texttt{EIF5A} and \texttt{ATP11A} for the high and low sparsity group shown in Figure~\ref{fig:ccle_synthY_30targets_box}. The detailed data-generating procedure is given in Appendix~\ref{app:ccle-details}.

\paragraph{Methods compared} We compare GCM tests using normal or permutation calibration and the same random forest nuisance estimators~\citep{wright2017ranger} for $\E[\bm Y\mid\bm X_{-j}]$ and $\E[\bm X_j\mid \bm X_{-j}]$. In other words, both calibrations use the same observed statistic and nuisance fits; the permutation reference additionally refits the forest after each marginal permutation, as prescribed by Algorithm~\ref{alg:gcm}. Permutation calibration uses the adaptive procedure in Algorithm~\ref{alg:adaptive}, with threshold $h=15$ and at most $B_{\max}=499$ draws. A marginal Pearson test is included as a baseline sensitive to confounding.

\begin{figure}[!ht]
	\centering
	\includegraphics[width=\textwidth]{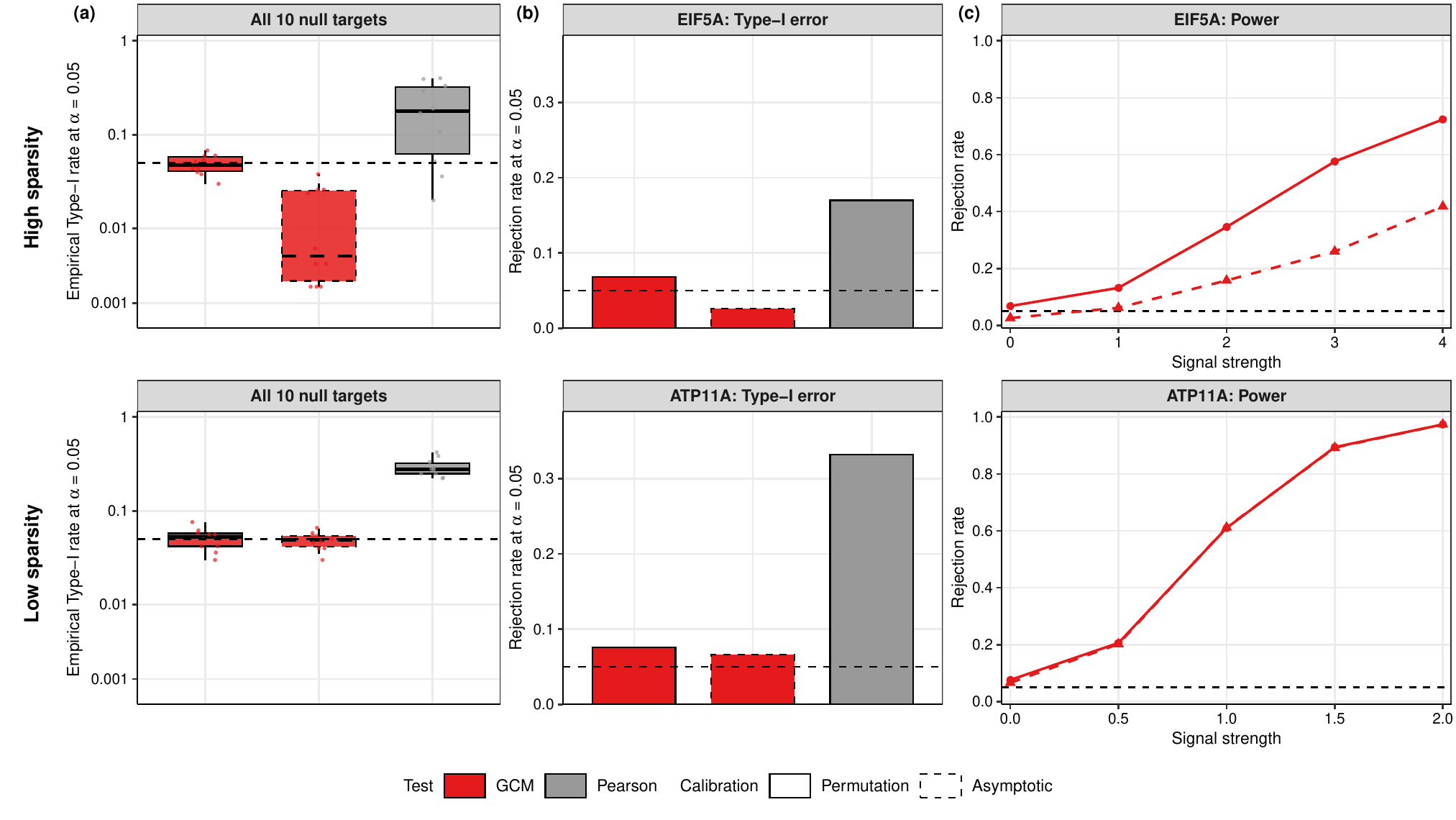}
	\caption{CCLE semisynthetic null calibration and power. Rows are the two groups defined by mutation sparsity, high sparsity (representative \texttt{EIF5A}, mutation frequency $0.008$) on top and low sparsity (representative \texttt{ATP11A}, frequency $0.040$) below. Column~(a) summarizes the $10$ conditional null targets in each sparsity group; column~(b) shows the representative target used to illustrate null calibration; column~(c) shows power of two selected mutations in column (b). Dashed black lines mark $0.05$. The rejection rates are averaged over $500$ independent draws of the synthetic outcome.}
	\label{fig:ccle_synthY_30targets_box}
\end{figure}

\paragraph{Results interpretation} Columns~(a)--(b) of Figure~\ref{fig:ccle_synthY_30targets_box} extend the EIF5A illustration of Figure~\ref{fig:ccle_eif5a_intro} to the groups with high and low sparsity, one per row. Averaged over the $20$ displayed targets across both groups, GCM with permutation calibration rejects at $0.050$ and asymptotic GCM at $0.031$, and the gap is confined to rare mutations: in the group with high sparsity asymptotic GCM rejects at $0.013$ against $0.049$ for permutation, whereas the two agree at $0.049$ and $0.051$ in the group with low sparsity. The intuitive reason is that the large amount of zeros precludes fast CLT convergence while permutation traces the sampling distribution much better. Pearson test rejects at $0.246$, confirming that the constructed CI null retains substantial marginal confounding. Column~(c) reports rejection at the nominal cutoff over the chosen effect grids. The cost of this conservatism appears under the alternative: in the row with high sparsity asymptotic GCM rejects less often than permutation at every effect size, $0.42$ against $0.72$ at the largest shift, whereas the two coincide in the row with low sparsity.

\subsection{FDR and power on Frangieh Perturb-seq data}
\label{sec:frangieh}

\paragraph{Genomic background} Single-cell CRISPR screens probe gene function by combining genetic perturbations with measurements of gene expression in individual cells. Guide RNAs (gRNAs) direct perturbations to selected genomic targets, allowing researchers to assess how disrupting one gene changes the expression of that gene or others. We analyze the control condition of the melanoma Perturb-CITE-seq study of \citet{frangieh2021perturbseq}, which combined pooled CRISPR--Cas9 perturbations with single-cell RNA and surface protein measurements to investigate mechanisms of cancer immune evasion. Our analysis uses the RNA measurements.

For each perturbation--gene pair, $\bm X$ indicates whether a cell carries the perturbation, $\bm Y$ is the gene's transcript count, and $\bm Z$ records technical covariates such as sequencing depth and batch. Testing for a perturbation--gene association therefore amounts to testing $\bm X\indep\bm Y\mid\bm Z$. Individual perturbations occur in only a small fraction of cells, and many genes have zero observed expression in most cells. This joint sparsity can leave few informative observations even in a large experiment, making accurate calibration particularly important when testing many pairs.

\paragraph{Evaluation with positive and negative controls} We construct $20$ replicates of an analysis containing $50$ positive control pairs and $950$ negative control pairs. The positive controls pair each of the $50$ most highly expressed targeted genes with the gRNAs targeting that same gene. Across replicates, we hold the observed data and these positive control pairs fixed and regenerate the $950$ negative control pairs by sampling response genes with replacement from a fixed eligible pool of non-targeted genes and pairing each with three distinct, randomly sampled non-targeting gRNAs. Appendix~\ref{app:frangieh-details} gives further details of the construction of control pairs.

\begin{figure}[!ht]
	\centering
	\includegraphics[width=0.98\textwidth]{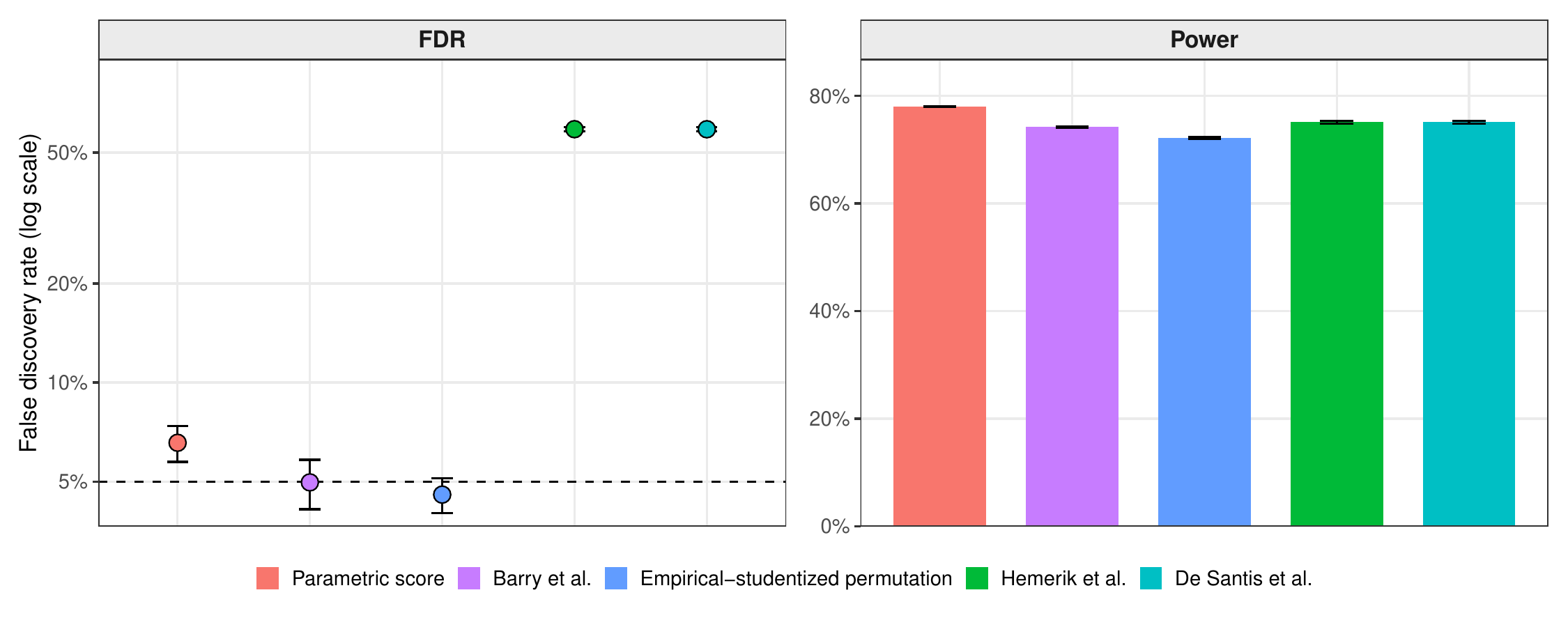}
	\caption{Empirical false discovery rate and power for five calibrations of the same negative binomial efficient score in the
	Frangieh Perturb-seq analysis, at target BH level $q=0.05$. The FDR
	panel uses point estimates on a logarithmic vertical scale, while the
	power panel uses bars on a linear scale. Error bars show $\pm1$ standard error, and the black dashed line marks the target FDR.}
	\label{fig:frangieh_summary}
\end{figure}

\paragraph{Methods compared} We compare the five score calibrations from Section~\ref{sec:nb-score-simulation}, using the same fitted negative binomial null model and efficient score contributions for every method. All tests adjust for $\log(1+M)$, where $M$ is the total gene UMI count, and for technical batch. We apply the Benjamini--Hochberg procedure to the parametric $p$-values and use its adaptive resampling implementation for the four randomization methods, all at target FDR $q=0.05$. The adaptive procedures use two-sided tests, threshold $h=15$, and at most $B_{\max}=25{,}000$ draws per pair. Within each replicate, the false discovery proportion is the fraction of discoveries belonging to the negative control set, and power is the fraction of positive controls detected; Figure~\ref{fig:frangieh_summary} reports averages across replicates.

\paragraph{Results interpretation} Both permutation calibrations achieve empirical FDR close to the $5\%$ target in this benchmark. The permutation tests using information and empirical studentization yield FDRs of $4.97\%$ and $4.56\%$, respectively, with corresponding power of $74.2\%$ and $72.2\%$. The parametric score detects $78.0\%$ of positive controls, but its empirical FDR of $6.56\%$ exceeds the target. The two sign-flip tests exhibit much greater FDR inflation, reaching $58.91\%$ and $58.89\%$. Their common positive control detection rate of $75.1\%$ therefore does not represent power at the target FDR.

The nearly identical results for the two sign-flip procedures show that standardizing each sign flip separately does not remedy their miscalibration in this sparse setting. This pattern mirrors the simulation with low SNR in Figure~\ref{fig:nb-score-ci}: permutation calibration remains accurate when sign-flip calibration fails. Figure~\ref{fig:frangieh_pv} in Appendix~\ref{app:frangieh-details} shows the null and positive-control $p$-value distributions.

\section{Conclusion}

This paper develops a framework using marginal permutation for conditional independence testing based on empirically studentized generalized score statistics. Although marginal permutation disrupts the dependence between $\bm X$ and $\bm Z$ permitted under the conditional independence null, the resulting procedures achieve two validity guarantees: finite-sample validity under the stronger marginal independence null and asymptotic validity under conditional independence, subject to conditions for each method. We instantiate this framework with an empirically studentized GLM score test and versions of the GCM and PCM calibrated by permutation. The GLM procedure remains asymptotically valid under $\E[\bm Y\mid\bm X,\bm Z]=\mu_\beta(\bm Z)$ despite arbitrary misspecification of the working variance, while the GCM retains its usual guarantee based on the product of nuisance error rates and the PCM retains the normalized nuisance conditions of the original PCM analysis, supplemented by a corresponding normalized condition on the stability of refitting after permutation. Across simulations, a semisynthetic CCLE experiment, and a single-cell CRISPR screen analysis, permutation calibration yields substantial improvements in finite-sample calibration in settings with sparse observations or low effective sample size. At the same time, our results for local power show that, in the regimes studied, these gains in robustness incur no first-order loss of power relative to the corresponding asymptotically calibrated procedures.

Several directions merit further study. First, it would be valuable to extend permutation calibration to other statistics in the class of generalized scores and to establish conditions under which it delivers superior finite-sample calibration relative to conventional asymptotic references. Second, although adaptive sampling and computational shortcuts reduce the cost of permutation, repeated permutation refitting can remain burdensome in large-scale applications. Efficient analytical approximations to the conditional permutation distribution, building on classical saddlepoint approximations for permutation tests and more recent conditional saddlepoint methodology~\citep{robinson1982saddlepoint,niu2024-spa}, could further accelerate these procedures. Finally, the theory developed here is first order: it establishes asymptotic validity and the absence of a first-order loss of local power, but does not explain the finite-sample gains produced by permutation calibration. Developing a higher-order theory that quantifies when and by how much permutation improves calibration would therefore be especially informative, although it is likely to be technically challenging because the permutation distribution depends on nuisance functions estimated and refitted from the data.

{\small
\bibliographystyle{plainnat}
\bibliography{references}
}

\clearpage
\begin{center}
  {\LARGE\bfseries Supplementary materials}
\end{center}
\bigskip

\appendix

\noindent\textbf{Supplementary conventions.}
In the PCM proofs of size validity and permutation stability,
$n\equiv|\cD_1|=n_\star$ denotes the size of the inference fold. The subsection on local power instead uses $n$ for the full sample size and $n_1,n_2$ for the two fold sizes.

\section{Practical implementation}\label{sec:practical}

The permutation calibration of Section~\ref{sec:method} is valid, but in
its na\"ive form it can be expensive. For the GCM and PCM, each permutation in principle requires refitting a nuisance regression: the regression of $X$ on $Z$, denoted $\hat\mu_x^{(\pi)}$, for the GCM, the projection fit on the first fold
$\hat\mu_f^{(\pi)}$ for the PCM. For the flexible learners used in practice this refit dominates the cost of the test, so the runtime of the
Monte~Carlo schemes with a fixed budget of $B$ in Algorithms~\ref{alg:gcm}--\ref{alg:vpcm}
is essentially $B\times(\text{cost of one refit})$. 

Two distinct inefficiencies are at play: the refit required for each permutation, and the fixed
Monte~Carlo budget, which draws far more permutations than a decision
typically requires. We address them in turn. The first vanishes outright
whenever the regression is a linear smoother: response linearity reduces every refit to a cached matrix--vector product, which we elaborate in Appendix~\ref{subsec:closed-form-refits}. The second is mitigated by an adaptive stopping rule that draws only as many permutations
as the data demand, which we describe in Appendix~\ref{subsec:adaptive-sampling}. 

\subsection{Closed-form refits for linear smoothers}
\label{subsec:closed-form-refits}

Suppose the learner for regressing $X$ on $Z$ is a \emph{linear smoother}
(Definition~\ref{def:linear-smoother}): its fitted values are
$\widehat{\boldsymbol\mu}_x=SX$ for
$X\equiv(X_1,\ldots,X_n)^\top$ and a hat matrix
$S\in\R^{n\times n}$ that depends on $Z$ alone (with a fixed seed). For $\pi\in\cG$, write
$X^{(\pi)}\equiv(X_{\pi(1)},\ldots,X_{\pi(n)})^\top$. Because permuting $X$ leaves $S$ untouched, the refit on the
permuted sample $(X_{\pi(i)},Z_i)_{i=1}^n$ is simply
$\widehat{\boldsymbol\mu}_x^{(\pi)}=SX^{(\pi)}$, and the
permuted GCM weight vector collapses to a fixed linear map:
\[
\boldsymbol\nu^{\mathrm{GCM},(\pi)}
  = X^{(\pi)}-SX^{(\pi)}
  = (I-S)X^{(\pi)}.
\]
The smoother is therefore fit \emph{once}; every subsequent permutation
reuses the cached residual operator $I-S$, replacing a full regression
fit by a single matrix--vector product. The total cost drops from $B$
refits to one fit plus $B$ matrix--vector products---typically a reduction
of orders of magnitude. This covers ordinary least squares, ridge
regression, kernel ridge regression, smoothing splines, and the
Nadaraya--Watson estimator. Appendix~\ref{app:proof-perm-stab} separately
characterizes when the resulting error measuring permutation stability is
negligible. The same reduction applies to the projection step of the PCM
whenever the projection learner is a linear smoother. 

For learners that are genuinely nonlinear in the response---random forests, gradient boosting, neural networks---no such shortcut exists, and the adaptive scheme below is aimed at exactly that regime.

\subsection{Adaptive permutation calibration}\label{subsec:adaptive-sampling}

Even when each permutation is cheap, the scheme with a fixed budget of $B$ is wasteful: it
draws all $B$ permutations regardless of how extreme the observed statistic
is. For a $T_n$ deep in the tail the test would already reject after a
handful of permutations, while for a $T_n$ near the centre the procedure
``knows'' early that it will not reject yet still pays the full budget.
Following the sequential Monte~Carlo testing idea of
\citet{besag1991sequential}, we replace the fixed budget by a sequential
rule that draws permutations one at a time and stops as soon as the data
permit a definitive decision. 

Fix an integer $h\ge 1$, a deterministic cap $B_{\max}\in\NN$, and a score map $\rho$
(one-sided $\rho(u)=u$, two-sided $\rho(u)=|u|$). Here $T_n$ denotes the observed statistic for the chosen test and $T_n^{(\Pi_t)}$ its refitted counterpart under the $t$th fresh permutation. Define the score stream $W_0\equiv\rho(T_n)$ and $W_t\equiv\rho(T_n^{(\Pi_t)})$ for $t\ge1$. Counting a
permutation at least as extreme as the observed value as a \emph{loss},
\[
K_t\equiv\sum_{s=1}^{t}\indc{W_s\ge W_0},
\qquad
t_h\equiv\inf\{t\ge 1:K_t=h\},
\]
the \emph{anytime-valid generalization of the Besag--Clifford
Monte~Carlo $p$-value} described by
\citet{fischer-barry-ramdas-EJS} is
\begin{equation}\label{eq:av-pvalue}
p_t^{\mathrm{AV}}\equiv
\begin{cases}
\displaystyle\frac{h}{t+h-K_t}, & K_t<h,\\[6pt]
\displaystyle\frac{h}{t_h}, & K_t\ge h.
\end{cases}
\end{equation}
The procedure rejects as soon as $p_t^{\mathrm{AV}}\le\alpha$, and stops for
futility once the $h$th loss occurs with $p_t^{\mathrm{AV}}>\alpha$, after
which the $p$-value freezes at $h/t_h$. Let
$\cG_t\equiv\sigma(W_0,\ldots,W_t)$ denote the score filtration, and let
$\tau\le B_{\max}$ denote the resulting $\cG_t$-stopping time, including
the deterministic stop at $B_{\max}$. 

Under the alternative the loss count
grows slowly and the test rejects after a few refits. Under the null the $h$th loss arrives quickly and the test halts. In both regimes the realised number of permutations---and hence nuisance refits---is typically far below
the fixed budget of $B$, while Type~I validity is preserved. Combined with the shortcut for linear smoothers above, this adaptive rule is what makes permutation calibration practical for the applications in Section~\ref{sec:realdata}, where the test is repeated thousands of times to assess differential expression across many genes or screen for significant variables.
The full procedure is stated in Algorithm~\ref{alg:adaptive}.

\begin{algorithm}[!ht]
  \caption{Adaptive permutation calibration for a generalized score statistic}
  \label{alg:adaptive}
  \KwIn{Data, level $\alpha\in(0,1)$, integer loss threshold $h\ge1$,
  finite integer cap $B_{\max}\ge1$, fixed measurable score map $\rho$,
  and a fixed measurable procedure for computing a real-valued $T_n$
  (including nuisance fitting, tuning, and preprocessing).}
  Use randomness independent of the data, with independent randomization for
  any sample split, learner seeds, and permutation draws\;
  If splitting is used, fix the split and reserve the training fold; permute only exposure labels in the inference fold below\;
  Fix the learner seeds and use them for every fit; compute $T_n$
  and set $W_0=\rho(T_n)$\;
  Initialise $t\gets0$, $K\gets0$, and $t_h\gets\infty$\;
  \While{$t<B_{\max}$}{
    $t\gets t+1$\;
    Draw $\Pi_t$ independently and uniformly with replacement from all
    permutations of the exposure labels in the sample being permuted\;
    Compute $T_n^{(\Pi_t)}$ by applying the same procedure and fixed seeds
    to the relabeled sample, holding $(Y,Z)$ and any training data fixed;
    reuse fitted quantities only when unchanged by the permutation\;
    Set $W_t=\rho(T_n^{(\Pi_t)})$\;
    \If{$W_t\ge W_0$}{$K\gets K+1$\;}
    \If{$K=h$ for the first time}{$t_h\gets t$\;}
    Compute $p_t^{\mathrm{AV}}$ from~\eqref{eq:av-pvalue}\;
    \lIf{$p_t^{\mathrm{AV}}\le\alpha$}{\textbf{reject} and \textbf{stop}}
    \lIf{$K=h$ and $p_t^{\mathrm{AV}}>\alpha$}{\textbf{stop} for futility}
  }
  \If{the loop ends without an earlier stop}{return
  $p_{B_{\max}}^{\mathrm{AV}}$ and the corresponding decision\;}
  \end{algorithm}

\subsection{Validity of the adaptive permutation calibration}
\label{subsec:adaptive-validity}

The adaptive rule recomputes the prescribed weight component at every
permutation, by either a nuisance refit or an exact cached update, yet it
inherits the validity of the procedures with a fixed budget.  Under the strong
marginal independence null, the entire score stream is conditionally
exchangeable and the process counting losses is finite-sample anytime valid.
Under the conditions of Theorem~\ref{thm:product-score-validity}, the observed
statistic and each fresh permutation statistic converge to the same continuous
limit. The asymptotic guarantee below keeps $h$ and $B_{\max}$ fixed as $n\to\infty$.

\begin{theorem}[Finite-sample validity of adaptive calibration under MI]
\label{thm:adaptive-strong}
Let $p_\tau^{\mathrm{AV}}$ be the $p$-value at the stopping time of
Algorithm~\ref{alg:adaptive}. If the observations
$(X_i,Y_i,Z_i)_{i=1}^n$ are i.i.d. under
$H_0^{\mathrm{MI}}:\bm X\indep(\bm Y,\bm Z)$, then, for every
sample size $n$ and level $\alpha\in(0,1)$, we have $\PP\bigl(p_\tau^{\mathrm{AV}}\le\alpha\bigr)\le\alpha$.
\end{theorem}

\begin{theorem}[Asymptotic validity of adaptive calibration]\label{thm:adaptive-target}
Let $p_\tau^{\mathrm{AV}}$ be the $p$-value at the stopping time of
Algorithm~\ref{alg:adaptive}, with $h$ and $B_{\max}$ fixed as
$n\to\infty$. Suppose the observed
and permuted statistics satisfy conditions~(i)--(iii) of Theorem~\ref{thm:product-score-validity}.
If $\rho$ is continuous and $\rho(\bm G)$ has a continuous distribution
for $\bm G\sim N(0,1)$, then, for every fixed $\alpha\in(0,1)$, we have $\PP\bigl(p_\tau^{\mathrm{AV}}\le\alpha\bigr)\le\alpha+o(1)$.
\end{theorem}

Theorems~\ref{thm:glm-perm-emp}, \ref{thm:gcm-valid}, and~\ref{thm:pcm-valid} verify the conditions of Theorem~\ref{thm:product-score-validity} for
the empirically studentized GLM score, GCM, and PCM, respectively.
Both score maps used here, $\rho(u)=u$ and $\rho(u)=|u|$, satisfy the
stated continuity requirements.

\subsection{Proofs of Theorem~\ref{thm:adaptive-strong} and Theorem~\ref{thm:adaptive-target}}

We present auxiliary results in Appendix~\ref{subsec:auxiliary-lemma-adaptive-sampling} and prove Theorems~\ref{thm:adaptive-strong} and~\ref{thm:adaptive-target} in Appendix~\ref{subsec:proof-adaptive-strong} and Appendix~\ref{subsec:proof-adaptive-target}, respectively.

\subsubsection{Auxiliary results and their proofs}\label{subsec:auxiliary-lemma-adaptive-sampling}

We first prove an exchangeability result for an infinite score stream.  An infinite formulation is convenient because the anytime-valid property is independent of any deterministic computational cap. Algorithm~\ref{alg:adaptive} simply stops this stream no later than $B_{\max}$.

\begin{lemma}[Anytime-valid bound from conditional exchangeability]
  \label{lem:av-from-exch}
  Fix an integer $h\ge1$. Let $\cH$ be a $\sigma$-algebra and let
  $(W_t)_{t\ge0}$ be a real-valued sequence that is conditionally
  exchangeable given $\cH$. Recall $\cG_t\equiv\sigma(W_0,\ldots,W_t), K_t\equiv\sum_{s=1}^t\indc{W_s\ge W_0}$ and $t_h\equiv\inf\{t\ge1:K_t=h\}$. Furthermore, consider $p_t^{\mathrm{AV}}$ in~\eqref{eq:av-pvalue}, with the convention
  $h/\infty=0$. Then, for every almost surely finite $(\cG_t)$-stopping
  time $\tau\ge1$ and every $\alpha\in(0,1)$,
  \[
  \PP\bigl(p_\tau^{\mathrm{AV}}\le\alpha\mid\cH\bigr)
  \le\alpha\quad\text{almost surely}.
  \]
  Consequently,
  $\PP(p_\tau^{\mathrm{AV}}\le\alpha)\le\alpha$.
  \end{lemma}

  \begin{proof}
    Fix an integer $k\ge h$. For each $j\in\{0,\ldots,k-1\}$, define
    \[
    R_{k,j}\equiv\sum_{s=0}^{k-1}\indc{W_s\ge W_j},
    \qquad
    A_k\equiv\{j\in\{0,\ldots,k-1\}:R_{k,j}\le h\}.
    \]
    We have $|A_k|\le h$ for every realization of the scores. Indeed, if
    $A_k$ is nonempty, choose $j_\star\in A_k$ with
    $W_{j_\star}=\min_{j\in A_k}W_j$. Every $s\in A_k$ then satisfies
    $W_s\ge W_{j_\star}$, so
    \[
    |A_k|\le R_{k,j_\star}\le h.
    \]
    Conditional exchangeability implies that
    $\PP(R_{k,j}\le h\mid\cH)$ is the same for every $j$, since permuting
    the score indices permutes the corresponding counts $R_{k,j}$.
    Since $R_{k,0}=1+K_{k-1}$, it follows that
    \begin{equation}\label{eq:rank-conservative}
    \begin{aligned}
    \PP(K_{k-1}\le h-1\mid\cH)
    &=\PP(R_{k,0}\le h\mid\cH)\\
    &=\frac1k\sum_{j=0}^{k-1}\PP(R_{k,j}\le h\mid\cH)
      =\frac1k\E[|A_k|\mid\cH]
    \le\frac hk.
    \end{aligned}
    \end{equation}

    If $t<t_h<\infty$, then
    \[
    h-K_t=K_{t_h}-K_t
    =\sum_{s=t+1}^{t_h}\indc{W_s\ge W_0}
    \le t_h-t.
    \]
    Thus $t_h\ge t+h-K_t$, which also holds when $t_h=\infty$.
    By~\eqref{eq:av-pvalue},
    \[
    p_t^{\mathrm{AV}}=\frac{h}{t+h-K_t}\ge\frac{h}{t_h}
    \quad\text{if }t<t_h,
    \qquad
    p_t^{\mathrm{AV}}=\frac{h}{t_h}
    \quad\text{if }t\ge t_h.
    \]
    Hence, for every almost surely finite stopping time $\tau\ge1$,
    \begin{equation}\label{eq:reduce-to-tstar}
    \{p_\tau^{\mathrm{AV}}\le\alpha\}
    \subseteq
    \left\{\frac{h}{t_h}\le\alpha\right\}
    =
    \left\{t_h\ge\left\lceil\frac{h}{\alpha}\right\rceil\right\}.
    \end{equation}
    Let $k=\lceil h/\alpha\rceil$. Because $K_t$ starts at zero and
    increases by either zero or one at each step, $t_h\ge k$ holds
    exactly when $K_{k-1}\le h-1$. Therefore,
    \eqref{eq:rank-conservative} and~\eqref{eq:reduce-to-tstar} give
    \[
    \PP\bigl(p_\tau^{\mathrm{AV}}\le\alpha\mid\cH\bigr)
    \le
    \PP(K_{k-1}\le h-1\mid\cH)
    \le \frac{h}{k}
    \le\alpha.
    \]
    Taking expectations proves the unconditional statement.
    \end{proof}

    \begin{lemma}[Joint convergence from a conditional deterministic limit]
      \label{lem:joint-conv}
      Let $\bm X_n\convd \bm X$ and let $\cF_n\supseteq\sigma(\bm X_n)$.  Fix
      $L\ge1$.  Conditional on $\cF_n$, let
      $\bm Y_{n,1},\ldots,\bm Y_{n,L}$ be i.i.d., and suppose
      \[
      \bm Y_{n,1}\mid\cF_n\convdp \bm Y,
      \]
      where the law of $\bm Y$ is deterministic.  Then
      \[
      (\bm X_n,\bm Y_{n,1},\ldots,\bm Y_{n,L})
      \convd
      (\bm X,\bm Y_1,\ldots,\bm Y_L),
      \]
      where $\bm Y_1,\ldots,\bm Y_L$ are i.i.d. copies of $\bm Y$, independent of $\bm X$.
      \end{lemma}
    
      \begin{proof}
        Let
        $\psi_n(u)\equiv\E[e^{iu\bm Y_{n,1}}\mid\cF_n]$ and
        $\psi(u)\equiv\E[e^{iu\bm Y}]$. Conditional weak convergence gives $\psi_n(u)\to\psi(u)$ in probability; since
        $|\psi_n|\le1$, the convergence is also in $L^1$. Conditional independence
        then yields, for fixed $(u_0,\ldots,u_L)$,
        \[
        \E e^{iu_0\bm X_n+i\sum_{t=1}^Lu_t\bm Y_{n,t}}
        =\E\!\left[e^{iu_0\bm X_n}\prod_{t=1}^L\psi_n(u_t)\right]
        \longrightarrow
        \E e^{iu_0\bm X}\prod_{t=1}^L\psi(u_t).
        \]
        L\'evy's continuity theorem proves the stated independent joint limit.
        \end{proof}
    
    \begin{proposition}[Transfer to adaptive calibration over a bounded horizon]
      \label{prop:oracle-validity}
      Fix $L\ge1$.  Let $W_{n,0}$ be $\cF_n$-measurable and, conditional on
      $\cF_n$, let $W_{n,1},\ldots,W_{n,L}$ be i.i.d.  Suppose that there is a
      random variable $W$ with a continuous distribution such that
      \[
      W_{n,0}\convd W,
      \qquad
      W_{n,1}\mid\cF_n\convdp W.
      \]
      Then, for every stopping time $\tau_n\le L$ adapted to
      $\sigma(W_{n,0},\ldots,W_{n,t})$,
      \[
      \PP\bigl(p_{\tau_n}^{\mathrm{AV}}\le\alpha\bigr)
      \le\alpha+o(1).
      \]
      \end{proposition}
    
      \begin{proof}
        Lemma~\ref{lem:joint-conv} gives
        \[
        (W_{n,0},W_{n,1},\ldots,W_{n,L})
        \convd
        (W_0^\star,W_1^\star,\ldots,W_L^\star),
        \]
        where $W_0^\star,\ldots,W_L^\star$ are i.i.d. copies of $W$. Write
        $p_L^{\star,\mathrm{AV}}$ for~\eqref{eq:av-pvalue} computed from
        this limiting score sequence.
        The process $t\mapsto p_t^{\mathrm{AV}}$ is nonincreasing: before the
        $h$th loss, a loss leaves the denominator $t+h-K_t$ unchanged and a
        non-loss increases it by one; after the $h$th loss the process is frozen.
        Hence
        \[
        \{p_{\tau_n}^{\mathrm{AV}}\le\alpha\}
        \subseteq
        \{p_L^{\mathrm{AV}}\le\alpha\}.
        \]
        
        The indicator of the event $\{p_L^{\mathrm{AV}}\le\alpha\}$ is locally
        constant whenever no coordinate $s_t$, $t\ge1$, equals $s_0$, because it
        depends only on the comparisons $\indc{s_t\ge s_0}$.  Since the distribution of $W$ is
        continuous,
        \[
        \PP\bigl(\exists t\in\{1,\ldots,L\}:W_t^\star=W_0^\star\bigr)=0.
        \]
        Thus $\{p_L^{\mathrm{AV}}\le\alpha\}$ is a continuity set for the
        limiting distribution, and the Portmanteau theorem yields
        \[
        \PP(p_L^{\mathrm{AV}}\le\alpha)
        \longrightarrow
        \PP(p_L^{\star,\mathrm{AV}}\le\alpha).
        \]
        The infinite i.i.d. extension of $(W_0^\star,\ldots,W_L^\star)$ is exchangeable, so
        Lemma~\ref{lem:av-from-exch} implies
        $\PP(p_L^{\star,\mathrm{AV}}\le\alpha)\le\alpha$.  Combining the preceding
        displays proves the result.
        \end{proof}

\subsubsection{Proof of Theorem~\ref{thm:adaptive-strong}}\label{subsec:proof-adaptive-strong}

\begin{proof}
Let $I$ be the indices permuted by Algorithm~\ref{alg:adaptive},
$X^{(0)}=(X_i)_{i\in I}$, and $X^{(t)}$ its $t$th independently
permuted version. Let $\cH$ be generated by the split, any training data,
$(Y_i,Z_i)_{i\in I}$, learner seeds, and the unordered collection of
$(X_i)_{i\in I}$, including multiplicities.

By i.i.d. sampling under MI and the algorithm's independent randomization,
$X^{(0)}\mid\cH$ is uniform over its distinct rearrangements.
Each fresh uniform index permutation samples this same distribution,
regardless of the original ordering. This remains true with repeated
values, since each distinct rearrangement corresponds to the same
number of index permutations. Thus, for every $t\ge1$,
\[
\mathcal L(X^{(t)}\mid\cH,X^{(0)},\ldots,X^{(t-1)})
=\mathcal L(X^{(0)}\mid\cH).
\]
Hence $X^{(0)},X^{(1)},\ldots$ are conditionally i.i.d. given $\cH$.
The algorithm applies the same score computation to each vector, with
all other inputs fixed by $\cH$, so $W_0,W_1,\ldots$ are conditionally
i.i.d. as well. Since $\tau\le B_{\max}$,
Lemma~\ref{lem:av-from-exch} gives
$\PP(p_\tau^{\mathrm{AV}}\le\alpha\mid\cH)\le\alpha$.
Taking expectations completes the proof.
\end{proof}

\subsubsection{Proof of Theorem~\ref{thm:adaptive-target}}\label{subsec:proof-adaptive-target}
\begin{proof}
Set $L=B_{\max}$ and let $\cF_n$ be the sigma-field generated by the
observed sample, any split, and the learner randomness held fixed by
Algorithm~\ref{alg:adaptive}.
For $W_{n,0}=\rho(T_n)$ and
$W_{n,t}=\rho(T_n^{(\Pi_t)})$, $1\le t\le L$, the independent permutation
draws and common computation procedure make the permuted scores
conditionally i.i.d. given $\cF_n$. The observed score is
$\cF_n$-measurable. Theorem~\ref{thm:product-score-validity} and continuity of $\rho$ give
\[
W_{n,0}\convd\rho(\bm G),\qquad
W_{n,1}\mid\cF_n\convdp\rho(\bm G),\qquad \bm G\sim N(0,1).
\]
Since $\rho(\bm G)$ has a continuous distribution and the algorithm's
stopping time satisfies $\tau\le L$, all hypotheses of
Proposition~\ref{prop:oracle-validity} hold. Its conclusion is
$\PP(p_\tau^{\mathrm{AV}}\le\alpha)\le\alpha+o(1)$.
\end{proof}

\section{Additional simulation results and details}
\label{app:additional-simulation-results}

This section provides additional simulation results and details that complement the main text, including details for the experiment generating Figure~\ref{fig:glm-variance-robustness}, details for negative binomial simulations and results under marginal independence.

\subsection{Data generation details in Figure~\ref{fig:glm-variance-robustness}}
\label{app:glm-variance-simulation}

This appendix records the complete procedure used to generate the data for
Figure~\ref{fig:glm-variance-robustness}.  Independently across observations,
generate
\begin{align*}
 Z_i^\circ\sim\mathrm{Ber}(3/4), D_i=(1,Z_i^\circ)^\top, X_i\mid Z_i^\circ=z\sim N(0,s_z^2), \V(Y_i\mid Z_i^\circ=z)&=c(z)\mu_y(z),
\end{align*}
where $s_0=1$ and $s_1=1/3$. Here we set $\mu_y(0)=20$ and $\mu_y(1)=30$. When $c(z)>1$, we generate $Y_i$ from a negative binomial distribution with mean $\mu_y(z)$ and size $\mu_y(z)/\{c(z)-1\}$, which gives variance $c(z)\mu_y(z)$ exactly. When $c(z)=1$, we draw $Y_i$ from a Poisson distribution with mean $\mu_y(z)$.
In every case we fit the Poisson log-mean model using $D_i$.  Because $Z_i^\circ$
is binary, this working mean model is saturated and exactly correct.  A fixed
global rescaling of $X$ multiplies the score numerator and either studentizer's
standard deviation by the same factor, so it leaves all rejection decisions
unchanged; the stated exposure scales can therefore be standardized if needed
for Assumption~\ref{assumption:glm-reg}.

The multiplicative design sets $c(0)=c(1)=5$.  The design with arbitrary misspecification changes only the second factor, setting $c(0)=5$ and $c(1)=1$.  Thus the
two experiments have identical laws for $(Z_i^\circ,X_i)$ and identical
conditional outcome means.  In particular, any contrast between them is due to the form of the variance misspecification.

\subsection{Negative binomial simulation details}
\label{app:nb-score-mi}

\paragraph{Simulation parameters}
All designs use $n=400$ and contain the null and four equally spaced
downregulation alternatives. The signal grids are
\begin{align*}
  \Gamma_{\mathrm{High}}&=\{0,-0.1125,-0.225,-0.3375,-0.45\},\\
  \Gamma_{\mathrm{Low}}&=\{0,-0.2125,-0.425,-0.6375,-0.85\},\\
  \Gamma_{\mathrm{Var,MI}}&=\{0,-0.055,-0.11,-0.165,-0.22\},\\
  \Gamma_{\mathrm{Var,CI}}&=\{0,-0.0675,-0.135,-0.2025,-0.27\}.
\end{align*}
The grids for high and low SNR are used under both MI and CI. 

\paragraph{Results under marginal independence}
We give the MI counterpart to the CI experiment in Section~\ref{sec:nb-score-simulation}. Under MI, $\bm X\independent(\bm Y,\bm Z)$, so label exchangeability makes permutation calibration finite-sample valid. Using the method order defined in Section~\ref{sec:nb-score-simulation}, the null rejection rates under high SNR are $(5.05,5.55,5.50,5.35,5.30)\%$. All five methods are therefore close to the nominal level when the exposure is common. When the exposure prevalence is $0.025$, the null rejection rates under low SNR are $(3.80,4.25,5.25,12.25,12.10)\%$. Our method remains near nominal, whereas both sign-flip procedures reject far too often. Thus, the failure of the sign-flip reference when scores are sparse persists even without association between $\bm X$ and $\bm Z$. When misspecification of the negative binomial variance depends on the covariates, all five procedures are therefore close to nominal under label exchangeability, with at most mild elevation of the parametric test. 

Together with the CI results, these findings show that empirically studentized permutation is the procedure whose rejection rate is most consistently close to nominal across all six null settings.

\begin{figure}[!ht]
  \centering
  \includegraphics[width=0.92\textwidth]{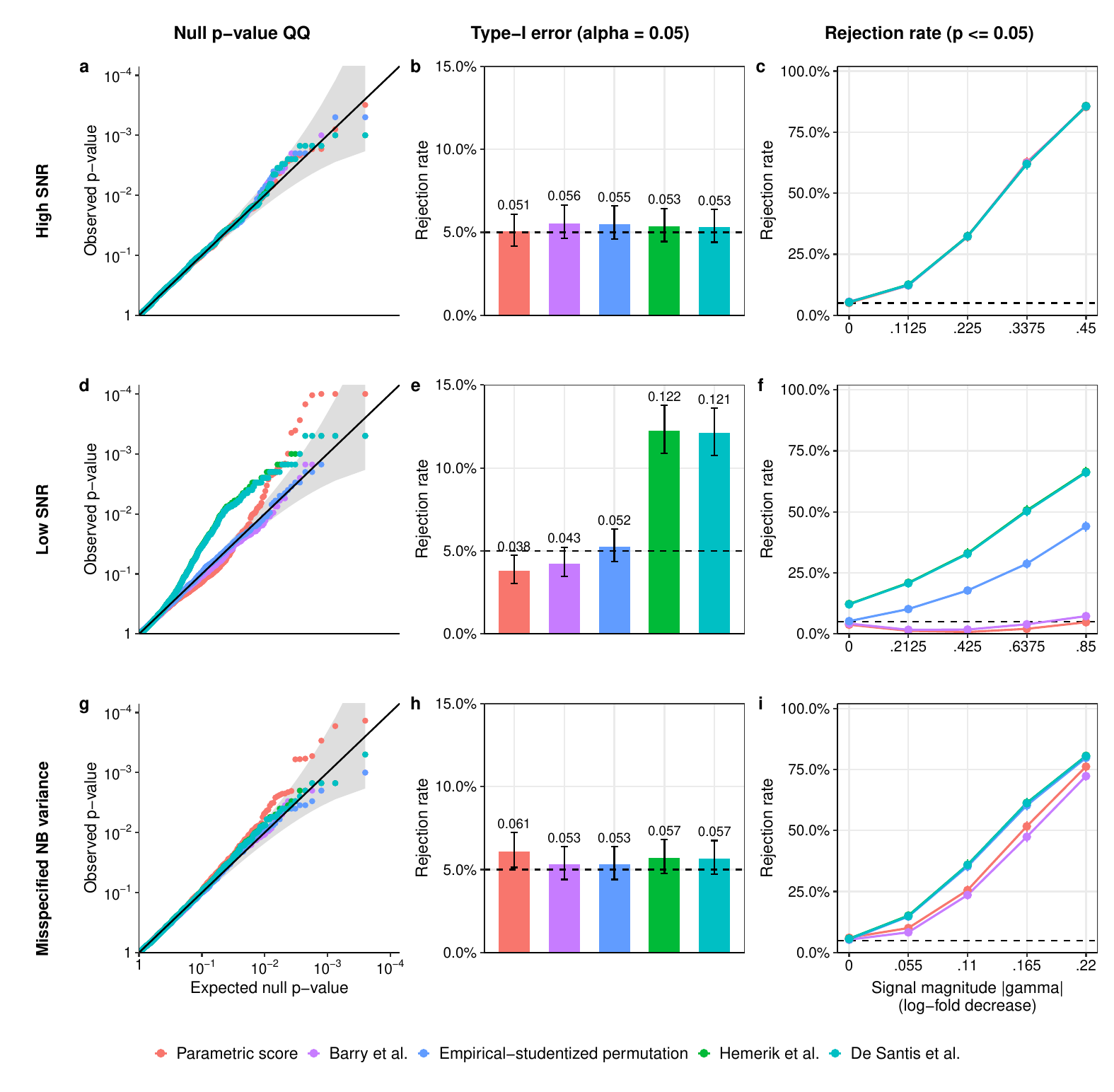}
  \caption{Score calibration under marginal independence. Rows show high SNR, low SNR, and misspecified negative binomial variance; columns show null $p$-value QQ plots, type~I error, and rejection rate at the nominal cutoff. Signal zero is the null. Grey regions are pointwise 95\% envelopes under a uniform null distribution, error bars are 95\% Monte Carlo intervals, and black dashed lines mark $0.05$. Each cell uses $n=400$ and $2{,}000$ datasets; resampling calibrations use $1{,}999$ draws.}
  \label{fig:nb-score-mi}
\end{figure}

\subsection{Parameters for the simulations in Section~\ref{sec:sim-additive}}
\label{app:sim-additive-parameters}

At $\gamma=0$, the $\bm X\independent \bm Z$ design uses
$n\in\{20,40,60,80,100\}$, and the two CI designs use
$n\in\{200,300,400,500,600\}$. Figure~\ref{fig:pcm_additive} reports the results when $n=100$ and $n=600$, respectively. At each setup, the signal grids are $\gamma\in\{0,0.2,0.4,0.6,0.8\}$ for $\bm X\independent \bm Z$, $\gamma\in\{0,0.075,0.15,0.225,0.30\}$ for Additive
Gaussian, and $\gamma\in\{0,0.05,0.10,0.15,0.20\}$ for Heteroscedastic. 

\section{Additional details for analyses in Section~\ref{sec:realdata}}
\label{app:empirical-details}

\subsection{CCLE semisynthetic analysis}
\label{app:ccle-details}

\paragraph{Data and targets}
We use the Cancer Cell Line Encyclopedia panels of drug responses and mutations distributed with \citet{kook2024comets}.  After restricting
to cell lines observed in both panels, the data contain the PLX4720
area under the dose-response curve and $1638$ binary mutation
indicators on $n=474$ cell lines.  Screening mutations by
$|\widehat{\mathrm{cor}}(Y_{\mathrm{real}},X_k)|>0.05$ leaves
$p=466$ indicators. We divide the mutations into three groups at the
empirical frequency terciles and sample $10$ targets from each group
(\texttt{set.seed(2026)}). We report the $20$ targets from the lowest- and
highest-frequency groups. Their frequency ranges are $[0.002,0.015]$ and
$[0.034,0.055]$ for high and low sparsity, respectively.  The representatives used to assess null calibration are
\texttt{EIF5A\_MUT} and
\texttt{ATP11A\_MUT} from high and low sparsity groups, respectively. 

\paragraph{Synthetic response}
For target $j$, the remaining $465$ mutations form $Z=X_{-j}$. For every replicate and target $j$, we fix $s=10$, discard the real outcome, and draw a fresh response from $Z$ alone. Let $S$ be a uniformly sampled subset of $s$ conditioning variables, draw $\beta_k\iidsim N(0,1)$ for $k\in S$, sample $|I|=\min\{2s,\binom{s}{2}\}$ interaction pairs, and draw
$\gamma_{k\ell}\iidsim N(0,1)$. We set
\[
\mu_i=\sum_{k\in S}\beta_kZ_{ik}
      +\sum_{(k,\ell)\in I}\gamma_{k\ell}Z_{ik}Z_{i\ell},
\qquad
  Y_i=\mu_i+\varepsilon_i.
\]
For each target and replicate, we calculate the empirical standard deviation
of the signal across the $n$ cell lines,
\[
\bar\mu\equiv\frac1n\sum_{i=1}^n\mu_i,
\qquad
\sigma_\mu\equiv\left\{\frac1{n-1}\sum_{i=1}^n(\mu_i-\bar\mu)^2\right\}^{1/2},
\qquad
\sigma\equiv
\begin{cases}
\sigma_\mu, & 0<\sigma_\mu<\infty,\\
1, & \text{otherwise}.
\end{cases}
\]
Conditional on $Z$ and the sampled signal components, we draw independent
errors $\varepsilon_i\sim N(0,\sigma^2)$. When $\sigma_\mu$ is positive and
finite, the empirical variance of the signal divided by the specified noise
variance is exactly one; the realized variance of the noise varies across
replicates. The response is generated from $Z$ alone, so
$\bm Y\independent \bm X_j\mid \bm Z$.

\paragraph{Representative example construction.} For the paired extension with conditional effects, we retain the $20$
reported $\delta=0$ settings and show nonzero effects for two fixed representatives: EIF5A (high sparsity/rare) and ATP11A (low sparsity/common). Within a representative target, we generate
\[
Y_i^{(\delta)}=Y_i^{(0)}+\delta\sigma X_{ij}.
\]
The EIF5A target with four carriers uses $\delta\in\{0,1,2,3,4\}$ and ATP11A uses
$\delta\in\{0,0.5,1,1.5,2\}$. 

\paragraph{Methods and replication}
GCM uses random forest regressions of $Y$ on $Z$ and of $X_j$ on $Z$.
For each replicate, one observed, empirically studentized GCM statistic
is calibrated in two ways: by the two-sided standard normal reference
and by adaptive marginal permutation. Both $p$-values therefore share
the observed statistic and fits obtained from the observed data. We also use adaptive permutation procedure with $h=15$ and $B_{\max}=499$ draws. Each of the $20$ reported null settings is run for $R=500$ replicates, and all tests use $\alpha=0.05$.

\subsection{Frangieh Perturb-seq analysis}
\label{app:frangieh-details}

\paragraph{Data and pair construction}
We use the control condition of the Perturb-seq experiment~\citep{frangieh2021perturbseq}. It includes $57{,}624$ cells with nonzero gene UMIs, $23{,}712$ genes, $717$ gene-targeting gRNAs spanning $239$ targets, and $74$ non-targeting (NT) gRNAs. The common control population consists of $4{,}148$ cells carrying only NT gRNAs. We consider $20$ replicates: the positive pairs are fixed and the $950$ null pairs
are resampled from the total pool of negative control pairs. The positive controls are the $50$ most highly expressed genes among the
available targets. Their treatment indicator identifies cells carrying any gRNA targeting the gene, with $238$ treated cells
on average.  For each null, we pair a random gene outside the positive control set, with mean
expression at least $0.005$ with three randomly sampled NT gRNAs.

\paragraph{Adaptive resampling hyperparameters.}
Each of the four randomization methods uses
two-sided adaptive BH calibration at $q=0.05$, with $h=15$ and at most $B_{\max}=25{,}000$ draws per pair, required in Algorithm~\ref{alg:adaptive}.

\begin{figure}[!ht]
  \centering
  \includegraphics[width=\textwidth]{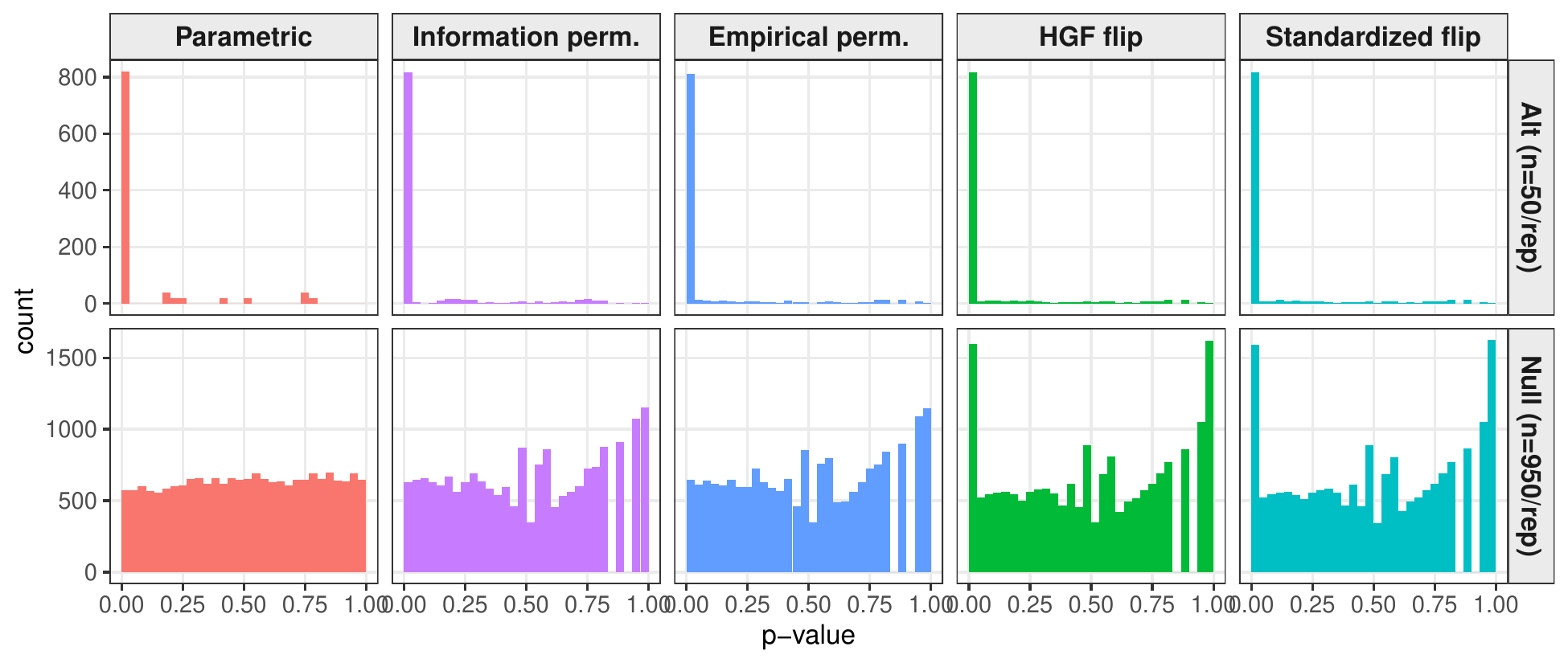}
  \caption{Pooled Frangieh $p$-value distributions for the five score
  calibrations across $20$ replicates, separated into positive controls
  (top) and synthetic null pairs formed from response genes outside the positive control set
  and non-targeting gRNAs (bottom). Colors match
  Figure~\ref{fig:nb-score-ci}.}
  \label{fig:frangieh_pv}
\end{figure}

\paragraph{Additional diagnostic plot.} Figure~\ref{fig:frangieh_pv} separates the $p$-values for positive controls
and null pairs. All five methods concentrate the positive-control $p$-values
near zero, while both sign-flip methods show an excess of small $p$-values
among null pairs. 

\section{General lemmas}\label{app:lemmas}

\begin{definition}[Conditional convergence in distribution; {\citet[Definition~1]{niu2023reconciling}}]
\label{def:cond-conv}
Let $W_n$ be a sequence of random variables and $\cF_n$ a sequence of
$\sigma$-algebras.
We say that $W_n$ converges in distribution to a random variable $W$
conditionally on $\cF_n$ if, for all $t\in\R$ at which the CDF of $W$ is
continuous,
$\PP(W_n\le t\mid\cF_n)\xrightarrow{\PP}\PP(W\le t)$.
We denote this by $W_n\mid\cF_n\convdp W$.
\end{definition}

\begin{lemma}[Conditional Slutsky; {\citet[Supplement, Theorem~3]{niu2023reconciling}}]\label{lem:slutsky}
Let $W_n$ be a sequence of random variables and $\cF_n$ a sequence of
$\sigma$-algebras.
Suppose $a_n\xrightarrow{\PP}1$ and $b_n\xrightarrow{\PP}0$.
If $W_n\mid\cF_n\convdp W$ for some $W$ with continuous CDF, then
$a_nW_n+b_n\mid\cF_n\convdp W$.
\end{lemma}

\begin{lemma}[Conditional weak convergence implies marginal convergence]
  \label{lem:cond-to-marg}
  Let $W_n$ be random variables and $\cF_n$ sigma-fields. If
  $W_n\mid\cF_n\convdp W$ for a nonrandom limit law $W$ with continuous
  distribution function, then $W_n\convd W$.
  \end{lemma}
  \begin{proof}
  Fix a continuity point $t$ of the distribution function of $W$.
  By conditional convergence,
  \[
  P_n(t)\equiv\PP(W_n\le t\mid\cF_n)
  \xrightarrow{\PP}\PP(W\le t).
  \]

  Since $0\le P_n(t)\le1$, this convergence also holds in expectation.
  The law of total expectation therefore gives
  \[
  \PP(W_n\le t)=\E[P_n(t)]\longrightarrow\PP(W\le t).
  \]
  As $t$ was arbitrary, $W_n\convd W$.
  \end{proof}

\begin{lemma}[Conditional quantile convergence; {\citet[Supplement, Lemma~1]{niu2023reconciling}}]
\label{lem:quantile}
Let $W_n$ be a sequence of random variables, $\cF_n$ a sequence of
$\sigma$-algebras, and $\alpha\in(0,1)$.
If $W_n\mid\cF_n\convdp W$ for some $W$ whose CDF is continuous and
strictly increasing at $Q_\alpha[W]$, then
$Q_\alpha[W_n\mid\cF_n]\xrightarrow{\PP}Q_\alpha[W]$.
\end{lemma}
\begin{lemma}[Conditional Lyapunov central limit theorem]
  \label{lem:conditional-lyapunov}
  For each $n$, let $\cG_n$ be a sigma-field and let
  $W_{n1},\ldots,W_{nk_n}$ be conditionally independent given
  $\cG_n$, with
  \[
  \E[W_{ni}\mid\cG_n]=0,
  \qquad i=1,\ldots,k_n.
  \]
  Write $s_n^2\equiv\sum_{i=1}^{k_n}\E[W_{ni}^2\mid\cG_n]$.
  Suppose that $0<s_n^2<\infty$ with probability tending to one and that, for
  some $\delta>0$,
  \[
  L_n
  \equiv
  \frac{
    \sum_{i=1}^{k_n}
    \E[|W_{ni}|^{2+\delta}\mid\cG_n]
  }{
    s_n^{2+\delta}
  }
  \xrightarrow{\PP}0.
  \]
  Then
  \[
  \frac{\sum_{i=1}^{k_n}W_{ni}}{s_n}
  \ \Bigm|\ \cG_n
  \ \convdp\ N(0,1).
  \]
  Both ratios are defined as zero on $\{s_n^2\notin(0,\infty)\}$.
  \end{lemma}
  \begin{proof}
  Fix $t\in\R$ and write $T_n=\sum_iW_{ni}/s_n$.
  Let $K_n(\omega,\cdot)$ be a regular conditional distribution of
  $(W_{ni})_{i=1}^{k_n}$ given $\cG_n$, and let $K_{n,i}$ be its
  $i$th marginal. These kernels exist because $\R^{k_n}$ is a standard
  Borel space. For each $n$, there is an event $\Omega_n$ with
  $\PP(\Omega_n)=1$ on which, for all $i$,
  \[
  K_n(\omega,\cdot)=\bigotimes_{i=1}^{k_n}K_{n,i}(\omega,\cdot),
  \qquad
  \int f(x)\,K_{n,i}(\omega,dx)
  =\E[f(W_{ni})\mid\cG_n](\omega),
  \]
  for $f(x)=x,x^2,|x|^{2+\delta}$, with the first conditional moment
  equal to zero.

  From any subsequence, choose a further subsequence $(n_j)$ such that
  \[
  \sum_{j=1}^\infty\left\{
  \PP(L_{n_j}>j^{-1})+
  \PP\bigl(s_{n_j}^2\notin(0,\infty)\bigr)
  \right\}<\infty.
  \]
  By Borel--Cantelli, there is an event $\Omega_{\mathrm{BC}}$ of
  probability one on which
  \[
  L_{n_j}(\omega)\longrightarrow0,
  \qquad 0<s_{n_j}^2(\omega)<\infty\quad\text{eventually}.
  \]
  Set
  \[
  \Omega_\star=\Omega_{\mathrm{BC}}\cap\bigcap_{n=1}^\infty\Omega_n,
  \qquad \PP(\Omega_\star)=1.
  \]

  Fix $\omega\in\Omega_\star$. On an auxiliary probability space with
  probability and expectation denoted by $\PP_\omega$ and $\E_\omega$,
  choose an array $(V_{j,i})$ such that, for every Borel set
  $B\subseteq\R^{k_{n_j}}$,
  \[
  \PP_\omega\!\left((V_{j,i})_{i=1}^{k_{n_j}}\in B\right)
  =K_{n_j}(\omega,B).
  \]
  The product identity on $\Omega_\star$ makes
  $V_{j,1},\ldots,V_{j,k_{n_j}}$ independent under $\PP_\omega$. For all sufficiently large $j$,
  \begin{align*}
  \E_\omega[V_{j,i}]
  &=\E[W_{n_j i}\mid\cG_{n_j}](\omega)=0,\\
  \sum_{i=1}^{k_{n_j}}\E_\omega[V_{j,i}^2]
  &=s_{n_j}^2(\omega)\in(0,\infty),\\
  \frac{\sum_{i=1}^{k_{n_j}}\E_\omega[|V_{j,i}|^{2+\delta}]}
  {s_{n_j}^{2+\delta}(\omega)}
  &=L_{n_j}(\omega)\longrightarrow0.
  \end{align*}
  The ordinary Lyapunov central limit theorem therefore gives, under
  $\PP_\omega$,
  \[
  \frac{\sum_{i=1}^{k_{n_j}}V_{j,i}}{s_{n_j}(\omega)}
  \convd N(0,1).
  \]
  Since $s_{n_j}$ is $\cG_{n_j}$-measurable, the conditional CDF
  version induced by $K_{n_j}$ satisfies
  \[
  \PP(T_{n_j}\le t\mid\cG_{n_j})(\omega)
  =\PP_\omega\!\left(
  \frac{\sum_iV_{j,i}}{s_{n_j}(\omega)}\le t\right)
  \longrightarrow\Phi(t).
  \]
  Since the initial subsequence was arbitrary, the subsequence criterion
  yields
  \[
  \PP(T_n\le t\mid\cG_n)\xrightarrow{\PP}\Phi(t),
  \]
  as required.
  \end{proof}

\begin{lemma}[Hájek central limit theorem; {\citet[Lemma~5]{barry2025permuted}}]\label{lem:hajek}
Let $X_1,X_2,\ldots$ be \iid\ with $\E[X_i]=0$ and $\E[X_i^2]=1$.
For each $n\ge2$, let $(c_{n,i})_{i=1}^n$ be a possibly data-dependent
vector of weights satisfying
\begin{enumerate}[(i)]
\item $\sum_{i=1}^n c_{n,i}=0$ a.s.;
\item $\sum_{i=1}^n c_{n,i}^2=1$ a.s.;
\item $\lim_{d\to\infty}\limsup_{n\to\infty}
\sum_{i=1}^n c_{n,i}^2\indc{|\sqrt n c_{n,i}|\ge d}=0$ a.s.;
\item $\max_{1\le i\le n}|X_i|/\sqrt n\xrightarrow{\PP}0$.
\end{enumerate}
Let $\cF_n\equiv\sigma(X_1,\ldots,X_n,c_{n,1},\ldots,c_{n,n})$,
and let $\pi$ be a uniform permutation of $\{1,\ldots,n\}$ independent
of $\cF_n$. Then
$\sum_{i=1}^n X_{\pi(i)}c_{n,i}\mid\cF_n\convdp N(0,1)$.
\end{lemma}

\begin{lemma}[Conditional Markov and Chebyshev transfer]
    \label{lem:cond-markov}\label{lem:quad-mean}
    Let $\cG_n$ be a sequence of sigma-fields.
    \begin{enumerate}[(a)]
    \item If $W_n\ge0$ and
    $\E[W_n\mid\cG_n]\xrightarrow{\PP}0$, then
    $W_n\xrightarrow{\PP}0$.
    \item If $\bm X_n$ is real-valued with
    $\E[\bm X_n^2\mid\cG_n]<\infty$ almost surely and
    \[
    \E[\bm X_n\mid\cG_n]\xrightarrow{\PP}0,
    \qquad
    \operatorname{Var}(\bm X_n\mid\cG_n)\xrightarrow{\PP}0,
    \]
    then $\bm X_n\xrightarrow{\PP}0$.
    \end{enumerate}
    \end{lemma}
    \begin{proof}
    For part~(a), fix $\varepsilon,\eta>0$ and write
    $m_n\equiv\E[W_n\mid\cG_n]$. Conditional Markov's inequality gives
    \begin{align*}
    \PP(W_n>\varepsilon)
    &\le
    \PP(m_n>\varepsilon\eta)
    +
    \E\!\left[
      \indc{m_n\le\varepsilon\eta}
      \PP(W_n>\varepsilon\mid\cG_n)
    \right]\\
    &\le
    \PP(m_n>\varepsilon\eta)+\eta.
    \end{align*}
    The first term tends to zero; then let $\eta\downarrow0$.
    
    For part~(b), all moments are taken under the conditional law given
    $\cG_n$. Let $m_n\equiv\E[\bm X_n\mid\cG_n]$. Since
    \[
    \E[(\bm X_n-m_n)^2\mid\cG_n]
    =
    \operatorname{Var}(\bm X_n\mid\cG_n)
    \xrightarrow{\PP}0,
    \]
    part~(a), applied to $(\bm X_n-m_n)^2$, yields
    $\bm X_n-m_n\xrightarrow{\PP}0$.  Together with $m_n\xrightarrow{\PP}0$
    this proves the claim.
    \end{proof}

\begin{lemma}[Maximal bound from a moment condition]
        \label{lem:max-moment}
        Let $\cG_n$ be a sigma-field and let
        $Z_{n1},\ldots,Z_{nn}$ be conditionally identically distributed given
        $\cG_n$. If $\E[|Z_{n1}|^q\mid\cG_n]=O_\PP(1)$ for some $q>0$, then
        $\max_{1\le i\le n}|Z_{ni}|=O_\PP(n^{1/q})$.
        The same conclusion holds without conditioning when
        $\E|Z_{n1}|^q=O(1)$.
        \end{lemma}
        \begin{proof}
        For $M>0$, the conditional union bound and Markov's inequality give
        \[
        \PP\left(
          \max_i|Z_{ni}|>Mn^{1/q}
          \middle|\cG_n
        \right)
        \le
        \frac1{M^q}\E[|Z_{n1}|^q\mid\cG_n].
        \]
        Given $\varepsilon>0$, choose $K<\infty$ such that the conditional
        $q$th moment exceeds $K$ with probability at most $\varepsilon$ for
        all sufficiently large $n$.  On the complementary event the preceding
        bound is at most $K/M^q$.  Choosing $M$ so that $K/M^q\le\varepsilon$
        proves the conditional claim after taking expectations.  The
        unconditional claim is the same argument without $\cG_n$.
        \end{proof}

\begin{lemma}[Hoeffding identity; {\citet[Lemma~6]{barry2025permuted}}]
\label{lem:hoeffding}
Let $n\ge2$, let $(a_i)_{i=1}^n$ and $(b_i)_{i=1}^n$ be random variables and let
$\Pi$ be a uniform random permutation of $\{1,\ldots,n\}$, independent
of $(a_i,b_i)_{i=1}^n$.
Then
\[
\E\Bigl[\frac1n\sum_{i=1}^n a_{\Pi(i)}b_i\Bigm|(a_i,b_i)_{i=1}^n\Bigr]
=\Bigl(\frac1n\sum_{i=1}^n a_i\Bigr)\Bigl(\frac1n\sum_{i=1}^n b_i\Bigr),
\]
and
\[
\var\Bigl(\frac1n\sum_{i=1}^n a_{\Pi(i)}b_i\Bigm|(a_i,b_i)_{i=1}^n\Bigr)
\le\frac{1}{n-1}\Bigl(\frac1n\sum_{i=1}^n a_i^2\Bigr)
\Bigl(\frac1n\sum_{i=1}^n b_i^2\Bigr).
\]
\end{lemma}

\begin{lemma}[Combinatorial central limit theorem; {\citet[Theorem~3]{hoeffding1951combinatorial}}]\label{lem:comb-clt}
  For each $n\ge2$, let $(c^{(n)}_{ik})_{1\le i,k\le n}$ be deterministic real numbers with all row sums zero: $\sum_{k=1}^n c^{(n)}_{ik}=0$ for each~$i$.
  Write $c^{(n)}_{\cdot k}\equiv n^{-1}\sum_i c^{(n)}_{ik}$ and
  $d^{(n)}_{ik}\equiv c^{(n)}_{ik}-c^{(n)}_{\cdot k}$.
  Let $\Pi$ be a uniform random permutation of $\{1,\ldots,n\}$. Then
  \begin{enumerate}[(a)]
  \item $\displaystyle\E\Bigl[\sum_{i=1}^n c^{(n)}_{i,\Pi(i)}\Bigr]=0$.
  \item $\displaystyle\var\Bigl(\sum_{i=1}^n c^{(n)}_{i,\Pi(i)}\Bigr)=\frac{1}{n-1}\sum_{i,k}(d^{(n)}_{ik})^2$.
  \item Suppose $\sum_{i,k}(d^{(n)}_{ik})^2>0$ for all sufficiently large $n$
  and, as $n\to\infty$,
  \begin{equation}\label{eq:comb-clt-neg}
  \frac{\max_{i,k}(d^{(n)}_{ik})^2}{\frac1n\sum_{i,k}(d^{(n)}_{ik})^2}\to 0,
  \end{equation}
then
\[
  \frac{\sum_{i=1}^n c^{(n)}_{i,\Pi(i)}}{\sqrt{\frac{1}{n-1}\sum_{i,k}(d^{(n)}_{ik})^2}}\xrightarrow{d} N(0,1).
\]
  \end{enumerate}
\end{lemma}

\begin{lemma}[Combinatorial CLT, conditional form (Lemma~\ref{lem:comb-clt}')]\label{lem:comb-clt-prime}
      Let $(\cF_n)_{n\ge1}$ be a sequence of $\sigma$-algebras.
      For each $n\ge2$, let $(c^{(n)}_{ik})_{1\le i,k\le n}$ be a $\cF_n$-measurable real array such that
      $\sum_{k=1}^n c^{(n)}_{ik}=0$ a.s.\ for each $i$.
      Write $c^{(n)}_{\cdot k}\equiv n^{-1}\sum_i c^{(n)}_{ik}$ and
      $d^{(n)}_{ik}\equiv c^{(n)}_{ik}-c^{(n)}_{\cdot k}$.
      Let $\Pi$ be a uniform random permutation of $\{1,\ldots,n\}$ independent of $\cF_n$.
      Assume $\PP\bigl(\sum_{i,k}(d^{(n)}_{ik})^2>0\bigr)\to1$ and
      \[
      \frac{\max_{i,k}(d^{(n)}_{ik})^2}{\frac1n\sum_{i,k}(d^{(n)}_{ik})^2}\xrightarrow{\PP}0.
      \]
      Then, in the sense of Definition~\ref{def:cond-conv},
      \[
      \frac{\sum_{i=1}^n c^{(n)}_{i,\Pi(i)}}{\sqrt{\frac{1}{n-1}\sum_{i,k}(d^{(n)}_{ik})^2}}\ \Big|\ \cF_n\ \convdp\ N(0,1).
      \]
      Both ratios are defined as zero on
      $\{\sum_{i,k}(d^{(n)}_{ik})^2=0\}$.
      \end{lemma}  

\begin{proof}
Fix $t\in\R$ and let $R_n$ denote the ratio in the assumption. Set
\[
s_n^2\equiv\frac1{n-1}\sum_{i,k}(d^{(n)}_{ik})^2,
\qquad
T_n^{(\pi)}\equiv\frac{\sum_i c^{(n)}_{i,\pi(i)}}{s_n},
\]
with $T_n^{(\pi)}=0$ when $s_n=0$.
Since the array is $\cF_n$-measurable and $\Pi$ is independent and
uniform, a regular conditional distribution of $T_n^{(\Pi)}$ given
$\cF_n$ is
\[
K_n(\omega,B)\equiv\frac1{n!}\sum_\pi
\indc{T_n^{(\pi)}(\omega)\in B},
\qquad B\subseteq\R\ \text{Borel},
\]
where the sum runs over all permutations of $\{1,\ldots,n\}$.

From any subsequence, choose a further subsequence $(n_j)$ such that
\[
\sum_{j=1}^\infty\left\{
\PP(R_{n_j}>j^{-1})+\PP(s_{n_j}^2=0)
\right\}<\infty.
\]
By Borel--Cantelli and the almost-sure row-sum assumptions, the event
\[
\begin{split}
\Omega_\star\equiv{}&
\{R_{n_j}\to0,\ s_{n_j}^2>0\text{ eventually}\}\\
&{}\cap\bigcap_{n\ge2}\bigcap_{i=1}^n
\left\{\sum_{k=1}^n c^{(n)}_{ik}=0\right\}
\end{split}
\]
satisfies $\PP(\Omega_\star)=1$.

Fix $\omega\in\Omega_\star$. The deterministic arrays
$(c^{(n_j)}_{ik}(\omega))_{i,k}$ have zero row sums and positive
permutation variance for all sufficiently large $j$, and
\[
\frac{\max_{i,k}(d^{(n_j)}_{ik}(\omega))^2}
{n_j^{-1}\sum_{i,k}(d^{(n_j)}_{ik}(\omega))^2}
=R_{n_j}(\omega)\longrightarrow0.
\]
Lemma~\ref{lem:comb-clt} therefore gives
\[
K_{n_j}(\omega,(-\infty,t])
=\frac1{n_j!}\sum_\pi
\indc{\frac{\sum_i c^{(n_j)}_{i,\pi(i)}(\omega)}{s_{n_j}(\omega)}\le t}
\longrightarrow\Phi(t).
\]
Since the initial subsequence was arbitrary, the subsequence criterion
and the version for the conditional distribution $K_n$ yield
\[
\PP(T_n^{(\Pi)}\le t\mid\cF_n)
=K_n(\cdot,(-\infty,t])\xrightarrow{\PP}\Phi(t),
\]
as required.
\end{proof}

\begin{lemma}[Moments of permutation averages within rows]
  \label{lem:perm-array-moments}
  Let $n\ge2$, let $\cF_n$ be a sigma-field and let
  $(h_{ik})_{1\le i,k\le n}$ be an $\cF_n$-measurable real array.
  Let $\Pi$ be a uniform permutation of $\{1,\ldots,n\}$, independent of $\cF_n$. Define
  \[
  h_{i\cdot}\equiv\frac1n\sum_kh_{ik},\qquad
  h_{\cdot k}\equiv\frac1n\sum_ih_{ik},\qquad
  h_{\cdot\cdot}\equiv\frac1{n^2}\sum_{i,k}h_{ik},
  \]
  and the doubly centered array
  \[
  \widetilde h_{ik}\equiv
  h_{ik}-h_{i\cdot}-h_{\cdot k}+h_{\cdot\cdot}.
  \]
  Then, conditional on $\cF_n$,
  \begin{align}
  \E\!\left[
    \frac1n\sum_i h_{i,\Pi(i)}
    \middle|\cF_n
  \right]
  &=h_{\cdot\cdot},
  \label{eq:perm-array-mean}\\
  \operatorname{Var}\!\left(
    \frac1n\sum_i h_{i,\Pi(i)}
    \middle|\cF_n
  \right)
  &=
  \frac1{n^2(n-1)}\sum_{i,k}\widetilde h_{ik}^2
  \le
  \frac1{n^2(n-1)}\sum_{i,k}h_{ik}^2.
  \label{eq:perm-array-var}
  \end{align}
  In particular, let $(q_i)_i$ and $(A_{ik})_{i,k}$ be $\cF_n$-measurable,
  write $A_{i\cdot}\equiv n^{-1}\sum_k A_{ik}$, and suppose
  $A_{i\cdot}=0$ for every $i$. If $h_{ik}=q_iA_{ik}$, then
  \[
  \E\!\left[
    \frac1{\sqrt n}\sum_iq_iA_{i,\Pi(i)}
    \middle|\cF_n
  \right]=0
  \]
  and
  \begin{equation}
  \operatorname{Var}\!\left(
    \frac1{\sqrt n}\sum_iq_iA_{i,\Pi(i)}
    \middle|\cF_n
  \right)
  \le
  \frac1{n(n-1)}\sum_{i,k}q_i^2A_{ik}^2.
  \label{eq:perm-array-linear-var}
  \end{equation}
  \end{lemma}
\begin{proof}
Put $c_{ik}\equiv h_{ik}-h_{i\cdot}$. Then
\[
\sum_k c_{ik}=0,\qquad
c_{\cdot k}=h_{\cdot k}-h_{\cdot\cdot},\qquad
c_{ik}-c_{\cdot k}=\widetilde h_{ik},
\]
and
\[
\frac1n\sum_i h_{i,\Pi(i)}
=h_{\cdot\cdot}+\frac1n\sum_i c_{i,\Pi(i)}.
\]
Conditional on $\cF_n$, the array $(c_{ik})$ and $h_{\cdot\cdot}$
are fixed, and $\Pi$ remains uniform. Applying
Lemma~\ref{lem:comb-clt}(a)--(b) to $(c_{ik})$ therefore gives
\eqref{eq:perm-array-mean} and the equality in
\eqref{eq:perm-array-var}.

For the inequality, the definition of $\widetilde h$ gives
\[
\sum_k\widetilde h_{ik}=0,\qquad
\sum_i\widetilde h_{ik}=0.
\]
Consequently,
\begin{align*}
\sum_{i,k}\widetilde h_{ik}
  (h_{i\cdot}+h_{\cdot k}-h_{\cdot\cdot})
&=\sum_i h_{i\cdot}\sum_k\widetilde h_{ik}
 +\sum_k h_{\cdot k}\sum_i\widetilde h_{ik}
 -h_{\cdot\cdot}\sum_{i,k}\widetilde h_{ik}\\
&=0.
\end{align*}
Expanding $h_{ik}=\widetilde h_{ik}+h_{i\cdot}+h_{\cdot k}-h_{\cdot\cdot}$
therefore gives
\[
\sum_{i,k}h_{ik}^2
=\sum_{i,k}\widetilde h_{ik}^2
 +\sum_{i,k}(h_{i\cdot}+h_{\cdot k}-h_{\cdot\cdot})^2
\ge\sum_{i,k}\widetilde h_{ik}^2.
\]

Finally, if $h_{ik}=q_iA_{ik}$ and $A_{i\cdot}=0$, then
\[
h_{i\cdot}=q_iA_{i\cdot}=0,\qquad h_{\cdot\cdot}=0.
\]
Writing $M_n\equiv n^{-1}\sum_i h_{i,\Pi(i)}$, we have
$n^{-1/2}\sum_iq_iA_{i,\Pi(i)}=\sqrt n\,M_n$, so
\begin{align*}
\E[\sqrt n\,M_n\mid\cF_n]
&=\sqrt n\,h_{\cdot\cdot}=0,\\
\operatorname{Var}(\sqrt n\,M_n\mid\cF_n)
&=n\operatorname{Var}(M_n\mid\cF_n)\\
&\le\frac{n}{n^2(n-1)}\sum_{i,k}h_{ik}^2
=\frac1{n(n-1)}\sum_{i,k}q_i^2A_{ik}^2.
\end{align*}
\end{proof}

    \begin{lemma}[Studentizer transfer under mean-square perturbations]
      \label{lem:studentizer-transfer}
      Let $R_{ni}\equiv U_{ni}+\Delta_{ni}$ and write
      $\widehat V_n(R)\equiv n^{-1}\sum_iR_{ni}^2-(n^{-1}\sum_iR_{ni})^2$.
      Let $\tau_n^2=O_\PP(1)$.  Suppose
      \[
      \frac1n\sum_iU_{ni}^2-\tau_n^2=o_\PP(1),
      \qquad
      \frac1{\sqrt n}\sum_iU_{ni}=O_\PP(1),
      \qquad
      \frac1n\sum_i\Delta_{ni}^2=o_\PP(1).
      \]
      Then
      \[
      \widehat V_n(R)-\tau_n^2=o_\PP(1).
      \]
      If, in addition, $\tau_n^2$ is bounded away from zero in probability,
      then $\widehat V_n(R)/\tau_n^2\xrightarrow{\PP}1$.
      \end{lemma}
      \begin{proof}
      By Cauchy--Schwarz,
      \[
      \left|\frac1n\sum_iU_{ni}\Delta_{ni}\right|
      \le
      \left(\frac1n\sum_iU_{ni}^2\right)^{1/2}
      \left(\frac1n\sum_i\Delta_{ni}^2\right)^{1/2}
      =o_\PP(1).
      \]
      Hence $n^{-1}\sum_iR_{ni}^2=\tau_n^2+o_\PP(1)$.  Moreover,
      \[
      \left|\frac1n\sum_i\Delta_{ni}\right|^2
      \le\frac1n\sum_i\Delta_{ni}^2=o_\PP(1),
      \qquad
      \frac1n\sum_iU_{ni}=O_\PP(n^{-1/2})=o_\PP(1).
      \]
      Thus $(n^{-1}\sum_iR_{ni})^2=o_\PP(1)$, proving the first claim.  The
      ratio statement follows from the assumed lower bound.
      \end{proof}

\section{Auxiliary regularity conditions}\label{app:aux-reg}

The conditions collected in this section are the technical regularity
requirements used in the asymptotic analysis of
Sections~\ref{sec:glm-warmup} and~\ref{sec:dr}.  We first state the
condition on the fourth moment of the outcome shared by all three procedures.
The remaining assumptions are then grouped according to the section in
which the corresponding procedure is introduced.

\subsection{Shared moment conditions on the outcome}
\label{subsec:aux-reg-outcome}

\begin{assumption}[Shared moment conditions on the outcome]
\label{assumption:reg-outcome}
Assume
\[
C_{Y,4}\equiv
\esssup_z \E\bigl[|\bm Y|^4\mid \bm Z=z\bigr]<\infty.
\]
\end{assumption}

\begin{remark}[Consequences of the shared outcome condition]
\label{rmk:shared-outcome-moments}
Let $m(\bm Z)\equiv\E[\bm Y\mid \bm Z]$ and
$b\equiv \bm Y-m(\bm Z)$.  Conditional Jensen's inequality and
$(u-v)^4\le 8(u^4+v^4)$ give
\[
\esssup_z \E[b^4\mid \bm Z=z]\le 16C_{Y,4},
\qquad
\esssup_z \E[b^2\mid \bm Z=z]\le 4C_{Y,4}^{1/2}.
\]
For later reference, set
\[
\bar m_{Y,4}\equiv16C_{Y,4},
\qquad
\bar v_Y\equiv4C_{Y,4}^{1/2}.
\]
Thus $\E[b^4\mid \bm Z]\le\bar m_{Y,4}$ and
$\E[b^2\mid \bm Z]\le\bar v_Y$ almost surely.  In particular,
$\E[\bm Y^4]<\infty$ and $\E[b^4]<\infty$.
\end{remark}

\subsection{Regularity conditions for Section~\ref{sec:glm-warmup}: the GLM score test}
\label{subsec:aux-reg-glm}

For $\mu$ in the mean space and a generic dispersion value $\vartheta$, define
\[
c_\vartheta(\mu)\equiv\frac{1}{V_\vartheta(\mu)g'(\mu)},
\qquad
W_\vartheta(\mu)\equiv\frac{1}{V_\vartheta(\mu)g'(\mu)^2},
\qquad
\kappa_\vartheta(\mu)\equiv\frac{\partial_\mu c_\vartheta(\mu)}{g'(\mu)}.
\]
Write $c_\star=c_{\vartheta_\star}$, $W_\star=W_{\vartheta_\star}$, and $\kappa_\star=\kappa_{\vartheta_\star}$.

\begin{assumption}[GLM auxiliary regularity]
\label{assumption:reg-glm}
In addition to Assumption~\ref{assumption:glm-reg} and
Assumption~\ref{assumption:reg-outcome}, assume:
\begin{enumerate}[(i)]
\item There is a constant $C_Z<\infty$ such that $\|Z_i\|\le C_Z$
almost surely, and one coordinate of $Z_i$ is identically one.
\item $\E[Z_iZ_i^\top]$ is positive definite.
\item Let
\[
K_\mu\equiv\{g^{-1}(z^\top\beta):\|z\|\le C_Z\}.
\]
There are an open neighbourhood $\cN$ of $K_\mu$ within the mean
space and a compact interval $\Theta_0$ of dispersion values containing
$\vartheta_\star$ in its interior. The inverse link $g^{-1}$ is continuously
differentiable with bounded derivative on $g(\cN)$, and
$\mu\mapsto c_\vartheta(\mu)$ is continuously differentiable on $\cN$
for every $\vartheta\in\Theta_0$.

There are deterministic constants $0<v_-\le v_+<\infty$ and
$0<g_-\le g_+<\infty$ such that, for all
$(\mu,\vartheta)\in\cN\times\Theta_0$,
\[
v_-\le V_\vartheta(\mu)\le v_+,
\qquad
g_-\le |g'(\mu)|\le g_+.
\]
There are also deterministic constants $C_0,L_0<\infty$ such that,
for every $f\in\{c,W,\kappa\}$ and all
$(\mu,\vartheta),(\mu',\vartheta')\in\cN\times\Theta_0$,
\[
|f_\vartheta(\mu)|\le C_0,
\qquad
|f_\vartheta(\mu)-f_{\vartheta'}(\mu')|
\le L_0\bigl(|\mu-\mu'|+|\vartheta-\vartheta'|\bigr).
\]
\item The limit of the profile information and the permutation score variance
are positive, respectively:
\[
I_\gamma\equiv\E[R_i^2]>0,
\qquad
\E[S_i]>0.
\]
\item The observed oracle score has finite fourth moment:
\[
\E[(\delta_i\xi_i)^4]<\infty.
\]
\end{enumerate}
\end{assumption}

\begin{remark}[Implications of the GLM regularity conditions]
\label{rmk:moments-Y}
Using the constants in Assumption~\ref{assumption:reg-glm}(iii), set
\[
C_c\equiv\frac1{v_-g_-},\qquad
w_-\equiv\frac1{v_+g_+^2},\qquad
w_+\equiv\frac1{v_-g_-^2}.
\]
The definitions of $c_\vartheta$ and $W_\vartheta$ give
\[
|c_{\vartheta_\star}(\mu_i)|\le C_c,
\qquad 0<w_-\le W_i\le w_+<\infty.
\]
Under the null mean model in Assumption~\ref{assumption:glm-reg},
$\xi_i=c_{\vartheta_\star}(\mu_i)(Y_i-\mu_i)$ and
$\mu_i=\E[Y_i\mid Z_i]$. Remark~\ref{rmk:shared-outcome-moments}
therefore gives
\begin{align*}
S_i=\E[\xi_i^2\mid Z_i]
&\le C_c^2\bar v_Y,\\
\E[\xi_i^4\mid Z_i]
&\le C_c^4\bar m_{Y,4},\\
\E[\|Z_i\xi_i\|^2]
&\le C_Z^2C_c^2\bar v_Y.
\end{align*}
In particular, $\E[\xi_i^4]\le C_c^4\bar m_{Y,4}<\infty$.

Let $\lambda_Z\equiv\lambda_{\min}\{\E[Z_iZ_i^\top]\}>0$,
which exists by Assumption~\ref{assumption:reg-glm}(ii). Then
\[
M=\E[W_iZ_iZ_i^\top]\succeq w_-\E[Z_iZ_i^\top],
\qquad \lambda_{\min}(M)\ge w_-\lambda_Z>0.
\]
Since $\E[X_i^2]=1$ by Assumption~\ref{assumption:glm-reg}(iii),
\[
\|r\|=\|\E[W_iZ_iX_i]\|\le w_+C_Z,
\qquad
\|\theta\|=\|M^{-1}r\|
\le\frac{w_+C_Z}{w_-\lambda_Z}<\infty.
\]
Finally, writing $\delta_i=X_i-\theta^\top Z_i$, the fourth-moment
assumption on $X_i$ and $\|Z_i\|\le C_Z$ imply
\[
\E[\delta_i^4]
\le8\bigl\{\E[X_i^4]+C_Z^4\|\theta\|^4\bigr\}<\infty.
\]
Cauchy--Schwarz now yields
\[
\E[\delta_i^2\xi_i^2]
\le\bigl\{\E[\delta_i^4]\,\E[\xi_i^4]\bigr\}^{1/2}<\infty.
\]
\end{remark}

\subsection{Regularity conditions for Section~\ref{sec:dr}: the GCM and PCM tests}
\label{subsec:aux-reg-dr}

\begin{assumption}[GCM auxiliary regularity]
\label{assumption:reg-gcm}
In addition to Assumption~\ref{assumption:reg-outcome}, assume:
\begin{enumerate}[(i)]
\item $\bm X$ has a uniformly bounded conditional fourth moment:
\[
C_{X,4}\equiv\esssup_z\E[|\bm X|^4\mid \bm Z=z]<\infty.
\]
\item The variance of the observed oracle summand is nondegenerate:
\[
\sigma_{\mathrm{GCM}}^2
\equiv\var\bigl((\bm X-\mu_x(\bm Z))(\bm Y-\mu_y(\bm Z))\bigr)>0.
\]
\end{enumerate}
\end{assumption}

\begin{remark}[Consequences of the GCM moment condition]
\label{rmk:gcm-weight-moments}
Assumption~\ref{assumption:reg-gcm}(i) and conditional Jensen's inequality give
$\E[\bm X^4]\le C_{X,4}$ and
\begin{align*}
\E[(\bm X-\mu_x(\bm Z))^4\mid\bm Z]
&\le 8\{\E[\bm X^4\mid\bm Z]+\mu_x(\bm Z)^4\}
\le16C_{X,4},\\
\E[(\bm X-\mu_x(\bm Z))^2\mid\bm Z]
&\le\{\E[(\bm X-\mu_x(\bm Z))^4\mid\bm Z]\}^{1/2}
\le4C_{X,4}^{1/2}.
\end{align*}
Moreover, Assumption~\ref{assumption:reg-gcm}(ii) implies
\[
\sigma_{\mathrm{GCM},0}^2
\equiv\var(\bm X)\,\E[(\bm Y-\mu_y(\bm Z))^2]>0,
\]
since either factor being zero would make the observed oracle summand
$(\bm X-\mu_x(\bm Z))(\bm Y-\mu_y(\bm Z))$ zero almost surely.
\end{remark}

\Needspace{8\baselineskip}
\begin{assumption}[PCM auxiliary regularity]\label{assumption:reg-pcm}
Let $\cH_n$ be the sigma-field generated by the training data $\cD_2$,
the independent split, and all learner seeds.
In addition to Assumption~\ref{assumption:reg-outcome}, assume:
\begin{enumerate}[(i)]
\item There exists a deterministic constant $c_Y>0$ such that
\[
\var(\bm Y\mid \bm Z=z)\ge c_Y
\qquad\text{for almost every }z.
\]

\item The observed scale of the learned direction satisfies $\PP(\sigma_{f,n}^2=0)\to0$, and, writing $a\equiv\hat f(\bm X,\bm Z)-\mu_f(\bm Z)$,
\begin{equation}\label{eq:pcm-observed-profile-l2l4}
\operatorname*{ess\,sup}_{z}
\frac{\E[a^2\mid \bm Z=z,\cH_n]}{\sigma_{f,n}^2}
=O_\PP(1),
\qquad
\frac{\E[a^4\mid\cH_n]}{\sigma_{f,n}^4}=O_\PP(1).
\end{equation}

\item The scale under the product law satisfies $\PP(\sigma_{f,n,\mathrm{perm}}^2=0)\to0$.  Moreover, there is a nonnegative $\cH_n$-measurable envelope $F$ such that, almost surely, for all $x,z$,
\begin{equation}\label{eq:pcm-perm-envelope}
\left|\hat f(x,z)-b^{\mathrm{perm}}(z)\right|\le F(x),
\qquad
\frac{\E[F(\bm X)^4\mid\cH_n]}{\sigma_{f,n,\mathrm{perm}}^4}=O_\PP(1).
\end{equation}
\end{enumerate}
\end{assumption}

\subsection{Consequences for the PCM permutation array}\label{subsec:pcm-product-law-bridge}

\begin{proposition}[Approximation under the product law for the PCM permutation array]\label{prop:pcm-product-law-bridge}
Under Assumptions~\ref{assumption:reg-outcome} and~\ref{assumption:reg-pcm}(iii), let
\[
n\equiv|\cD_1|,\qquad
B_{ik}\equiv\hat f(X_k,Z_i),\qquad
B_{i\cdot}\equiv\frac1n\sum_{k\in\cD_1}B_{ik}.
\]
All sums and maxima below are over indices in $\cD_1$. Then
\begin{align}
\frac{1}{n^2\sigma_{f,n,\mathrm{perm}}^2}
\sum_{i,k\in\cD_1}(B_{ik}-B_{i\cdot})^2
&\xrightarrow{\PP}1,\label{eq:pcm-envelope-bridge}\\
\frac{\displaystyle\max_i n^{-1}\sum_k(B_{ik}-B_{i\cdot})^2}{\sigma_{f,n,\mathrm{perm}}^2}
&=O_\PP(1),\label{eq:pcm-envelope-row}\\
\frac{\displaystyle\max_{i,k}\{Y_i-\mu_y(Z_i)\}^2(B_{ik}-B_{i\cdot})^2}{n\sigma_{f,n,\mathrm{perm}}^2}
&=o_\PP(1).\label{eq:pcm-envelope-cell}
\end{align}
Moreover,
\begin{equation}\label{eq:pcm-target-empirical-population}
\frac1{n\sigma_{f,n,\mathrm{perm}}^2}
\sum_{i\in\cD_1}\{B_{i\cdot}-b^{\mathrm{perm}}(Z_i)\}^2
=O_\PP(n^{-1}),
\end{equation}
and if $E_{n,y}^{\mathrm{PCM}}=o_\PP(1)$, then
\begin{equation}\label{eq:pcm-perm-weighted-outcome-mspe-derived}
\frac{1}{n^2\sigma_{f,n,\mathrm{perm}}^2}
\sum_{i,k\in\cD_1}
\{\mu_y(Z_i)-\hat\mu_y(Z_i)\}^2(B_{ik}-B_{i\cdot})^2
=o_\PP(1).
\end{equation}
\end{proposition}

\begin{proof}
Write $s_n^2=\sigma_{f,n,\mathrm{perm}}^2$ and work on $\{s_n^2>0\}$,
whose probability tends to one by Assumption~\ref{assumption:reg-pcm}(iii).
Set
\[
h(x,z)\equiv\hat f(x,z)-b^{\mathrm{perm}}(z),
\qquad H_{ik}\equiv h(X_k,Z_i),
\qquad \bar H_i\equiv\frac1n\sum_kH_{ik}.
\]
Given $\cH_n$, the inference observations are i.i.d. with their original law, and
\[
\int h(x,z)\,dP_X(x)=0,\qquad
\iint h(x,z)^2\,dP_X(x)\,dP_Z(z)=s_n^2.
\]
Since $B_{ik}-B_{i\cdot}=H_{ik}-\bar H_i$,
\begin{equation}\label{eq:pcm-fp-decomposition}
\frac1{n^2}\sum_{i,k}(B_{ik}-B_{i\cdot})^2
=\frac1{n^2}\sum_{i,k}H_{ik}^2-\frac1n\sum_i\bar H_i^2.
\end{equation}
By~\eqref{eq:pcm-perm-envelope} and conditional Jensen's inequality,
for $r\in\{2,4\}$,
\begin{equation}\label{eq:pcm-envelope-moments}
\frac{\E[F(\bm X)^r\mid\cH_n]}{s_n^r}
\le
\left\{\frac{\E[F(\bm X)^4\mid\cH_n]}{s_n^4}\right\}^{r/4}
=O_\PP(1).
\end{equation}
Let
\[
U_n\equiv\frac1{n(n-1)}\sum_{i\ne k}H_{ik}^2.
\]
For $i\ne k$, $X_k$ and $Z_i$ are independent given $\cH_n$, so
$\E[U_n\mid\cH_n]=s_n^2$.
Two summands $H_{ik}^2$ and $H_{j\ell}^2$ are independent given $\cH_n$
whenever $\{i,k\}\cap\{j,\ell\}=\varnothing$.
For each ordered pair $(i,k)$ with $i\ne k$, at most $4(n-1)$ ordered
pairs $(j,\ell)$, $j\ne\ell$, share an index with $(i,k)$.
The envelope and conditional Cauchy--Schwarz give
\[
\left|\cov(H_{ik}^2,H_{j\ell}^2\mid\cH_n)\right|
\le 2\E[F(\bm X)^4\mid\cH_n].
\]
Consequently,
\begin{align}
\var(U_n\mid\cH_n)
&\le
\frac{n(n-1)\,4(n-1)\,2\E[F(\bm X)^4\mid\cH_n]}{n^2(n-1)^2}\notag\\
&=\frac8n\E[F(\bm X)^4\mid\cH_n]
=O_\PP(s_n^4/n).
\label{eq:pcm-product-law-un-var}
\end{align}
Thus $\E[U_n/s_n^2-1\mid\cH_n]=0$ and
$\var(U_n/s_n^2-1\mid\cH_n)=O_\PP(n^{-1})$.
Lemma~\ref{lem:cond-markov}(b) gives $U_n/s_n^2\xrightarrow{\PP}1$.
For the diagonal contribution, conditional Markov and the $r=2$ case
of~\eqref{eq:pcm-envelope-moments} give
\[
\frac{n^{-1}\sum_iF(X_i)^2}{s_n^2}=O_\PP(1),
\qquad
\frac{n^{-2}\sum_iH_{ii}^2}{s_n^2}
\le\frac1n\frac{n^{-1}\sum_iF(X_i)^2}{s_n^2}
=O_\PP(n^{-1}).
\]
For the row means, conditional on $(X_i,Z_i,\cH_n)$, the variables
$H_{ik}$, $k\ne i$, are independent and
\[
\E[H_{ik}\mid X_i,Z_i,\cH_n]
=\int h(x,Z_i)\,dP_X(x)=0.
\]
Since $H_{ii}$ is measurable with respect to this conditioning information,
expanding $\bar H_i^2$ and taking expectations gives
\begin{align}
\E[\bar H_i^2\mid\cH_n]
&=\frac1{n^2}\E[H_{ii}^2\mid\cH_n]
+\frac1{n^2}\sum_{k\ne i}\E[H_{ik}^2\mid\cH_n]\notag\\
&=\frac1{n^2}\E[H_{ii}^2\mid\cH_n]
+\frac{n-1}{n^2}s_n^2\notag\\
&\le\frac1{n^2}\E[F(\bm X)^2\mid\cH_n]
+\frac{n-1}{n^2}s_n^2.
\label{eq:pcm-rowmean-explicit}
\end{align}
Dividing by $s_n^2$, averaging over $i$, and using the $r=2$ case
of~\eqref{eq:pcm-envelope-moments}, conditional Markov yields
\[
\frac1{ns_n^2}\sum_i\bar H_i^2=O_\PP(n^{-1}).
\]
Since $\bar H_i=B_{i\cdot}-b^{\mathrm{perm}}(Z_i)$, this proves
\eqref{eq:pcm-target-empirical-population}. Substituting into
\eqref{eq:pcm-fp-decomposition} gives
\[
\frac{\sum_{i,k}(B_{ik}-B_{i\cdot})^2}{n^2s_n^2}
=\frac{n-1}{n}\frac{U_n}{s_n^2}
+\frac{\sum_iH_{ii}^2}{n^2s_n^2}
-\frac{\sum_i\bar H_i^2}{ns_n^2}
\xrightarrow{\PP}1,
\]
which proves~\eqref{eq:pcm-envelope-bridge}.

For the row bound,
\[
\max_i\frac1n\sum_k(B_{ik}-B_{i\cdot})^2
=\max_i\left\{\frac1n\sum_kH_{ik}^2-\bar H_i^2\right\}
\le\frac1n\sum_kF(X_k)^2=O_\PP(s_n^2),
\]
proving~\eqref{eq:pcm-envelope-row}. If $E_{n,y}^{\mathrm{PCM}}=o_\PP(1)$,
put $e_{y,i}=\mu_y(Z_i)-\hat\mu_y(Z_i)$. Then
\[
\frac1{n^2s_n^2}\sum_{i,k}e_{y,i}^2(B_{ik}-B_{i\cdot})^2
\le E_{n,y}^{\mathrm{PCM}}
\frac{\max_i n^{-1}\sum_k(B_{ik}-B_{i\cdot})^2}{s_n^2}
=o_\PP(1),
\]
which proves~\eqref{eq:pcm-perm-weighted-outcome-mspe-derived}.

Finally, $|H_{ik}-\bar H_i|\le2\max_kF(X_k)$. Put
$b\equiv\bm Y-\mu_y(\bm Z)$ and $b_i\equiv Y_i-\mu_y(Z_i)$.
By Remark~\ref{rmk:shared-outcome-moments}, $\E[b^4]<\infty$.
For every $\varepsilon>0$,
\begin{align*}
\PP\!\left(\max_i|b_i|>\varepsilon n^{1/4}\right)
&\le n\PP\!\left(|b|>\varepsilon n^{1/4}\right)\\
&\le\varepsilon^{-4}
\E\!\left[b^4\indc{|b|>\varepsilon n^{1/4}}\right]
\longrightarrow0.
\end{align*}
Thus $\max_i b_i^2=o_\PP(n^{1/2})$.
Lemma~\ref{lem:max-moment}, applied to $F(X_k)/s_n$ conditional on $\cH_n$
with $q=4$, gives
\[
\frac{\max_kF(X_k)^2}{s_n^2}=O_\PP(n^{1/2}).
\]
Therefore
\[
\frac{\max_{i,k}b_i^2(B_{ik}-B_{i\cdot})^2}{ns_n^2}
\le4\frac{\max_i b_i^2}{n^{1/2}}
\frac{\max_kF(X_k)^2}{n^{1/2}s_n^2}
=o_\PP(1),
\]
proving~\eqref{eq:pcm-envelope-cell}.
\end{proof}

\section{Proofs of size validity}\label{app:proof-theory}

\subsection{Finite-sample validity}
\subsubsection{Proof of Proposition~\ref{prop:mi-exact}}\label{app:proof-mi}
\begin{proof}
By hypothesis, the joint distribution of the data is invariant under
$\cG$. For each $\pi\in\cG$, $T_n^{(\pi)}$ is the same statistic
evaluated on the transformed data. Theorem~1 of \citet{hemerik2018}
therefore gives
\[
\PP\!\left\{T_n>\mathbb Q_{1-\alpha}(T_n^{(\Pi)}\mid\cF_n)\right\}\le\alpha.
\]
For Monte Carlo calibration, take the identity together with $B$ independent
uniform draws $\Pi_1,\ldots,\Pi_B$ from $\cG$. By Theorem~2 and Section~3.4 of
\citet{hemerik2018},
\[
p=\frac{1+\sum_{b=1}^B\indc{T_n^{(\Pi_b)}\ge T_n}}{B+1}
\]
is super-uniform, so $\PP(p\le\alpha)\le\alpha$.
Under $H_0^{\mathrm{MI}}$, the i.i.d. sample satisfies
$(X^{(\pi)},Y,Z)\stackrel d=(X,Y,Z)$ for every permutation $\pi$,
which verifies the stated special case.
\end{proof}

\Needspace{8\baselineskip}
\subsection{Oracle approximation and calibration}
\subsubsection{Proof of Theorem~\ref{thm:product-score-validity}}
\label{app:proof-product-score}

\begin{proof}
All ratios below are evaluated on the events where their scales
are positive; arbitrary fixed conventions on the complementary events
do not affect any limit, since their probabilities tend to zero.
Define the oracle sums and second moments by
\begin{align*}
S_{s,n}&\equiv\frac1{\sqrt n}\sum_iU_{ni},
&M_{s,n}&\equiv\frac1n\sum_iU_{ni}^2,\\
S_{p,n}&\equiv\frac1{\sqrt n}\sum_iC_{n,i\Pi(i)},
&M_{p,n}&\equiv\frac1n\sum_iC_{n,i\Pi(i)}^2.
\end{align*}

\paragraph{Step 1: Observed oracle}
Let $\cH_n\subseteq\cF_n$ be the auxiliary conditioning sigma-field in
condition~(i). Conditional on $\cH_n$, the variables $U_{n1},\ldots,U_{nn}$
are i.i.d.\ with mean zero and variance $\sigma_{s,n}^2$. By condition~(i),
\[
\frac{\sum_{i=1}^n\E[|U_{ni}|^4\mid\cH_n]}
{\{\sum_{i=1}^n\E[U_{ni}^2\mid\cH_n]\}^2}
=
\frac{\E[U_{n1}^4\mid\cH_n]}{n\sigma_{s,n}^4}
=o_\PP(1).
\]
Lemma~\ref{lem:conditional-lyapunov} gives
\[
\frac{S_{s,n}}{\sigma_{s,n}}\ \Bigm|\ \cH_n\ \convdp\ N(0,1),
\qquad
\frac{S_{s,n}}{\sigma_{s,n}}\convd N(0,1)
\]
by Lemma~\ref{lem:cond-to-marg}. Moreover,
\[
\E\!\left[\frac{M_{s,n}}{\sigma_{s,n}^2}-1\Bigm|\cH_n\right]=0,
\qquad
\operatorname{Var}\!\left(\frac{M_{s,n}}{\sigma_{s,n}^2}\Bigm|\cH_n\right)
\le\frac{\E[U_{n1}^4\mid\cH_n]}{n\sigma_{s,n}^4}
=o_\PP(1).
\]
Applying Lemma~\ref{lem:cond-markov}(b) to
$M_{s,n}/\sigma_{s,n}^2-1$ with $\cG_n=\cH_n$ yields
\[
\frac{M_{s,n}}{\sigma_{s,n}^2}\convp1.
\]

\paragraph{Step 2: Permutation oracle}
Put $D_{ik}\equiv C_{n,ik}-C_{n,\cdot k}$, using the column means
$C_{n,\cdot k}$ from condition~(ii).
Centering within each row implies that $(D_{ik})$ has zero row and column sums.
Conditionally on $\cF_n$, Lemma~\ref{lem:comb-clt}(a)--(b) gives
\begin{align*}
\E[S_{p,n}\mid\cF_n]&=0,\\
v_{p,n}^2\equiv\operatorname{Var}(S_{p,n}\mid\cF_n)
&=\frac1{n(n-1)}\sum_{i,k}D_{ik}^2
=\frac n{n-1}\sigma_{p,n}^2
-\frac1{n-1}\sum_kC_{n,\cdot k}^2.
\end{align*}
The assumption on the column mean gives $v_{p,n}^2/\sigma_{p,n}^2\convp1$;
in particular, $v_{p,n}^2>0$ with probability tending to one.
Since $\max_{i,k}D_{ik}^2\le4\max_{i,k}C_{n,ik}^2$,
\[
\frac{\max_{i,k}D_{ik}^2}{n^{-1}\sum_{i,k}D_{ik}^2}
\le\frac{4n}{n-1}
\frac{\max_{i,k}C_{n,ik}^2}{n\sigma_{p,n}^2}
\frac{\sigma_{p,n}^2}{v_{p,n}^2}
=o_\PP(1).
\]
Lemma~\ref{lem:comb-clt-prime} and conditional Slutsky
(Lemma~\ref{lem:slutsky}) consequently give
\[
\frac{S_{p,n}}{\sigma_{p,n}}\ \Bigm|\ \cF_n
\ \convdp\ N(0,1).
\]
Apply Lemma~\ref{lem:perm-array-moments} to $h_{ik}=C_{n,ik}^2$.  Since
$M_{p,n}=n^{-1}\sum_i C_{n,i\Pi(i)}^2$ and
$\sigma_{p,n}^2=n^{-2}\sum_{i,k}C_{n,ik}^2$, \eqref{eq:perm-array-mean} gives
\[
\E[M_{p,n}/\sigma_{p,n}^2\mid\cF_n]=1.
\]
Applying the variance bound in~\eqref{eq:perm-array-var} with
$h_{ik}=C_{n,ik}^2$ gives
\begin{align}
\operatorname{Var}(M_{p,n}/\sigma_{p,n}^2\mid\cF_n)
&=\frac{1}{\sigma_{p,n}^4}\operatorname{Var}(M_{p,n}\mid\cF_n)\notag\\
&\le\frac{\sum_{i,k}C_{n,ik}^4}{n^2(n-1)\sigma_{p,n}^4}\notag\\
&\le\frac{(\max_{i,k}C_{n,ik}^2)\sum_{i,k}C_{n,ik}^2}
{n^2(n-1)\sigma_{p,n}^4}\notag\\
&=\frac n{n-1}\frac{\max_{i,k}C_{n,ik}^2}{n\sigma_{p,n}^2}
=o_\PP(1),
\label{eq:perm-oracle-second-moment-var}
\end{align}
where the last equality uses $\sum_{i,k}C_{n,ik}^2=n^2\sigma_{p,n}^2$.
Applying Lemma~\ref{lem:cond-markov}(b) to
$M_{p,n}/\sigma_{p,n}^2-1$ with $\cG_n=\cF_n$ yields
\[
\frac{M_{p,n}}{\sigma_{p,n}^2}\convp1.
\]
\paragraph{Step 3: Transfer to fitted statistics}
Write
\[
\widehat S_{s,n}=\frac1{\sqrt n}\sum_i\hat R_i,
\qquad
\widehat S_{p,n}=\frac1{\sqrt n}\sum_i\hat R_i^{(\Pi)},
\qquad
\widehat V_{s,n}=\widehat V_n,
\quad
\widehat V_{p,n}=\widehat V_n^{(\Pi)}.
\]
Steps~1--2, together with Lemma~\ref{lem:cond-to-marg} for the permutation
oracle, give, for $r=s,p$,
\begin{equation}\label{eq:oracle-transfer-inputs}
\frac{M_{r,n}}{\sigma_{r,n}^2}\convp1,
\qquad
\frac{S_{r,n}}{\sigma_{r,n}}=O_\PP(1).
\end{equation}
The second bound in condition~(iii) of Theorem~\ref{thm:product-score-validity} gives
\begin{equation}\label{eq:fitted-ms-transfer}
\frac1n\sum_i
\left(\frac{\Delta_{r,ni}}{\sigma_{r,n}}\right)^2=o_\PP(1).
\end{equation}
Apply Lemma~\ref{lem:studentizer-transfer}, with $\tau_n^2=1$, to
\[
\left(\frac{U_{ni}}{\sigma_{s,n}},
      \frac{\Delta_{s,ni}}{\sigma_{s,n}}\right)
\quad\text{and}\quad
\left(\frac{C_{n,i\Pi(i)}}{\sigma_{p,n}},
      \frac{\Delta_{p,ni}}{\sigma_{p,n}}\right).
\]
Equations~\eqref{eq:oracle-transfer-inputs}--\eqref{eq:fitted-ms-transfer}
then yield
\begin{equation}\label{eq:fitted-var-transfer}
\frac{\widehat V_{r,n}}{\sigma_{r,n}^2}\convp1,
\qquad r=s,p.
\end{equation}

For the observed numerator, the first bound in condition~(iii) and Step~1 give
\[
\frac{\widehat S_{s,n}}{\sigma_{s,n}}
=
\frac{S_{s,n}}{\sigma_{s,n}}
+
\frac{\widehat S_{s,n}-S_{s,n}}{\sigma_{s,n}}
\convd N(0,1).
\]
Together with~\eqref{eq:fitted-var-transfer}, ordinary Slutsky gives
\begin{equation}\label{eq:fitted-observed-clt}
T_n
=
\frac{\widehat S_{s,n}/\sigma_{s,n}}
{(\widehat V_{s,n}/\sigma_{s,n}^2)^{1/2}}
\convd N(0,1).
\end{equation}

For the permutation numerator, Step~2 and the first bound in condition~(iii), followed by
conditional Slutsky (Lemma~\ref{lem:slutsky}), give
\[
\frac{\widehat S_{p,n}}{\sigma_{p,n}}
\ \Bigm|\ \cF_n
\ \convdp\ N(0,1).
\]
By~\eqref{eq:fitted-var-transfer},
$(\widehat V_{p,n}/\sigma_{p,n}^2)^{-1/2}\convp1$.
Another application of conditional Slutsky (Lemma~\ref{lem:slutsky}) yields
\begin{equation}\label{eq:fitted-perm-clt}
T_n^{(\Pi)}
=
\frac{\widehat S_{p,n}}{\sigma_{p,n}}
\left(\frac{\widehat V_{p,n}}{\sigma_{p,n}^2}\right)^{-1/2}
\ \Bigm|\ \cF_n
\ \convdp\ N(0,1).
\end{equation}

\paragraph{Step 4: Calibration}
Let
\[
q_{n,1-\alpha}\equiv
\mathbb Q_{1-\alpha}(T_n^{(\Pi)}\mid\cF_n).
\]
By~\eqref{eq:fitted-perm-clt} and Lemma~\ref{lem:quantile},
\begin{equation}\label{eq:perm-critical-value-limit}
q_{n,1-\alpha}\convp z_{1-\alpha}.
\end{equation}
Together with~\eqref{eq:fitted-observed-clt}, Slutsky's theorem gives
\[
T_n-q_{n,1-\alpha}\convd N(-z_{1-\alpha},1).
\]
Since the limiting distribution is continuous at zero,
\[
\PP\!\left(T_n>q_{n,1-\alpha}\right)
=\PP\!\left(T_n-q_{n,1-\alpha}>0\right)
\longrightarrow1-\Phi(z_{1-\alpha})=\alpha.
\]
\end{proof}

\subsection{Asymptotic validity}
\subsubsection{GLM score test}
For $b$ near $\beta$ and $\vartheta\in\Theta_0$, set
\[
s_i(b;\vartheta)\equiv(Y_i-\mu_i(b))c_\vartheta(\mu_i(b)),\qquad
W_i(b;\vartheta)\equiv W_\vartheta(\mu_i(b)).
\]
The oracle quantities satisfy $s_i(\beta;\vartheta_\star)=\xi_i$ and $W_i(\beta;\vartheta_\star)=W_i$, whereas all fitted quantities use $(\hat\beta,\hat\vartheta)$. A direct calculation using the chain rule gives
\begin{equation}\label{eq:glm-s-derivative}
\nabla_b s_i(b;\vartheta)
=
\left\{-W_i(b;\vartheta)
+(Y_i-\mu_i(b))\kappa_\vartheta(\mu_i(b))\right\}Z_i.
\end{equation}
Lemmas~\ref{lem:glm-lipschitz}--\ref{lem:r-hat-sums} assume
Assumptions~\ref{assumption:glm-reg}, \ref{assumption:reg-outcome}, and~\ref{assumption:reg-glm}.

\begin{lemma}[Local plug-in bounds]
\label{lem:glm-lipschitz}
There are a neighbourhood $\cB$ of $\beta$ and a constant $C<\infty$ such that, for all $b,b'\in\cB$, $\vartheta,\vartheta'\in\Theta_0$, and every $i$,
\begin{align*}
|\mu_i(b)-\mu_i(b')|&\le C\|b-b'\|,\\
|s_i(b;\vartheta)-s_i(b';\vartheta')|
&\le C(1+|Y_i|)\{\|b-b'\|+|\vartheta-\vartheta'|\},\\
|s_i(b;\vartheta)^m-s_i(b';\vartheta')^m|
&\le C(1+|Y_i|^m)\{\|b-b'\|+|\vartheta-\vartheta'|\},\quad m\in\{2,4\},\\
|W_i(b;\vartheta)^m-W_i(b';\vartheta')^m|
&\le C\{\|b-b'\|+|\vartheta-\vartheta'|\},\quad m\in\{1,2\}.
\end{align*}
The same bound, with envelope $C(1+|Y_i|)$, holds for the difference of $\nabla_b s_i(b;\vartheta)$ at two such parameter pairs.
\end{lemma}

\begin{proof}
Choose a closed neighbourhood $\cB$ of $\beta$ small enough that
\[
\mu_i(b)=g^{-1}(Z_i^\top b)\in\cN_0\subset\cN
\qquad
\text{for all }b\in\cB\text{ and all }i.
\]
This is possible because $\|Z_i\|\le C_Z$ and $K_\mu\subset\cN$ is compact.
Choose $C_\mu<\infty$ with $|\mu|\le C_\mu$ on $\cN_0$.
Assumption~\ref{assumption:reg-glm}(iii) gives, for
$f\in\{c,W,\kappa\}$,
\[
|f_\vartheta(\mu)|\le C_0,\qquad
|f_\vartheta(\mu)-f_{\vartheta'}(\mu')|
\le L_0\{|\mu-\mu'|+|\vartheta-\vartheta'|\}.
\]
Let
\[
\Delta\equiv\|b-b'\|+|\vartheta-\vartheta'|,
\qquad
\mu\equiv\mu_i(b),\quad \mu'\equiv\mu_i(b').
\]
Since $(g^{-1})'$ is bounded on $g(\cN_0)$,
\begin{equation}
\label{eq:local-mu-bound}
|\mu-\mu'|
\le \sup_{\eta\in g(\cN_0)}|(g^{-1})'(\eta)|\,
|Z_i^\top(b-b')|
\le C\|b-b'\|.
\end{equation}
For the score contribution,
\begin{align*}
s_i(b;\vartheta)-s_i(b';\vartheta')
&=(Y_i-\mu)\{c_\vartheta(\mu)-c_{\vartheta'}(\mu')\}
 +(\mu'-\mu)c_{\vartheta'}(\mu'),
\end{align*}
so, using \eqref{eq:local-mu-bound},
\begin{align*}
|s_i(b;\vartheta)-s_i(b';\vartheta')|
&\le (|Y_i|+C_\mu)L_0\{|\mu-\mu'|+|\vartheta-\vartheta'|\}
   +C_0|\mu-\mu'|\\
&\le C(1+|Y_i|)\Delta.
\end{align*}
Also
\[
|s_i(b;\vartheta)|
\le (|Y_i|+C_\mu)C_0
\le C(1+|Y_i|),
\]
and the same bound holds at $(b',\vartheta')$. Therefore, for
$m\in\{2,4\}$,
\begin{align*}
|s_i(b;\vartheta)^m-s_i(b';\vartheta')^m|
&\le m\max\{|s_i(b;\vartheta)|,|s_i(b';\vartheta')|\}^{m-1}
   |s_i(b;\vartheta)-s_i(b';\vartheta')|\\
&\le C(1+|Y_i|)^m\Delta
\le C(1+|Y_i|^m)\Delta.
\end{align*}
Likewise,
\[
|W_i(b;\vartheta)-W_i(b';\vartheta')|
\le L_0\{|\mu-\mu'|+|\vartheta-\vartheta'|\}
\le C\Delta,
\]
and hence, for $m\in\{1,2\}$,
\begin{align*}
|W_i(b;\vartheta)^m-W_i(b';\vartheta')^m|
&\le mC_0^{m-1}|W_i(b;\vartheta)-W_i(b';\vartheta')|\\
&\le C\Delta.
\end{align*}
Finally, by \eqref{eq:glm-s-derivative},
\begin{align*}
&\|\nabla_b s_i(b;\vartheta)-\nabla_b s_i(b';\vartheta')\|\\
&\quad\le \|Z_i\|\Bigl(
 |W_i(b;\vartheta)-W_i(b';\vartheta')|
 +|(Y_i-\mu)\kappa_\vartheta(\mu)
   -(Y_i-\mu')\kappa_{\vartheta'}(\mu')|
\Bigr).
\end{align*}
The second term satisfies
\begin{align*}
&|(Y_i-\mu)\kappa_\vartheta(\mu)
   -(Y_i-\mu')\kappa_{\vartheta'}(\mu')|\\
&\quad\le (|Y_i|+C_\mu)
 |\kappa_\vartheta(\mu)-\kappa_{\vartheta'}(\mu')|
 +C_0|\mu-\mu'|\\
&\quad\le C(1+|Y_i|)\Delta.
\end{align*}
Since $\|Z_i\|\le C_Z$, this gives
\[
\|\nabla_b s_i(b;\vartheta)-\nabla_b s_i(b';\vartheta')\|
\le C(1+|Y_i|)\Delta,
\]
which proves all the stated bounds.
\end{proof}

  \begin{lemma}[Empirical GLM quantities]
    \label{lem:glm-info-consistency}

    \begin{align*}
    \frac1nZ^\top\hat WZ&\convp M,
    &
    \left(\frac1nZ^\top\hat WZ\right)^{-1}
    &\convp M^{-1},\\
    \frac1nX^\top\hat WX&\convp \E[W_iX_i^2],
    &
    \frac1nZ^\top\hat WX&\convp r,\\
    \frac1n\sum_{i=1}^n\hat W_i&\convp\E[W_i],
    &
    \frac1n\sum_{i=1}^n\hat W_i^2&\convp\E[W_i^2].
    \end{align*}
    For each fixed coordinate $j$,
    \[
    \frac1n\sum_{i=1}^n\hat W_i^2Z_{ij}^2
    \convp
    \E[W_i^2Z_{ij}^2].
    \]
    \end{lemma}

\begin{proof}
      By Lemma~\ref{lem:glm-lipschitz}, on an event whose probability
      tends to one,
      \[
      \left\|
      \frac1n\sum_iZ_iZ_i^\top(\hat W_i-W_i)
      \right\|
      \le C\bigl(\|\hat\beta-\beta\|+|\hat\vartheta-\vartheta_\star|\bigr)
      \frac1n\sum_i\|Z_i\|^2=o_\PP(1).
      \]
      The oracle average $n^{-1}\sum_iW_iZ_iZ_i^\top$ converges almost
      surely to $M$, proving the first convergence; positive definiteness of
      $M$ and continuity of inversion prove the second, with the inverse
      defined as zero when singular. Indeed,
      \[
      \PP(Z^\top\hat WZ\text{ is singular})
      \le\PP\!\left\{\left\|n^{-1}Z^\top\hat WZ-M\right\|
      \ge\lambda_{\min}(M)\right\}\longrightarrow0.
      \]
      Similarly,
      \begin{align*}
      \left|\frac1n\sum_iX_i^2(\hat W_i-W_i)\right|
      &\le C\bigl(\|\hat\beta-\beta\|+|\hat\vartheta-\vartheta_\star|\bigr)\frac1n\sum_iX_i^2=o_\PP(1),\\
      \left\|\frac1n\sum_iZ_iX_i(\hat W_i-W_i)\right\|
      &\le C\bigl(\|\hat\beta-\beta\|+|\hat\vartheta-\vartheta_\star|\bigr)
      \frac1n\sum_i\|Z_i\||X_i|=o_\PP(1).
      \end{align*}
      The corresponding oracle averages converge almost surely to
      $\E[W_iX_i^2]$ and $r$, respectively.  Finally,
      \[
      \frac1n\sum_i|\hat W_i^m-W_i^m|
      \le C\bigl(\|\hat\beta-\beta\|+|\hat\vartheta-\vartheta_\star|\bigr)=o_\PP(1),
      \qquad m\in\{1,2\},
      \]
      and, because $Z_{ij}$ is bounded,
      \[
      \frac1n\sum_iZ_{ij}^2|\hat W_i^2-W_i^2|
      =o_\PP(1).
      \]
      The strong law applied to the oracle terms establishes the remaining
      claims.
      \end{proof}

      \begin{lemma}[Dispersion plug-in negligibility]\label{lem:glm-dispersion-plugin}

      \[
      \frac1{\sqrt n}\sum_{i=1}^n
      \begin{pmatrix}X_i\\ Z_i\end{pmatrix}
      \bigl\{s_i(\beta;\hat\vartheta)-s_i(\beta;\vartheta_\star)\bigr\}
      =o_\PP(1).
      \]
      \end{lemma}

\begin{proof}
      It is enough to prove the claim for one coordinate. Let $q(\bm X,\bm Z)$ denote either $\bm X$ or one coordinate of $\bm Z$, and for $\vartheta\in\Theta_0$ define
      \[
      f_\vartheta(O)
      \equiv q(\bm X,\bm Z)\{\bm Y-\mu_\beta(\bm Z)\}
        c_\vartheta\!\bigl(\mu_\beta(\bm Z)\bigr),
      \qquad O=(\bm X,\bm Y,\bm Z).
      \]
      Let $P$ be the law of $O$, and write $Ph=\E[h(O)]$.
      Under~\eqref{eq:glm-null-mean}, $P f_\vartheta=0$ for every $\vartheta\in\Theta_0$ by conditioning on $(\bm X,\bm Z)$:
      \[
      P f_\vartheta
      =\E\!\left[q(\bm X,\bm Z)c_\vartheta(\mu_\beta(\bm Z))
      \E[\bm Y-\mu_\beta(\bm Z)\mid\bm X,\bm Z]\right]=0.
      \]
      Assumption~\ref{assumption:reg-glm}(iii) yields, for a constant $C$ independent of $\vartheta,\vartheta'$,
      \begin{equation}\label{eq:glm-disp-lipschitz-class}
      |f_\vartheta(O)-f_{\vartheta'}(O)|
      \le L(O)|\vartheta-\vartheta'|,
      \qquad
      L(O)\equiv C|q(\bm X,\bm Z)|(1+|\bm Y|).
      \end{equation}
      The envelope is square integrable. This is immediate when $q$ is a bounded coordinate of $\bm Z$. For $q=\bm X$, Assumptions~\ref{assumption:glm-reg} and~\ref{assumption:reg-outcome}, together with Cauchy--Schwarz, give
      \[
      \E[\bm X^2\bm Y^2]
      \le\{\E[\bm X^4]\E[\bm Y^4]\}^{1/2}<\infty,
      \]
      and hence $P L^2<\infty$.

      Since $\Theta_0$ is a compact subset of $\mathbb R$, the Lipschitz bound~\eqref{eq:glm-disp-lipschitz-class} and $P L^2<\infty$ imply that
      $\mathcal F\equiv\{f_\vartheta:\vartheta\in\Theta_0\}$ is $P$-Donsker; see \citet[Example~19.7]{vanderVaart1998}. Moreover,
      \[
      P\bigl(f_{\hat\vartheta}-f_{\vartheta_\star}\bigr)^2
      \le P L^2\,|\hat\vartheta-\vartheta_\star|^2
      =o_\PP(1),
      \]
      by consistency of $\hat\vartheta$. Lemma~19.24 of \citet{vanderVaart1998} therefore gives
      \[
      \mathbb G_n\bigl(f_{\hat\vartheta}-f_{\vartheta_\star}\bigr)=o_\PP(1),
      \qquad
      \mathbb G_n h\equiv\frac1{\sqrt n}\sum_i\{h(X_i,Y_i,Z_i)-Ph\}.
      \]
      Since $P f_\vartheta=0$ for all $\vartheta\in\Theta_0$, the last display is exactly
      \[
      \frac1{\sqrt n}\sum_{i=1}^n q(X_i,Z_i)
      \{s_i(\beta;\hat\vartheta)-s_i(\beta;\vartheta_\star)\}
      =o_\PP(1).
      \]
      Applying the same argument to $q=\bm X$ and to the finitely many coordinates of $\bm Z$ proves the vector statement.
      \end{proof}

      \begin{lemma}[Asymptotic linearity of the null estimator]
        \label{lem:glm-linearity}

        \[
        \sqrt n(\hat\beta-\beta)
        =M^{-1}\frac1{\sqrt n}\sum_{i=1}^nZ_i\xi_i+o_\PP(1).
        \]
        \end{lemma}

\begin{proof}
Set $J_n(b;\vartheta)\equiv n^{-1}\sum_iZ_i\nabla_b s_i(b;\vartheta)^\top$. The integral mean-value formula gives
\[
0=U_n(\hat\beta;\hat\vartheta)
=U_n(\beta;\hat\vartheta)+\bar J_n(\hat\beta-\beta),\qquad
\bar J_n\equiv\int_0^1J_n(\beta+t(\hat\beta-\beta);\hat\vartheta)\,dt.
\]
Lemma~\ref{lem:glm-dispersion-plugin} gives
\[
\sqrt n\,U_n(\beta;\hat\vartheta)
=\frac1{\sqrt n}\sum_iZ_i\xi_i+o_\PP(1).
\]
By~\eqref{eq:glm-s-derivative},
\[
J_n(\beta;\vartheta_\star)
=-\frac1n\sum_iW_iZ_iZ_i^\top
+\frac1n\sum_i(Y_i-\mu_i)\kappa_\star(\mu_i)Z_iZ_i^\top
\convp-M.
\]
Here the second average has mean zero because $\E[Y_i-\mu_i\mid Z_i]=0$;
both averages have integrable summands by Assumption~\ref{assumption:reg-glm}
and Remark~\ref{rmk:shared-outcome-moments}, so the law of large numbers applies.
On an event whose probability tends to one,
Lemma~\ref{lem:glm-lipschitz} and $\|Z_i\|\le C_Z$ give
\begin{align*}
\|\bar J_n-J_n(\beta;\vartheta_\star)\|
&\le\int_0^1\frac1n\sum_i\|Z_i\|
\bigl\|\nabla_b s_i(\beta+t(\hat\beta-\beta);\hat\vartheta)
-\nabla_b s_i(\beta;\vartheta_\star)\bigr\|\,dt\\
&\le CC_Z\{\|\hat\beta-\beta\|+|\hat\vartheta-\vartheta_\star|\}
\frac1n\sum_i(1+|Y_i|)=o_\PP(1).
\end{align*}
Thus $\bar J_n\convp-M$. Since $M$ is positive definite by
Remark~\ref{rmk:moments-Y}, $\bar J_n$ is invertible with probability tending
to one and $\bar J_n^{-1}\convp-M^{-1}$.

The vectors $Z_i\xi_i$ are independent and centered. By
Remark~\ref{rmk:moments-Y},
\[
\E\left\|\frac1{\sqrt n}\sum_iZ_i\xi_i\right\|^2
=\E\|Z_i\xi_i\|^2
\le C_Z^2C_c^2\bar v_Y<\infty.
\]
Hence $n^{-1/2}\sum_iZ_i\xi_i=O_\PP(1)$, and rearranging the score equation gives
\begin{align*}
\sqrt n(\hat\beta-\beta)
&=-\bar J_n^{-1}\left\{\frac1{\sqrt n}\sum_iZ_i\xi_i+o_\PP(1)\right\}\\
&=M^{-1}\frac1{\sqrt n}\sum_iZ_i\xi_i+o_\PP(1).
\end{align*}
\end{proof}

        \begin{lemma}[Moments of the fitted working residuals]
          \label{lem:r-hat-sums}
          For $m\in\{2,4\}$,
          \[
          \frac1n\sum_{i=1}^n\hat\epsilon_i^m
          \longrightarrow
          \E[\xi_i^m]
          \qquad\text{almost surely}.
          \]
          In particular,
          \[
          \liminf_n\frac1n\sum_i\hat\epsilon_i^2>0,
          \qquad
          \limsup_n\frac1n\sum_i\hat\epsilon_i^4<\infty
          \qquad\text{almost surely}.
          \]
          \end{lemma}
          
          \begin{proof}
By Lemma~\ref{lem:glm-lipschitz} and the almost-sure consistency of $(\hat\beta,\hat\vartheta)$, on an event of probability one and for all sufficiently large $n$,
\[
\left|\frac1n\sum_i(\hat\epsilon_i^m-\xi_i^m)\right|
\le C\bigl(\|\hat\beta-\beta\|+|\hat\vartheta-\vartheta_\star|\bigr)
\frac1n\sum_i(1+|Y_i|^m),\qquad m\in\{2,4\}.
\]
The first factor tends to zero almost surely and the second converges almost surely to a finite constant. The strong law also gives $n^{-1}\sum_i\xi_i^m\to\E[\xi_i^m]$. Finally, $\E[\xi_i^2]=\E[S_i]>0$ by Assumption~\ref{assumption:reg-glm}(iv), and $\E[\xi_i^4]<\infty$ by Remark~\ref{rmk:moments-Y}.
\end{proof}

\subsubsection{Proof of Proposition~\ref{prop:glm-unperm}}\label{app:proof-glm-unperm}

\begin{proof}
  Because $U_n(\hat\beta;\hat\vartheta)=0$,
  $\sum_iZ_i\hat\epsilon_i=0$, and hence
  $\sum_i\hat\nu_i^{\mathrm{GLM}}\hat\epsilon_i
  =\sum_iX_i\hat\epsilon_i$. Write
  $N_n(b;\vartheta)=n^{-1/2}\sum_iX_i s_i(b;\vartheta)$.
  The numerator of~\eqref{eq:glm-tilde-Tn} is $N_n(\hat\beta;\hat\vartheta)$.
  
  \paragraph{Step 1: Expansion of the numerator}
  Define $K_n(b;\vartheta)\equiv n^{-1}\sum_iX_i\nabla_b s_i(b;\vartheta)^\top$.
  The integral mean value formula gives
  \[
  N_n(\hat\beta;\hat\vartheta)
  =N_n(\beta;\hat\vartheta)+\bar K_n\sqrt n(\hat\beta-\beta),
  \qquad
  \bar K_n\equiv\int_0^1K_n(\beta+t(\hat\beta-\beta);\hat\vartheta)\,dt.
  \]
  Lemma~\ref{lem:glm-dispersion-plugin} implies
  \[
  N_n(\beta;\hat\vartheta)
  =\frac1{\sqrt n}\sum_iX_i\xi_i+o_\PP(1).
  \]
  At $(b,\vartheta)=(\beta,\vartheta_\star)$, equation~\eqref{eq:glm-s-derivative} gives
  \[
  K_n(\beta;\vartheta_\star)
  =-\frac1n\sum_iX_iW_iZ_i^\top
  +\frac1n\sum_iX_i(Y_i-\mu_i)\kappa_\star(\mu_i)Z_i^\top.
  \]
  The first average converges to $-r^\top$.  For the second,~\eqref{eq:glm-null-mean} gives
  \[
  \E[X_i(Y_i-\mu_i)\mid Z_i]
  =\E[X_i\E[Y_i-\mu_i\mid X_i,Z_i]\mid Z_i]=0.
  \]
  Its summands are integrable, so the strong law sends it to zero.
  Furthermore, by Lemma~\ref{lem:glm-lipschitz},
  \[
  \begin{aligned}
  &\sup_{0\le t\le1}
  \|K_n(\beta+t(\hat\beta-\beta);\hat\vartheta)-K_n(\beta;\vartheta_\star)\|\\
  &\qquad\le
  C\bigl(\|\hat\beta-\beta\|+|\hat\vartheta-\vartheta_\star|\bigr)
  \frac1n\sum_i|X_i|(1+|Y_i|)
  =o_\PP(1),
  \end{aligned}
  \]
  where the sample average is $O_\PP(1)$ by Cauchy--Schwarz.  Hence
  $\bar K_n\convp-r^\top$.  Lemma~\ref{lem:glm-linearity} now yields
  \begin{align}
  N_n(\hat\beta;\hat\vartheta)
  &=\frac1{\sqrt n}\sum_iX_i\xi_i
  -r^\top M^{-1}\frac1{\sqrt n}\sum_iZ_i\xi_i
  +o_\PP(1)\notag\\
  &=\frac1{\sqrt n}\sum_i
  (X_i-\theta^\top Z_i)\xi_i+o_\PP(1).
  \label{eq:glm-numerator-linear}
  \end{align}
  
  \paragraph{Step 2: Limit of the oracle numerator}
  Let $U_{ni}=\delta_i\xi_i$, with $\delta_i$ as in Section~\ref{sec:glm-warmup}. Under the null,
  $\E[\xi_i\mid X_i,Z_i]=\E[Y_i-\mu_i\mid X_i,Z_i]/(v_i g_i')=0$, so
  $\E[U_{ni}]=0$.  Moreover,
  \[
  \E[U_{ni}^2]=\E[\delta_i^2\xi_i^2]=\tau_s^2<\infty
  \]
  by Remark~\ref{rmk:moments-Y}. Therefore the
  i.i.d. central limit theorem and~\eqref{eq:glm-numerator-linear} give
  \[
  N_n(\hat\beta;\hat\vartheta)
  \convd
  N(0,\tau_s^2).
  \]
  
  \paragraph{Step 3: Limit of the profile studentizer}
  The fitted weights in $\tilde V_n^{\mathrm{GLM}}$ are $\hat W_i=W_i(\hat\beta;\hat\vartheta)$. Lemma~\ref{lem:glm-info-consistency} gives
  \[
  \frac1nX^\top\hat W X\convp\E[W_iX_i^2],\qquad
  \frac1nZ^\top\hat W X\convp r,\qquad
  \frac1nZ^\top\hat W Z\convp M.
  \]
  Since $M$ is positive definite, continuity of matrix inversion yields
  \begin{align*}
  \tilde V_n^{\mathrm{GLM}}
  &=\frac1nX^\top\hat W X
  -\left(\frac1nX^\top\hat WZ\right)
   \left(\frac1nZ^\top\hat WZ\right)^{-1}
   \left(\frac1nZ^\top\hat WX\right)\\
  &\convp
  \E[W_iX_i^2]-r^\top M^{-1}r.
  \end{align*}
  Since $\theta=M^{-1}r$,
  \begin{align*}
  \E[W_i(X_i-\theta^\top Z_i)^2]
  &=\E[W_iX_i^2]-2\theta^\top r+\theta^\top M\theta\\
  &=\E[W_iX_i^2]-r^\top M^{-1}r
  =\E[R_i^2].
  \end{align*}
  Thus $\tilde V_n^{\mathrm{GLM}}\convp\E[R_i^2]>0$.  Slutsky's theorem now yields
  \[
  \tilde T_n^{\mathrm{GLM}}
  \convd
  N\!\left(0,
  \frac{\tau_s^2}{\E[R_i^2]}
  \right).
  \]
  When $\var(Y_i\mid X_i,Z_i)=v_i$, one has $\E[\xi_i^2\mid X_i,Z_i]=W_i$,
  so $\tau_s^2=\E[W_i\delta_i^2]=\E[R_i^2]$ and the displayed variance is one.
  \end{proof}

\subsubsection{Proof of Proposition~\ref{prop:glm-unperm-emp}}
\label{app:proof-glm-unperm-emp}

\begin{proof}
Write $\Delta_{s,ni}=\hat R_i^{\mathrm{GLM}}-U_{ni}$, where
$U_{ni}=\delta_i\xi_i$.

\paragraph{Step 1: Numerator linearization}
The nuisance score equation implies
$\sum_iZ_i\hat\epsilon_i=0$, and therefore
\[
\frac1{\sqrt n}\sum_i\hat R_i^{\mathrm{GLM}}
=\frac1{\sqrt n}\sum_iX_i\hat\epsilon_i.
\]
Steps~1--2 in the proof of Proposition~\ref{prop:glm-unperm} give
\begin{equation}\label{eq:glm-emp-numerator-limit}
\frac1{\sqrt n}\sum_{i=1}^n\hat R_i^{\mathrm{GLM}}
=
\frac1{\sqrt n}\sum_{i=1}^nU_{ni}+o_\PP(1)
\convd N(0,\tau_s^2).
\end{equation}
In particular, $n^{-1/2}\sum_iU_{ni}=O_\PP(1)$ and
$n^{-1/2}\sum_i\hat R_i^{\mathrm{GLM}}=O_\PP(1)$.

\paragraph{Step 2: Mean-square approximation of the fitted score contributions}
Lemma~\ref{lem:glm-info-consistency} gives
\begin{equation}\label{eq:glm-a-consistency}
\hat\theta
=\left(\frac1nZ^\top\hat WZ\right)^{-1}
  \left(\frac1nZ^\top\hat WX\right)
\convp M^{-1}r=\theta.
\end{equation}
Because $\hat\nu_i^{\mathrm{GLM}}=X_i-Z_i^\top\hat\theta$ and
$\delta_i=X_i-Z_i^\top\theta$,
\begin{align*}
\Delta_{s,ni}
&=(\delta_i-Z_i^\top(\hat\theta-\theta))\hat\epsilon_i-\delta_i\xi_i\\
&=-Z_i^\top(\hat\theta-\theta)\hat\epsilon_i
  +\delta_i(\hat\epsilon_i-\xi_i).
\end{align*}
Hence $(u+v)^2\le2u^2+2v^2$ and $\|Z_i\|\le C_Z$ imply
\begin{align}
\frac1n\sum_i\Delta_{s,ni}^2
&\le
2C_Z^2\|\hat\theta-\theta\|^2\frac1n\sum_i\hat\epsilon_i^2
+\frac2n\sum_i\delta_i^2(\hat\epsilon_i-\xi_i)^2.
\label{eq:glm-emp-l2-bound}
\end{align}
The first term is $o_\PP(1)$ by~\eqref{eq:glm-a-consistency} and Lemma~\ref{lem:r-hat-sums}.  For the second term, Lemma~\ref{lem:glm-lipschitz} gives, eventually almost surely,
\[
|\hat\epsilon_i-\xi_i|
\le C(1+|Y_i|)
\{\|\hat\beta-\beta\|+|\hat\vartheta-\vartheta_\star|\}.
\]
Therefore
\begin{align*}
\frac1n\sum_i\delta_i^2(\hat\epsilon_i-\xi_i)^2
&\le
C\{\|\hat\beta-\beta\|+|\hat\vartheta-\vartheta_\star|\}^2
\frac1n\sum_i\delta_i^2(1+|Y_i|)^2.
\end{align*}
The last sample average is $O_\PP(1)$.  Indeed, boundedness of $\bm Z$ gives
$\delta_i^2\le C(1+X_i^2)$, while Cauchy--Schwarz and the fourth-moment assumptions imply
\[
\E[X_i^2Y_i^2]
\le\{\E[X_i^4]\E[Y_i^4]\}^{1/2}<\infty.
\]
Thus the consistency of $(\hat\beta,\hat\vartheta)$ yields
\begin{equation}\label{eq:glm-emp-l2-transfer}
\frac1n\sum_i\Delta_{s,ni}^2=o_\PP(1).
\end{equation}

\paragraph{Step 3: Consistency of the empirical studentizer}
By the law of large numbers,
\begin{equation}\label{eq:glm-oracle-second-moment}
\frac1n\sum_iU_{ni}^2
\convp
\E[\delta_i^2\xi_i^2]
=\tau_s^2.
\end{equation}
By~\eqref{eq:glm-emp-numerator-limit}, $n^{-1/2}\sum_iU_{ni}=O_\PP(1)$.
Thus~\eqref{eq:glm-oracle-second-moment} and~\eqref{eq:glm-emp-l2-transfer}
verify the remaining hypotheses of Lemma~\ref{lem:studentizer-transfer}
with oracle scores $U_{ni}$, perturbations $\Delta_{s,ni}$, and $\tau_n^2=\tau_s^2$.
It follows that
\begin{equation}\label{eq:glm-emp-studentizer-limit}
\widehat V_n^{\mathrm{GLM}}\convp\tau_s^2.
\end{equation}
Because $\tau_s^2>0$, Slutsky's theorem applied to~\eqref{eq:glm-emp-numerator-limit} proves
\[
T_n^{\mathrm{GLM}}\convd N(0,1).
\]
\end{proof}

\subsubsection{Proof of Theorem~\ref{thm:glm-perm-emp}}
\label{app:proof-glm-perm-emp}

\begin{proof}
We verify Theorem~\ref{thm:product-score-validity} with observed oracle
$U_{ni}=\delta_i\xi_i$ and permutation array
\[
C_{n,ik}\equiv(X_k-\bar X)\xi_i,
\qquad
s_{X,n}^2\equiv\frac1n\sum_k(X_k-\bar X)^2.
\]
Take the trivial training field in Theorem~\ref{thm:product-score-validity}(i). Under~\eqref{eq:glm-null-mean},
$\E[\xi_i\mid X_i,Z_i]=0$, so $\E[U_{ni}]=0$; independence across
observations makes the $U_{ni}$ i.i.d.  The definition of $\tau_s^2$ in
Section~\ref{sec:glm-warmup} and the hypothesis of positive variance of Proposition~\ref{prop:glm-unperm-emp} give
$\sigma_{s,n}^2=\E[\delta_i^2\xi_i^2]=\tau_s^2>0$.
By Assumption~\ref{assumption:reg-glm}(v),
\[
\frac{\E[U_{ni}^4]}{n\sigma_{s,n}^4}
=\frac{\E[(\delta_i\xi_i)^4]}{n\tau_s^4}
\longrightarrow0.
\]

For the two requirements for the observed score in Theorem~\ref{thm:product-score-validity}(iii), set
$\Delta_{s,ni}\equiv\hat R_i^{\mathrm{GLM}}-U_{ni}$. Equation~\eqref{eq:glm-emp-numerator-limit}
states that $n^{-1/2}\sum_i\Delta_{s,ni}=o_\PP(1)$, and
\eqref{eq:glm-emp-l2-transfer} states that
$n^{-1}\sum_i\Delta_{s,ni}^2=o_\PP(1)$. Because
$\sigma_{s,n}=\tau_s>0$, these give, respectively,
\[
\frac1{\sqrt n\,\sigma_{s,n}}\sum_i\Delta_{s,ni}=o_\PP(1),
\qquad
\frac1{n\sigma_{s,n}^2}\sum_i\Delta_{s,ni}^2=o_\PP(1).
\]

\paragraph{Permutation oracle}
Every row of $(C_{n,ik})$ has mean zero and
$\sigma_{p,n}^2=n^{-2}\sum_{i,k}C_{n,ik}^2
=s_{X,n}^2n^{-1}\sum_i\xi_i^2$.
The law of large numbers and $\E X_i=0$, $\E X_i^2=1$ give
$s_{X,n}^2\convp1$ and $n^{-1}\sum_i\xi_i^2\convp\tau_p^2>0$;
thus $\PP(\sigma_{p,n}^2>0)\to1$.
Writing $\bar\xi=n^{-1}\sum_i\xi_i$,
\[
\frac{\sum_k C_{n,\cdot k}^2}{(n-1)\sigma_{p,n}^2}
=\frac{n}{n-1}\frac{\bar\xi^2}{n^{-1}\sum_i\xi_i^2}
=o_\PP(1).
\]
For any i.i.d.\ $V_i$ with $\E|V_i|^4<\infty$,
$n\PP(|V_1|^4>\varepsilon n)=o(1)$; the union bound yields
$\max_iV_i^2=o_\PP(\sqrt n)$. Apply this to $X_i$ and $\xi_i$.
Since $\bar X=O_\PP(n^{-1/2})$,
\[
\frac{\max_{i,k}C_{n,ik}^2}{n\sigma_{p,n}^2}
=\frac{\max_k(X_k-\bar X)^2\max_i\xi_i^2}{n\sigma_{p,n}^2}
=o_\PP(1).
\]
These equations verify both array conditions in Theorem~\ref{thm:product-score-validity}(ii).

\paragraph{Permutation nuisance remainder}
Put $e_i\equiv\hat\epsilon_i-\xi_i$. By Lemma~\ref{lem:glm-lipschitz},
consistency of $(\hat\beta,\hat\vartheta)$ and the outcome moment bound,
\begin{equation}\label{eq:glm-perm-response-mspe}
\frac1n\sum_i e_i^2
\le C\{\|\hat\beta-\beta\|+|\hat\vartheta-\vartheta_\star|\}^2
       \frac1n\sum_i(1+|Y_i|)^2=o_\PP(1).
\end{equation}
Because the nuisance design contains an intercept,
$\sum_i\hat\epsilon_i=0$ and $\sum_iZ_i\hat\epsilon_i=0$. Hence the
normalized sum of
$\Delta_{p,ni}\equiv\hat R_i^{\mathrm{GLM},(\Pi)}-C_{n,i\Pi(i)}$
is exactly
\[
\frac1{\sqrt n}\sum_i\Delta_{p,ni}
=\frac1{\sqrt n}\sum_i(X_{\Pi(i)}-\bar X)e_i.
\]
Conditional on $\cF_n$, this expression has mean zero and variance at most
\[
\frac{n}{n-1}s_{X,n}^2\frac1n\sum_i e_i^2=o_\PP(1)
\]
by Lemma~\ref{lem:hoeffding}. Lemma~\ref{lem:cond-markov}(b) applied
to this zero-mean conditional sum proves
$n^{-1/2}\sum_i\Delta_{p,ni}=o_\PP(1)$, the $r=p$ sum condition
in Theorem~\ref{thm:product-score-validity}(iii).

For the squared remainder, define
\[
\widehat M_n\equiv\frac1nZ^\top\hat WZ,
\qquad
\hat r_n^{(\Pi)}\equiv\frac1n\sum_iX_{\Pi(i)}\hat W_iZ_i.
\]
Conditional on $\cF_n$, Lemma~\ref{lem:hoeffding} gives, for
each coordinate $j$,
\begin{align*}
\E[\hat r_{n,j}^{(\Pi)}\mid\cF_n]
&=\bar X\frac1n\sum_i\hat W_iZ_{ij}=o_\PP(1),\\
\var(\hat r_{n,j}^{(\Pi)}\mid\cF_n)
&\le\frac1{n-1}
\left(\frac1n\sum_iX_i^2\right)
\left(\frac1n\sum_i\hat W_i^2Z_{ij}^2\right)
=O_\PP(n^{-1}).
\end{align*}
Here $\bar X=o_\PP(1)$ and $n^{-1}\sum_iX_i^2\convp1$ by
Assumption~\ref{assumption:glm-reg}(iii); Assumption~\ref{assumption:reg-glm}(i),(iii)
bounds $\|Z_i\|$ and, eventually, $\hat W_i$. Lemma~\ref{lem:cond-markov}(b)
applied to $\hat r_{n,j}^{(\Pi)}-\E[\hat r_{n,j}^{(\Pi)}\mid\cF_n]$ gives
$\hat r_{n,j}^{(\Pi)}=o_\PP(1)$. Because $p$ is fixed,
$\hat r_n^{(\Pi)}=o_\PP(1)$. Lemma~\ref{lem:glm-info-consistency} gives
$\widehat M_n\convp M$, and $M$ is invertible by
Assumption~\ref{assumption:reg-glm}(ii),(iii). Therefore
\begin{equation}\label{eq:glm-emp-perm-a}
\hat\theta^{(\Pi)}=\widehat M_n^{-1}\hat r_n^{(\Pi)}=o_\PP(1).
\end{equation}
Using
\[
\Delta_{p,ni}
=(X_{\Pi(i)}-\bar X)e_i
 +(\bar X-Z_i^\top\hat\theta^{(\Pi)})\hat\epsilon_i,
\]
we obtain
\begin{align}
\frac1n\sum_i\{\Delta_{p,ni}\}^2
&\le\frac2n\sum_i(X_{\Pi(i)}-\bar X)^2e_i^2\notag\\
&\quad+4\{\bar X^2+C_Z^2\|\hat\theta^{(\Pi)}\|^2\}
             \frac1n\sum_i\hat\epsilon_i^2
=o_\PP(1).\label{eq:glm-emp-perm-l2}
\end{align}
Indeed, the first term has conditional expectation
$2s_{X,n}^2n^{-1}\sum_i e_i^2=o_\PP(1)$, and the second is
$o_\PP(1)$ by \eqref{eq:glm-emp-perm-a}, $\bar X=o_\PP(1)$, and
Lemma~\ref{lem:r-hat-sums}. Since
$\sigma_{p,n}^2\convp\tau_p^2>0$, the last two bounds give
\[
\frac1{\sqrt n\,\sigma_{p,n}}\sum_i\Delta_{p,ni}=o_\PP(1),
\qquad
\frac1{n\sigma_{p,n}^2}\sum_i\{\Delta_{p,ni}\}^2=o_\PP(1).
\]
These are the two $r=p$ requirements of Theorem~\ref{thm:product-score-validity}(iii); the
$r=s$ requirements and conditions~(i)--(ii) were checked above.
Theorem~\ref{thm:product-score-validity} now proves Theorem~\ref{thm:glm-perm-emp}.
\end{proof}

\subsubsection{Proof of Theorem~\ref{thm:glm-perm}}\label{app:proof-glm-perm}

\begin{proof}
  Conditional on $\cF_n$, the fitted quantities are fixed.
  The nuisance score equation gives
  $\sum_i\hat\nu_i^{\mathrm{GLM},(\Pi)}\hat\epsilon_i
  =\sum_iX_{\Pi(i)}\hat\epsilon_i$.
  When both denominators below are positive, write
  $\tilde T_n^{\mathrm{GLM},(\Pi)}=H_nD_n$, where
  $H_n\equiv\sum_iX_{\Pi(i)}\hat\epsilon_i/\sqrt{\sum_i\hat\epsilon_i^2}$ and
  \[
  D_n^2\equiv
  \frac{n^{-1}\sum_i\hat\epsilon_i^2}
  {n^{-1}{X^{(\Pi)}}^\top\hat WX^{(\Pi)}
  -\left(n^{-1}{X^{(\Pi)}}^\top\hat WZ\right)
   \left(n^{-1}Z^\top\hat WZ\right)^{-1}
   \left(n^{-1}Z^\top\hat WX^{(\Pi)}\right)}.
  \]

  By Lemma~\ref{lem:r-hat-sums}, $\sum_i\hat\epsilon_i^2>0$
  eventually almost surely. On this event, set
  $c_{ni}=\hat\epsilon_i/\sqrt{\sum_j\hat\epsilon_j^2}$.
  The intercept score equation gives $\sum_ic_{ni}=0$, and by definition
  $\sum_ic_{ni}^2=1$. For $d>0$,
  \begin{align*}
  \sum_i c_{ni}^2
  \indc{|\sqrt n c_{ni}|\ge d}
  &\le \frac{n}{d^2}\sum_i c_{ni}^4\\
  &=\frac1{d^2}
  \frac{n^{-1}\sum_i\hat\epsilon_i^4}
       {(n^{-1}\sum_i\hat\epsilon_i^2)^2}.
  \end{align*}
  By Lemma~\ref{lem:r-hat-sums}, the ratio on the right is almost surely
  bounded for all sufficiently large $n$.  Hence the Hájek truncation
  condition holds almost surely after first taking $n\to\infty$ and then
  $d\to\infty$.  Also
  \[
  \PP\left(\max_{i\le n}|X_i|>\varepsilon\sqrt n\right)
  \le
  \frac{\E[X_i^4]}{\varepsilon^4 n}
  \longrightarrow0.
  \]
  Since $\Pi$ is independent of $\cF_n$, the conditional law of $H_n$
  given $\cF_n$ is the same as that given $(X,c_n)$.
  Lemma~\ref{lem:hajek} therefore gives
  \begin{equation}\label{eq:glm-perm-H-clt}
  H_n\mid\cF_n\convdp N(0,1).
  \end{equation}

It remains to show that $D_n^2\convp\sigma_p^2$.
Lemma~\ref{lem:r-hat-sums} gives
\begin{equation}\label{eq:glm-perm-resid-second}
\frac1n\sum_i\hat\epsilon_i^2\convp\E[\xi_i^2]=\E[S_i].
\end{equation}

Let $Q_n^{(\Pi)}\equiv n^{-1}\sum_i\hat W_iX_{\Pi(i)}^2$.
Lemma~\ref{lem:hoeffding} gives
\[
\E[Q_n^{(\Pi)}\mid\cF_n]
=\left(\frac1n\sum_i\hat W_i\right)
 \left(\frac1n\sum_iX_i^2\right)
\convp\E[W_i].
\]
By Lemma~\ref{lem:hoeffding},
\begin{align*}
\var(Q_n^{(\Pi)}\mid\cF_n)
&\le\frac1{n-1}
\left(\frac1n\sum_i\hat W_i^2\right)
\left(\frac1n\sum_iX_i^4\right)
\convp0.
\end{align*}
Therefore
\[
\E[Q_n^{(\Pi)}-\E[W_i]\mid\cF_n]\convp0,
\qquad
\var(Q_n^{(\Pi)}-\E[W_i]\mid\cF_n)\convp0.
\]
Lemma~\ref{lem:quad-mean}(b) gives
\begin{equation}\label{eq:glm-perm-quadratic}
Q_n^{(\Pi)}\convp\E[W_i].
\end{equation}

The calculation in~\eqref{eq:glm-emp-perm-a}, established in the proof
of Theorem~\ref{thm:glm-perm-emp}, gives $\hat\theta^{(\Pi)}=o_\PP(1)$. Write
$\widehat M_n=n^{-1}Z^\top\hat WZ$ and
$\hat r_n^{(\Pi)}=n^{-1}Z^\top\hat WX^{(\Pi)}=\widehat M_n\hat\theta^{(\Pi)}$.
Since $\widehat M_n\convp M$, the cross term in the denominator is
\begin{equation}\label{eq:glm-perm-cross}
\bigl(\hat r_n^{(\Pi)}\bigr)^\top
 \widehat M_n^{-1}
\hat r_n^{(\Pi)}
=\bigl(\hat\theta^{(\Pi)}\bigr)^\top \widehat M_n\hat\theta^{(\Pi)}
=o_\PP(1).
\end{equation}

Combining~\eqref{eq:glm-perm-resid-second},
\eqref{eq:glm-perm-quadratic}, and~\eqref{eq:glm-perm-cross} gives
\[
D_n^2\convp\frac{\E[S_i]}{\E[W_i]}=\sigma_p^2.
\]
Because $\sigma_p^2>0$, $D_n\convp\sigma_p$.  Conditional Slutsky,
together with~\eqref{eq:glm-perm-H-clt}, yields
\[
\tilde T_n^{\mathrm{GLM},(\Pi)}\mid\cF_n
\convdp N(0,\sigma_p^2).
\]
Lemma~\ref{lem:quantile} then gives
\[
\mathbb Q_{1-\alpha}
\bigl(\tilde T_n^{\mathrm{GLM},(\Pi)}\mid\cF_n\bigr)
\convp\sigma_pz_{1-\alpha}.
\]
If $\var(Y_i\mid X_i,Z_i)=cv_i$ almost surely for a constant $c>0$, then
$\E[\xi_i^2\mid X_i,Z_i]=cW_i$. Hence $\tau_s^2=c\E[R_i^2]$ and
$\tau_p^2=c\E[W_i]$, so $\sigma_s^2=\sigma_p^2=c$.
Proposition~\ref{prop:glm-unperm} gives
$\tilde T_n^{\mathrm{GLM}}\convd N(0,c)$, while the permutation critical
value converges to $\sqrt c\,z_{1-\alpha}$.  Continuity of the normal
distribution therefore gives asymptotic rejection probability $\alpha$
for the ideal test using conditional quantiles.
\end{proof}

\subsubsection{Proof of Theorem~\ref{thm:gcm-valid}}
\label{app:proof-gcm-valid}

\begin{proof}
Write $\nu_i=\nu_i^{\mathrm{GCM}}$ and
$\epsilon_i=\epsilon_i^{\mathrm{GCM}}$ for the oracle residuals in Section~\ref{subsec:gcm}, and set
\[
e_{x,i}=\mu_x(Z_i)-\hat\mu_x(Z_i),\qquad
e_{y,i}=\mu_y(Z_i)-\hat\mu_y(Z_i).
\]
By
Assumption~\ref{assumption:reg-gcm}(i) and
Remark~\ref{rmk:shared-outcome-moments},
\begin{equation}\label{eq:gcm-oracle-moment-bounds}
\begin{aligned}
\E[\nu_i^2\mid Z_i]&\le4C_{X,4}^{1/2},
&\qquad\E[\nu_i^4\mid Z_i]&\le16C_{X,4},\\
\E[\epsilon_i^2\mid Z_i]&\le\bar v_Y,
&\E[\epsilon_i^4\mid Z_i]&\le16C_{Y,4}.
\end{aligned}
\end{equation}

\paragraph{Condition (i): observed oracle}
Take the trivial training field and $U_{ni}=\nu_i\epsilon_i$. The observations are
i.i.d.; under $H_0^{\mathrm{CI}}$,
$\E[U_{ni}\mid X_i,Z_i]=\nu_i\E[\epsilon_i\mid X_i,Z_i]=0$. Conditional
independence of $X_i$ and $Y_i$ given $Z_i$ and
\eqref{eq:gcm-oracle-moment-bounds} give
\begin{equation}\label{eq:gcm-observed-fourth-check}
\sigma_{s,n}^2=\E[U_{ni}^2]=\sigma_{\mathrm{GCM}}^2>0,\qquad
\frac{\E[U_{ni}^4]}{n\sigma_{s,n}^4}
=\frac{\E\{\E[\nu_i^4\mid Z_i]\E[\epsilon_i^4\mid Z_i]\}}
 {n\sigma_{\mathrm{GCM}}^4}
\le\frac{256C_{X,4}C_{Y,4}}
 {n\sigma_{\mathrm{GCM}}^4}\longrightarrow0.
\end{equation}
The positive variance is Assumption~\ref{assumption:reg-gcm}(ii).

\paragraph{Condition (ii): permutation oracle}
Take
\[
C_{n,ik}=(X_k-\bar X)\epsilon_i,\qquad
s_{X,n}^2=\frac1n\sum_k(X_k-\bar X)^2.
\]
The array is $\cF_n$-measurable and $\sum_k C_{n,ik}=0$ for every
$i$. Its scale is
\begin{equation}\label{eq:gcm-perm-scale-check}
\sigma_{p,n}^2=\frac1{n^2}\sum_{i,k}C_{n,ik}^2
=s_{X,n}^2\frac1n\sum_i\epsilon_i^2
\convp\var(\bm X)\E[\epsilon_i^2]
=\sigma_{\mathrm{GCM},0}^2>0.
\end{equation}
The law of large numbers gives the convergence; positivity follows from
Remark~\ref{rmk:gcm-weight-moments}. For
$\bar\epsilon=n^{-1}\sum_i \epsilon_i$, $C_{n,\cdot k}=(X_k-\bar X)\bar\epsilon$, so
\begin{equation}\label{eq:gcm-column-check}
\frac{\sum_k C_{n,\cdot k}^2}{(n-1)\sigma_{p,n}^2}
=\frac{n}{n-1}\frac{\bar\epsilon^2}{\frac1n\sum_i\epsilon_i^2}
=o_\PP(1),
\end{equation}
because $\bar\epsilon=O_\PP(n^{-1/2})$ and
$\frac1n\sum_i\epsilon_i^2\convp\E[\epsilon_i^2]>0$. Moreover,
$\E|X_i|^4\le C_{X,4}$ and $\E|\epsilon_i|^4\le16C_{Y,4}$.
For either i.i.d.\ sequence $V_i=X_i$ or $V_i=\epsilon_i$ and every
$\varepsilon>0$, $n\PP(|V_1|^4>\varepsilon n)\to0$; a union
bound therefore gives $\max_iV_i^2=o_\PP(\sqrt n)$.
Since $|\bar X|\le\max_k|X_k|$,
\[
\max_k(X_k-\bar X)^2\le4\max_kX_k^2=o_\PP(\sqrt n).
\]
Together with $\max_i \epsilon_i^2=o_\PP(\sqrt n)$ and the positive limit
in~\eqref{eq:gcm-perm-scale-check}, this yields
\begin{equation}\label{eq:gcm-cell-check}
\frac{\max_{i,k}C_{n,ik}^2}{n\sigma_{p,n}^2}
=\frac{\max_k(X_k-\bar X)^2\max_i \epsilon_i^2}
 {n\sigma_{p,n}^2}=o_\PP(1).
\end{equation}
Equations~\eqref{eq:gcm-column-check}--\eqref{eq:gcm-cell-check}
are the two array requirements of Theorem~\ref{thm:product-score-validity}(ii).

\paragraph{Condition (iii): observed fitted scores}
Write $\hat R_i^{\mathrm{GCM}}=U_{ni}+\Delta_{s,ni}$, where
$\Delta_{s,ni}=e_{x,i}\epsilon_i+\nu_ie_{y,i}+e_{x,i}e_{y,i}$.
For the learner seeds $(\omega_x,\omega_y)$ specified in
Appendix~\ref{sec:practical}, set
$\cH_x=\sigma((X_i,Z_i)_{i\le n},\omega_x)$ and
$\cH_y=\sigma((Y_i,Z_i)_{i\le n},\omega_y)$.
By Algorithm~\ref{alg:gcm}, $e_{x,i}$ and $e_{y,i}$ are respectively
$\cH_x$- and $\cH_y$-measurable.
Under $H_0^{\mathrm{CI}}$, the $\epsilon_i$ are conditionally independent and
centered given $\cH_x$, as are the $\nu_i$ given $\cH_y$. Thus
\begin{align}
\var\!\left(\frac1{\sqrt n}\sum_i e_{x,i}\epsilon_i\middle|\cH_x\right)
&=\frac1n\sum_i e_{x,i}^2\E[\epsilon_i^2\mid Z_i]
\le \bar v_YE_{n,x}=o_\PP(1),\notag\\
\var\!\left(\frac1{\sqrt n}\sum_i \nu_ie_{y,i}\middle|\cH_y\right)
&=\frac1n\sum_i e_{y,i}^2\E[\nu_i^2\mid Z_i]
\le 4C_{X,4}^{1/2}E_{n,y}=o_\PP(1).
\label{eq:gcm-observed-sum-bounds}
\end{align}
Their conditional means are zero, so
Lemma~\ref{lem:cond-markov}(b) makes both sums $o_\PP(1)$.
For the corresponding mean squares, conditional expectation gives
\begin{align}
\E\!\left[\frac1n\sum_i e_{x,i}^2\epsilon_i^2\middle|\cH_x\right]
&\le \bar v_YE_{n,x}=o_\PP(1),\notag\\
\E\!\left[\frac1n\sum_i \nu_i^2e_{y,i}^2\middle|\cH_y\right]
&\le 4C_{X,4}^{1/2}E_{n,y}=o_\PP(1).
\label{eq:gcm-observed-ms-bounds}
\end{align}
Lemma~\ref{lem:cond-markov}(a) applies to both nonnegative
averages. The product term satisfies
\begin{equation}\label{eq:gcm-observed-product-bounds}
\left|\frac1{\sqrt n}\sum_i e_{x,i}e_{y,i}\right|^2
\le nE_{n,x}E_{n,y}=o_\PP(1),\qquad
\frac1n\sum_i e_{x,i}^2e_{y,i}^2
\le nE_{n,x}E_{n,y}=o_\PP(1).
\end{equation}
The $o_\PP(1)$ bounds in
\eqref{eq:gcm-observed-sum-bounds}--\eqref{eq:gcm-observed-product-bounds}
use exactly $E_{n,x},E_{n,y}=o_\PP(1)$ and
$nE_{n,x}E_{n,y}=o_\PP(1)$ from Assumption~\ref{assumption:unperm}.
Combining~\eqref{eq:gcm-observed-sum-bounds},
\eqref{eq:gcm-observed-ms-bounds}, and~\eqref{eq:gcm-observed-product-bounds}
by the triangle inequality and $(u+v+w)^2\le3(u^2+v^2+w^2)$ gives
\begin{align*}
\left|\frac1{\sqrt n}\sum_i\Delta_{s,ni}\right|
&\le\left|\frac1{\sqrt n}\sum_i e_{x,i}\epsilon_i\right|
 +\left|\frac1{\sqrt n}\sum_i \nu_ie_{y,i}\right|
 +\left|\frac1{\sqrt n}\sum_i e_{x,i}e_{y,i}\right|
 =o_\PP(1),\\
\frac1n\sum_i\Delta_{s,ni}^2
&\le3\left\{\frac1n\sum_i e_{x,i}^2\epsilon_i^2
 +\frac1n\sum_i \nu_i^2e_{y,i}^2
 +\frac1n\sum_i e_{x,i}^2e_{y,i}^2\right\}
 =o_\PP(1).
\end{align*}
Therefore
\begin{equation}\label{eq:gcm-observed-reductions}
\frac1{\sqrt n}\sum_i\hat R_i^{\mathrm{GCM}}
=\frac1{\sqrt n}\sum_iU_{ni}+o_\PP(1),
\qquad
\frac1n\sum_i\Delta_{s,ni}^2=o_\PP(1).
\end{equation}
Since $\sigma_{s,n}^2=\sigma_{\mathrm{GCM}}^2>0$, these are the
two normalized conditions required for $r=s$:
\[
\frac1{\sqrt n\,\sigma_{s,n}}\sum_i\Delta_{s,ni}=o_\PP(1),
\qquad
\frac1{n\sigma_{s,n}^2}\sum_i\Delta_{s,ni}^2=o_\PP(1).
\]
The ordinary law of large numbers gives
$n^{-1}\sum_iU_{ni}^2\convp\sigma_{\mathrm{GCM}}^2$; the i.i.d.\ CLT
gives $n^{-1/2}\sum_iU_{ni}=O_\PP(1)$. Lemma~\ref{lem:studentizer-transfer}
applied with~\eqref{eq:gcm-observed-reductions} therefore also yields
\begin{equation}\label{eq:gcm-observed-denominator}
\widehat V_n^{\mathrm{GCM}}\convp\sigma_{\mathrm{GCM}}^2.
\end{equation}

\paragraph{Condition (iii): permuted fitted scores}
Write
\[
\tilde e_{x,i}=\bar X-\hat\mu_x^{(\Pi)}(Z_i),\quad
\Delta_{p,ni}=(X_{\Pi(i)}-\bar X)e_{y,i}+\tilde e_{x,i}\epsilon_i+\tilde e_{x,i}e_{y,i}.
\]
The first term in the normalized sum has conditional mean zero and,
by Lemma~\ref{lem:hoeffding},
\begin{equation}\label{eq:gcm-perm-e-sum}
\var_\Pi\!\left(\frac1{\sqrt n}\sum_i (X_{\Pi(i)}-\bar X)e_{y,i}
 \middle|\cF_n\right)
\le\frac{n}{n-1}s_{X,n}^2E_{n,y}=o_\PP(1).
\end{equation}
For the second term, condition on
$\cG_n=\sigma(\cH_x,\Pi)$.
The $\tilde e_{x,i}$ are $\cG_n$-measurable and the $\epsilon_i$ are conditionally
independent and centered. Hence
\begin{equation}\label{eq:gcm-perm-g-sum}
\var\!\left(\frac1{\sqrt n}\sum_i \tilde e_{x,i}\epsilon_i\middle|\cG_n\right)
=\frac1n\sum_i \tilde e_{x,i}^2\E[\epsilon_i^2\mid Z_i]
\le \bar v_Y\tilde E_{n,x}=o_\PP(1).
\end{equation}
Lemma~\ref{lem:cond-markov}(b) applies to both sums. The product
term satisfies
\begin{equation}\label{eq:gcm-perm-product-sum}
\left|\frac1{\sqrt n}\sum_i \tilde e_{x,i}e_{y,i}\right|^2
\le n\tilde E_{n,x}E_{n,y}=o_\PP(1).
\end{equation}
For the mean-square requirement, conditional expectation and
Lemma~\ref{lem:cond-markov}(a) give
\begin{align}
\E_\Pi\!\left[\frac1n\sum_i (X_{\Pi(i)}-\bar X)^2e_{y,i}^2
 \middle|\cF_n\right]
&=s_{X,n}^2E_{n,y}=o_\PP(1),\notag\\
\E\!\left[\frac1n\sum_i \tilde e_{x,i}^2\epsilon_i^2
 \middle|\cG_n\right]
&\le \bar v_Y\tilde E_{n,x}=o_\PP(1).
\label{eq:gcm-perm-ms-bounds}
\end{align}
Finally,
\begin{equation}\label{eq:gcm-perm-product-ms}
\frac1n\sum_i \tilde e_{x,i}^2e_{y,i}^2
\le n\tilde E_{n,x}E_{n,y}=o_\PP(1).
\end{equation}
Here Assumption~\ref{assumption:perm} supplies
$\tilde E_{n,x}=o_\PP(1)$ and
$nE_{n,y}\tilde E_{n,x}=o_\PP(1)$; Assumption~\ref{assumption:unperm}
supplies $E_{n,y}=o_\PP(1)$.
Combining~\eqref{eq:gcm-perm-e-sum}, \eqref{eq:gcm-perm-g-sum},
\eqref{eq:gcm-perm-product-sum}, \eqref{eq:gcm-perm-ms-bounds}, and
\eqref{eq:gcm-perm-product-ms} gives
\begin{align*}
\left|\frac1{\sqrt n}\sum_i\Delta_{p,ni}\right|
&\le\left|\frac1{\sqrt n}\sum_i (X_{\Pi(i)}-\bar X)e_{y,i}\right|
 +\left|\frac1{\sqrt n}\sum_i \tilde e_{x,i}\epsilon_i\right|
 +\left|\frac1{\sqrt n}\sum_i \tilde e_{x,i}e_{y,i}\right|
 =o_\PP(1),\\
\frac1n\sum_i\Delta_{p,ni}^2
&\le3\left\{\frac1n\sum_i (X_{\Pi(i)}-\bar X)^2e_{y,i}^2
 +\frac1n\sum_i \tilde e_{x,i}^2\epsilon_i^2
 +\frac1n\sum_i \tilde e_{x,i}^2e_{y,i}^2\right\}
 =o_\PP(1).
\end{align*}
By~\eqref{eq:gcm-perm-scale-check},
$\sigma_{p,n}^{-1}=O_\PP(1)$ and
$\sigma_{p,n}^{-2}=O_\PP(1)$. Multiplying the two displayed
bounds by these factors, respectively, gives
\[
\frac1{\sqrt n\,\sigma_{p,n}}\sum_i\Delta_{p,ni}=o_\PP(1),
\qquad
\frac1{n\sigma_{p,n}^2}\sum_i\Delta_{p,ni}^2=o_\PP(1).
\]
All conditions of Theorem~\ref{thm:product-score-validity} are now verified, and its conclusion
gives the two normal limits and the level statement of Theorem~\ref{thm:gcm-valid}.
\end{proof}


\subsubsection{Proof of Theorem~\ref{thm:pcm-valid}}
\label{app:proof-pcm-valid}

\begin{proof}
Apply Theorem~\ref{thm:product-score-validity} with sample size $n_\star=|\cD_1|$, as in Section~\ref{subsec:pcm}.
All sums and maxima below are over $\cD_1$. Let $\cH_n$ be as in
Assumption~\ref{assumption:reg-pcm}. Since the split and seeds are independent of the observations,
the inference observations are i.i.d. with their original law given
$\cH_n$. Write $\nu_i=\nu_i^{\mathrm{PCM}}$ and
$\epsilon_i=\epsilon_i^{\mathrm{PCM}}$ for the oracle residuals, and set
\[
e_{y,i}=\mu_y(Z_i)-\hat\mu_y(Z_i),\qquad
e_{f,i}=\mu_f(Z_i)-\hat\mu_f(Z_i).
\]
Assumption~\ref{assumption:reg-pcm}(i) and
Remark~\ref{rmk:shared-outcome-moments} give the explicit bounds
\begin{equation}\label{eq:pcm-proof-outcome-bounds}
c_Y\le V_Y(Z_i)\le \bar v_Y,\qquad
\E[\epsilon_i^4\mid Z_i]\le \bar m_{Y,4}
\quad\text{almost surely}.
\end{equation}
All ratios below are evaluated on the events where their denominators
are positive; Assumption~\ref{assumption:reg-pcm}(ii)--(iii) makes
those events have probability tending to one.

\paragraph{Condition (i): observed oracle scores}
Take $U_{ni}=\nu_i\epsilon_i$. Conditional on $\cH_n$, the observations in the inference fold are i.i.d. Under $H_0^{\mathrm{CI}}$,
$\E[\epsilon_i\mid X_i,Z_i,\cH_n]=0$, so the $U_{ni}$ are centered.
Writing $\sigma_{s,n}^2=\E[U_{ni}^2\mid\cH_n]$ and using
$\sigma_{f,n}^2=\E[\nu_i^2\mid\cH_n]$, we obtain
\begin{equation}\label{eq:pcm-proof-observed-scale}
\sigma_{s,n}^2=\E[\nu_i^2V_Y(Z_i)\mid\cH_n],\qquad
c_Y\sigma_{f,n}^2\le\sigma_{s,n}^2
\le\bar v_Y\sigma_{f,n}^2.
\end{equation}
In particular $\PP(\sigma_{s,n}^2>0)\to1$.
Conditional independence and~\eqref{eq:pcm-observed-profile-l2l4} give
\begin{equation}\label{eq:pcm-proof-lyapunov}
\frac{\E[U_{ni}^4\mid\cH_n]}{n_\star\sigma_{s,n}^4}
=\frac{\E\!\left[\nu_i^4\E[\epsilon_i^4\mid Z_i]\mid\cH_n\right]}
 {n_\star\sigma_{s,n}^4}
\le\frac{\bar m_{Y,4}}{n_\star c_Y^2}
\frac{\E[\nu_i^4\mid\cH_n]}{\sigma_{f,n}^4}
=o_\PP(1).
\end{equation}
This verifies every part of Theorem~\ref{thm:product-score-validity}(i).

\paragraph{Condition (ii): permutation oracle array}
Put $B_{ik}=\hat f(X_k,Z_i)$,
$B_{i\cdot}=n_\star^{-1}\sum_k B_{ik}$, and
\[
A_{ik}=B_{ik}-B_{i\cdot},\qquad
r_i=\frac1{n_\star}\sum_k A_{ik}^2,\qquad
D_n=\frac1{n_\star}\sum_i r_i,\qquad
C_{n,ik}=\epsilon_iA_{ik}.
\]
This array is $\cF_n$-measurable and each row sums to zero. Its
Theorem~\ref{thm:product-score-validity} scale is
\begin{equation}\label{eq:pcm-proof-perm-scale}
\sigma_{p,n}^2=\frac1{n_\star^2}\sum_{i,k}C_{n,ik}^2
=\frac1{n_\star}\sum_i \epsilon_i^2r_i.
\end{equation}
Let
$\cH_{X,n}=\sigma(\cH_n,(X_i,Z_i)_{i\in\cD_1})$.
Proposition~\ref{prop:pcm-product-law-bridge} gives
\begin{equation}\label{eq:pcm-proof-bridge-inputs}
D_n/\sigma_{f,n,\mathrm{perm}}^2\convp1,\qquad
\max_i r_i/\sigma_{f,n,\mathrm{perm}}^2=O_\PP(1),\qquad
\frac{\max_{i,k}\epsilon_i^2A_{ik}^2}{n_\star \sigma_{f,n,\mathrm{perm}}^2}=o_\PP(1)
\end{equation}
by~\eqref{eq:pcm-envelope-bridge}--\eqref{eq:pcm-envelope-cell}.
Conditional on $\cH_{X,n}$, the $\epsilon_i$ are independent and
$\E[\epsilon_i^2\mid\cH_{X,n}]=V_Y(Z_i)$. Hence
\begin{align}
\E[\sigma_{p,n}^2\mid\cH_{X,n}]&=\frac1{n_\star}\sum_i V_Y(Z_i)r_i
\in[c_YD_n,\bar v_YD_n],\notag\\
\var(\sigma_{p,n}^2\mid\cH_{X,n})
&\le\frac1{n_\star^2}\sum_i r_i^2\E[\epsilon_i^4\mid Z_i]
\le\frac{\bar m_{Y,4}}{n_\star}(\max_i r_i)D_n.
\label{eq:pcm-proof-q-moments}
\end{align}
The last bound divided by $\sigma_{f,n,\mathrm{perm}}^4$ is $O_\PP(n_\star^{-1})$.
Conditional Chebyshev, in the form of
Lemma~\ref{lem:cond-markov}(b), therefore gives
$(\sigma_{p,n}^2-\E[\sigma_{p,n}^2\mid\cH_{X,n}])/\sigma_{f,n,\mathrm{perm}}^2=o_\PP(1)$.
Together with~\eqref{eq:pcm-proof-bridge-inputs}, this proves
\begin{equation}\label{eq:pcm-proof-scale-comparison}
\PP\!\left(\frac{c_Y}{4}\le\frac{\sigma_{p,n}^2}{\sigma_{f,n,\mathrm{perm}}^2}
\le2\bar v_Y\right)\longrightarrow1,
\qquad \sigma_{f,n,\mathrm{perm}}^2/\sigma_{p,n}^2=O_\PP(1),\qquad
\PP(\sigma_{p,n}^2>0)\longrightarrow1.
\end{equation}
For $C_{n,\cdot k}=n_\star^{-1}\sum_i \epsilon_iA_{ik}$, the conditional
mean is zero. Independence and~\eqref{eq:pcm-proof-outcome-bounds}
give
\begin{equation}\label{eq:pcm-proof-column}
\E\!\left[
\frac1{n_\star-1}\sum_k C_{n,\cdot k}^2
\middle|\cH_{X,n}\right]
=\frac1{n_\star^2(n_\star-1)}\sum_{i,k}V_Y(Z_i)A_{ik}^2
\le\frac{\bar v_YD_n}{n_\star-1}.
\end{equation}
After division by $\sigma_{f,n,\mathrm{perm}}^2$, the right side is $o_\PP(1)$.
Lemma~\ref{lem:cond-markov}(a) and
\eqref{eq:pcm-proof-scale-comparison} imply
$\sum_k C_{n,\cdot k}^2/\{(n_\star-1)\sigma_{p,n}^2\}=o_\PP(1)$.
The other array condition follows without another approximation:
\begin{equation}\label{eq:pcm-proof-max-cell}
\frac{\max_{i,k}C_{n,ik}^2}{n_\star\sigma_{p,n}^2}
=\frac{\max_{i,k}\epsilon_i^2A_{ik}^2}{n_\star \sigma_{f,n,\mathrm{perm}}^2}
  \frac{\sigma_{f,n,\mathrm{perm}}^2}{\sigma_{p,n}^2}=o_\PP(1)
\end{equation}
by~\eqref{eq:pcm-proof-bridge-inputs} and
\eqref{eq:pcm-proof-scale-comparison}. Thus both requirements in
Theorem~\ref{thm:product-score-validity}(ii) hold.

\paragraph{Condition (iii): observed fitted scores}
The score difference is exactly
\begin{equation}\label{eq:pcm-proof-observed-decomp}
\Delta_{s,ni}=\hat R_i^{\mathrm{PCM}}-U_{ni}
=e_{f,i}\epsilon_i+\nu_ie_{y,i}+e_{f,i}e_{y,i}.
\end{equation}
Let $\cH_{Y,n}=\sigma(\cH_n,(Y_i,Z_i)_{i\in\cD_1})$.
Algorithm~\ref{alg:vpcm} and the fitting conventions in Appendix~\ref{sec:practical}
make $e_{f,i}$ measurable with respect to $\cH_{X,n}$ and $e_{y,i}$
measurable with respect to $\cH_{Y,n}$, including tuning and preprocessing.
Under $H_0^{\mathrm{CI}}$, the $\epsilon_i$ are independent and centered given
$\cH_{X,n}$, and the $\nu_i$ are independent and centered given $\cH_{Y,n}$.
\begin{equation}\label{eq:pcm-proof-observed-two-bounds}
\begin{aligned}
\var\!\left(\frac{\sum_i e_{f,i}\epsilon_i}{\sqrt{n_\star}\,\sigma_{f,n}}
\middle|\cH_{X,n}\right)
&=\E\!\left[\frac{\sum_i e_{f,i}^2\epsilon_i^2}{n_\star\sigma_{f,n}^2}
\middle|\cH_{X,n}\right]\\
&=\frac1{n_\star\sigma_{f,n}^2}\sum_i e_{f,i}^2V_Y(Z_i)
\le\bar v_Y\frac{E_{n,f}^{\mathrm{PCM}}}{\sigma_{f,n}^2}=o_\PP(1),\\[1ex]
\var\!\left(\frac{\sum_i \nu_ie_{y,i}}{\sqrt{n_\star}\,\sigma_{f,n}}
\middle|\cH_{Y,n}\right)
&=\E\!\left[\frac{\sum_i \nu_i^2e_{y,i}^2}{n_\star\sigma_{f,n}^2}
\middle|\cH_{Y,n}\right]\\
&=\frac1{n_\star\sigma_{f,n}^2}\sum_i
 e_{y,i}^2\E[\nu_i^2\mid Z_i,\cH_n]\\
&\le\left\{\operatorname*{ess\,sup}_z
 \frac{\E[\nu_i^2\mid Z_i=z,\cH_n]}{\sigma_{f,n}^2}\right\}
 E_{n,y}^{\mathrm{PCM}}=o_\PP(1).
\end{aligned}
\end{equation}
The last bounds use~\eqref{eq:pcm-observed-profile-l2l4} and Assumption~\ref{assumption:pcm-unperm}. Lemma~\ref{lem:cond-markov}(a)--(b)
therefore makes both normalized sums and both normalized mean squares
$o_\PP(1)$. For the product term, Cauchy--Schwarz and nonnegativity give
\begin{equation}\label{eq:pcm-proof-observed-product}
\begin{aligned}
\frac{\bigl|n_\star^{-1/2}\sum_i e_{f,i}e_{y,i}\bigr|^2}
 {\sigma_{f,n}^2}
&\le\frac{n_\star E_{n,f}^{\mathrm{PCM}}E_{n,y}^{\mathrm{PCM}}}{\sigma_{f,n}^2}=o_\PP(1),\\
\frac1{n_\star\sigma_{f,n}^2}\sum_i e_{f,i}^2e_{y,i}^2
&\le\frac{n_\star E_{n,f}^{\mathrm{PCM}}E_{n,y}^{\mathrm{PCM}}}{\sigma_{f,n}^2}=o_\PP(1).
\end{aligned}
\end{equation}
Combining~\eqref{eq:pcm-proof-observed-two-bounds} and
\eqref{eq:pcm-proof-observed-product} in~\eqref{eq:pcm-proof-observed-decomp},
the triangle inequality and $(u+v+w)^2\le3(u^2+v^2+w^2)$ give
\begin{align*}
\frac{\bigl|\sum_i\Delta_{s,ni}\bigr|}{\sqrt{n_\star}\,\sigma_{f,n}}
&\le\frac{
 \bigl|\sum_i e_{f,i}\epsilon_i\bigr|
 +\bigl|\sum_i\nu_i e_{y,i}\bigr|
 +\bigl|\sum_i e_{f,i}e_{y,i}\bigr|}
 {\sqrt{n_\star}\,\sigma_{f,n}}=o_\PP(1),\\
\frac{\sum_i\Delta_{s,ni}^2}{n_\star\sigma_{f,n}^2}
&\le\frac3{n_\star\sigma_{f,n}^2}
 \sum_i\bigl(e_{f,i}^2\epsilon_i^2+\nu_i^2e_{y,i}^2
 +e_{f,i}^2e_{y,i}^2\bigr)=o_\PP(1).
\end{align*}
Finally,
$\sigma_{f,n}/\sigma_{s,n}\le c_Y^{-1/2}$ and
$\sigma_{f,n}^2/\sigma_{s,n}^2\le c_Y^{-1}$ by
\eqref{eq:pcm-proof-observed-scale}. Multiplying the two displayed
bounds by these factors gives
\[
\frac{\sum_i\Delta_{s,ni}}{\sqrt{n_\star}\,\sigma_{s,n}}
=o_\PP(1),\qquad
\frac{\sum_i\Delta_{s,ni}^2}{n_\star\sigma_{s,n}^2}
=o_\PP(1).
\]
These are the requirements for the observed score of Theorem~\ref{thm:product-score-validity}(iii).

\paragraph{Condition (iii): permuted fitted scores}
Set $\tilde e_{f,i}=B_{i\cdot}-\hat\mu_f^{(\Pi)}(Z_i)$, so that
$\tilde E_{n,f}^{\mathrm{PCM}}=n_\star^{-1}\sum_i\tilde e_{f,i}^2$. Expanding the fitted score gives
\begin{equation}\label{eq:pcm-proof-perm-decomp}
\Delta_{p,ni}=\hat R_i^{\mathrm{PCM},(\Pi)}-C_{n,i\Pi(i)}
=e_{y,i}A_{i,\Pi(i)}+\epsilon_i \tilde e_{f,i}+e_{y,i}\tilde e_{f,i}.
\end{equation}
For the first term, apply the final assertion of
Lemma~\ref{lem:perm-array-moments} with $q_i=e_{y,i}$ and the array $(A_{ik})$, centered within each row. It gives conditional mean zero and
\begin{equation}\label{eq:pcm-proof-perm-first-sum}
\begin{aligned}
&\var_\Pi\!\left(
\frac{\sum_i e_{y,i}A_{i,\Pi(i)}}
 {\sqrt{n_\star}\,\sigma_{f,n,\mathrm{perm}}}\middle|\cF_n\right)\\
&\qquad\le\frac1{n_\star(n_\star-1)\sigma_{f,n,\mathrm{perm}}^2}
\sum_{i,k}e_{y,i}^2A_{ik}^2=o_\PP(1).
\end{aligned}
\end{equation}
The last step is~\eqref{eq:pcm-perm-weighted-outcome-mspe-derived},
whose right side differs by the factor $n_\star/(n_\star-1)$.
For $\cG_n=\sigma(\cH_{X,n},\Pi)$,
the refit in Algorithm~\ref{alg:vpcm} makes $\tilde e_{f,i}$
$\cG_n$-measurable, while the $\epsilon_i$ are conditionally independent and centered. Hence
\begin{equation}\label{eq:pcm-proof-perm-second-sum}
\begin{aligned}
\var\!\left(
\frac{\sum_i\epsilon_i\tilde e_{f,i}}
 {\sqrt{n_\star}\,\sigma_{f,n,\mathrm{perm}}}\middle|\cG_n\right)
&=\frac1{n_\star\sigma_{f,n,\mathrm{perm}}^2}
\sum_i V_Y(Z_i)\tilde e_{f,i}^2\\
&\le\bar v_Y\frac{\tilde E_{n,f}^{\mathrm{PCM}}}
 {\sigma_{f,n,\mathrm{perm}}^2}=o_\PP(1).
\end{aligned}
\end{equation}
The last bound is Assumption~\ref{assumption:pcm-perm}.
The third sum is controlled deterministically by its other part:
\begin{equation}\label{eq:pcm-proof-perm-third-sum}
\frac{\bigl|n_\star^{-1/2}\sum_i e_{y,i}\tilde e_{f,i}\bigr|^2}{\sigma_{f,n,\mathrm{perm}}^2}
\le\frac{n_\star E_{n,y}^{\mathrm{PCM}}\tilde E_{n,f}^{\mathrm{PCM}}}{\sigma_{f,n,\mathrm{perm}}^2}=o_\PP(1)
\end{equation}
by the same assumption. Applying Lemma~\ref{lem:cond-markov}(b) to
\eqref{eq:pcm-proof-perm-first-sum} and~\eqref{eq:pcm-proof-perm-second-sum},
and combining with~\eqref{eq:pcm-proof-perm-third-sum} in
\eqref{eq:pcm-proof-perm-decomp}, gives
\begin{align*}
\frac{\bigl|\sum_i\Delta_{p,ni}\bigr|}
 {\sqrt{n_\star}\,\sigma_{f,n,\mathrm{perm}}}
&\le\frac{
 \bigl|\sum_i e_{y,i}A_{i,\Pi(i)}\bigr|
 +\bigl|\sum_i\epsilon_i\tilde e_{f,i}\bigr|
 +\bigl|\sum_i e_{y,i}\tilde e_{f,i}\bigr|}
 {\sqrt{n_\star}\,\sigma_{f,n,\mathrm{perm}}}
=o_\PP(1).
\end{align*}

For the requirement for the squared score, conditional permutation
expectation and~\eqref{eq:pcm-perm-weighted-outcome-mspe-derived}
give
\begin{equation}\label{eq:pcm-proof-perm-first-ms}
\E_\Pi\!\left[
\frac1{n_\star \sigma_{f,n,\mathrm{perm}}^2}\sum_i e_{y,i}^2A_{i,\Pi(i)}^2
\middle|\cF_n\right]
=\frac1{n_\star^2\sigma_{f,n,\mathrm{perm}}^2}\sum_{i,k}e_{y,i}^2A_{ik}^2=o_\PP(1).
\end{equation}
Conditional on $\cG_n$,
\begin{equation}\label{eq:pcm-proof-perm-second-ms}
\begin{aligned}
\E\!\left[
\frac{\sum_i\epsilon_i^2\tilde e_{f,i}^2}
 {n_\star\sigma_{f,n,\mathrm{perm}}^2}\middle|\cG_n\right]
&=\frac1{n_\star\sigma_{f,n,\mathrm{perm}}^2}
\sum_i V_Y(Z_i)\tilde e_{f,i}^2\\
&\le\bar v_Y\frac{\tilde E_{n,f}^{\mathrm{PCM}}}
 {\sigma_{f,n,\mathrm{perm}}^2}=o_\PP(1).
\end{aligned}
\end{equation}
The remaining square satisfies
\begin{equation}\label{eq:pcm-proof-perm-third-ms}
\frac1{n_\star \sigma_{f,n,\mathrm{perm}}^2}\sum_i e_{y,i}^2\tilde e_{f,i}^2
\le\frac{n_\star E_{n,y}^{\mathrm{PCM}}\tilde E_{n,f}^{\mathrm{PCM}}}{\sigma_{f,n,\mathrm{perm}}^2}=o_\PP(1).
\end{equation}
Lemma~\ref{lem:cond-markov}(a),
\eqref{eq:pcm-proof-perm-first-ms}, \eqref{eq:pcm-proof-perm-second-ms}, and
\eqref{eq:pcm-proof-perm-third-ms}, together with
$(u+v+w)^2\le3(u^2+v^2+w^2)$, give
\begin{align*}
\frac{\sum_i\Delta_{p,ni}^2}{n_\star\sigma_{f,n,\mathrm{perm}}^2}
&\le\frac3{n_\star\sigma_{f,n,\mathrm{perm}}^2}
 \sum_i\bigl(e_{y,i}^2A_{i,\Pi(i)}^2
 +\epsilon_i^2\tilde e_{f,i}^2
 +e_{y,i}^2\tilde e_{f,i}^2\bigr)
=o_\PP(1).
\end{align*}
Finally, $\sigma_{f,n,\mathrm{perm}}^2/\sigma_{p,n}^2=O_\PP(1)$ by
\eqref{eq:pcm-proof-perm-scale}--\eqref{eq:pcm-proof-scale-comparison}. Therefore the conversion to
the Theorem~\ref{thm:product-score-validity} scale is
\begin{align*}
\frac{\sum_i\Delta_{p,ni}}{\sqrt{n_\star}\,\sigma_{p,n}}
&=\frac{\sigma_{f,n,\mathrm{perm}}}{\sigma_{p,n}}
 \frac{\sum_i\Delta_{p,ni}}{\sqrt{n_\star} \sigma_{f,n,\mathrm{perm}}}=o_\PP(1),\\
\frac{\sum_i\Delta_{p,ni}^2}{n_\star\sigma_{p,n}^2}
&=\frac{\sigma_{f,n,\mathrm{perm}}^2}{\sigma_{p,n}^2}
 \frac{\sum_i\Delta_{p,ni}^2}{n_\star \sigma_{f,n,\mathrm{perm}}^2}=o_\PP(1).
\end{align*}
These verify the two requirements for the permutation score of Theorem~\ref{thm:product-score-validity}(iii).
All conditions of Theorem~\ref{thm:product-score-validity} are now verified, so its conclusion gives
Theorem~\ref{thm:pcm-valid}.
\end{proof}


\section{Stability of permutation refitting}
\label{app:proof-perm-stab}

This appendix gives sufficient conditions for Assumptions~\ref{assumption:perm} and~\ref{assumption:pcm-perm}.

\subsection{Learner setup and complexity measures}
\label{subsec:perm-stab-learner-setup}

Let $n_\star=n$ for GCM and $n_\star=|\cD_1|$ for PCM denote the refit
sample size. Assume $n_\star\ge2$.
All vectors, matrices, sums, and permutations in this appendix use these observations, relabelled $1,\ldots,n_\star$. The feature map and kernel are fixed, or for PCM are determined by $\cH_n$. For each penalized learner, the tuning parameter $\lambda_n$ is $\cF_n$-measurable and held fixed across permutations. All learners use an unpenalized intercept term.
Throughout this appendix, $\mathbf u=(u_i)_{i=1}^{n_\star}$ denotes
the permuted response vector, with its dependence on $\Pi$ suppressed.
For a feature map $\phi:\mathcal Z\to\R^d$, write
\[
\Phi=(\phi(Z_1)^\top,\ldots,\phi(Z_{n_\star})^\top),\qquad
P_1=n_\star^{-1}\mathbf1\mathbf1^\top,\qquad H=I-P_1,\qquad \Phi_c=H\Phi.
\]
\begin{definition}[Linear smoother]\label{def:linear-smoother}
A regression rule is a \emph{linear smoother} if its fitted vector is
$S\mathbf u$ for the same $\cF_n$-measurable matrix $S$ under every permutation. It reproduces constants if $S\mathbf 1=\mathbf 1$.
Write $\mathcal C(S)=\|S\|_F^2$ for its variance complexity.
\end{definition}

Now we instantiate the linear smoother definition with common learners.

\begin{itemize}
  \item \textbf{Ordinary least squares.} Let $\widetilde\Phi=[\mathbf1,\Phi]$ have full column rank $d+1$. OLS solves
  \[
  \min_{\alpha\in\R,\,\boldsymbol\gamma\in\R^d}\|\mathbf u-\alpha\mathbf1-\Phi\boldsymbol\gamma\|_2^2.
  \]
  Its fitted vector is $S_{\mathrm{OLS}}\mathbf u$, where $S_{\mathrm{OLS}}=\widetilde\Phi(\widetilde\Phi^\top\widetilde\Phi)^{-1}\widetilde\Phi^\top$ and
  \[
  \mathcal C_{\mathrm{OLS}}=\mathcal C(S_{\mathrm{OLS}})=d+1.
  \]
  \item\textbf{Ridge regression.} For $\lambda_n>0$, ridge solves
  \[
  \min_{\alpha\in\R,\,\boldsymbol\gamma\in\R^d}\left\{\|\mathbf u-\alpha\mathbf1-\Phi\boldsymbol\gamma\|_2^2 +\lambda_n\|\boldsymbol\gamma\|_2^2\right\}.
  \]
  Its fitted vector is $S_{\mathrm{ridge}}(\lambda_n)\mathbf u$, where $S_{\mathrm{ridge}}(\lambda_n)=P_1+\Phi_c(\Phi_c^\top\Phi_c+\lambda_n I_d)^{-1}\Phi_c^\top$. If $\rho_1,\ldots,\rho_d$ are the eigenvalues of $\Phi_c^\top\Phi_c$, then
  \begin{equation}\label{eq:ridge-frob}
    \mathcal C_{\mathrm{ridge}}(\lambda_n)=\mathcal C\{S_{\mathrm{ridge}}(\lambda_n)\}=1+\sum_{j=1}^d\left(\frac{\rho_j}{\rho_j+\lambda_n}\right)^2\le d+1.
  \end{equation}
  \item\textbf{Kernel ridge regression.} Let $k$ be a positive-definite kernel on $\mathcal Z$ with RKHS $\mathcal H_k$, $K_{i\ell}=k(Z_i,Z_\ell)$, and $K_c=HKH$, where $H$ is the centering matrix $H=I-n_\star^{-1}\mathbf1\mathbf1^\top$. For $\lambda_n>0$ and $\mathbf h=(h(Z_i))_{i=1}^{n_\star}$, KRR solves
  \[
  \min_{\alpha\in\R,\,h\in\mathcal H_k}\left\{\|\mathbf u-\alpha\mathbf1-\mathbf h\|_2^2+\lambda_n\|h\|_{\mathcal H_k}^2\right\}.
  \]
  The unpenalized intercept and the representer theorem give the fitted vector $S_{\mathrm{KRR}}(\lambda_n)\mathbf u$, where $S_{\mathrm{KRR}}(\lambda_n)=P_1+K_c(K_c+\lambda_n I)^{-1}H$. If $\nu_1,\ldots,\nu_{n_\star}$ are the eigenvalues of $K_c$, then
  \begin{equation}\label{eq:eff-dim}
    \mathcal C_{\mathrm{KRR}}(\lambda_n)=\mathcal C\{S_{\mathrm{KRR}}(\lambda_n)\}=1+\sum_{j=1}^{n_\star}\left(\frac{\nu_j}{\nu_j+\lambda_n}\right)^2.
  \end{equation}
\end{itemize}

\paragraph{Beyond linear smoothers: lasso.} For $\lambda_n>0$, lasso solves
\[
  \min_{\alpha\in\R,\,\boldsymbol\gamma\in\R^d}\left\{\frac1{2n_\star}\|\mathbf u-\alpha\mathbf1-\Phi\boldsymbol\gamma\|_2^2+\lambda_n\|\boldsymbol\gamma\|_1\right\}.
\]

\begin{remark}[Eliminating the intercept]
The unpenalized intercept $\alpha$ can be eliminated by centering the
response vector and the fitted values. For OLS, ridge, and lasso, let
$a_n>0$ denote the loss normalization and let
$\mathcal P_n(\boldsymbol\gamma)$ be a penalty independent of $\alpha$,
including the zero penalty. The equivalent centered problem is
\begin{equation}\label{eq:centered-objective}
\hat{\boldsymbol\gamma}\in\argmin{\boldsymbol\gamma\in\R^d}
\left\{a_n
\|H\mathbf u-\Phi_c\boldsymbol\gamma\|_2^2
+\mathcal P_n(\boldsymbol\gamma)\right\}.
\end{equation}
Here $a_n=1$ for OLS and ridge, and $a_n=(2n_\star)^{-1}$ for lasso and
the respective penalties are zero, $\lambda_n\|\boldsymbol\gamma\|_2^2$,
and $\lambda_n\|\boldsymbol\gamma\|_1$. The intercept is recovered as
$\hat\alpha=\bar u-\bar\phi^\top\hat{\boldsymbol\gamma}$,
where $\bar u=n_\star^{-1}\sum_i u_i$ and
$\bar\phi=n_\star^{-1}\sum_i\phi(Z_i)$.
For KRR, the same argument replaces $\Phi_c\boldsymbol\gamma$ by
$H\mathbf h$, where $\mathbf h=(h(Z_i))_{i=1}^{n_\star}$, and minimizes
over $h\in\mathcal H_k$ with $a_n=1$ and penalty
$\lambda_n\|h\|_{\mathcal H_k}^2$. Its intercept is
$\bar u-n_\star^{-1}\sum_i\hat h(Z_i)$.
In each case, centering leaves the loss normalization and the penalty
unchanged.
\end{remark}

\paragraph{Additional notation.} For a $\cF_n$-measurable array $A$, set $u_i=A_{i,\Pi(i)}$, $m_i=A_{i\cdot}=n_\star^{-1}\sum_{k=1}^{n_\star} A_{ik}$, and $\phi_c(Z_i)=\phi(Z_i)-\bar\phi$. Define
\begin{equation}\label{eq:lasso-score-generic}
  \Delta_n(A)=\left\|\frac1{n_\star}\sum_i\phi_c(Z_i)(u_i-m_i)\right\|_\infty,\qquad v_{\phi,\max}=\max_{\ell\le d}\frac1{n_\star}\sum_i\phi_{c,\ell}(Z_i)^2.
\end{equation}
For GCM, $A_{ik}=X_k$. For PCM, $A_{ik}=B_{ik}$.

\subsection{Main stability results}
\label{subsec:perm-stab-main-results}

We divide the analysis into two parts: the GCM and the PCM.

\subsubsection{GCM}
\label{subsec:perm-stab-results-gcm}

For the GCM, the conditional permutation target is the constant vector $\bar X\mathbf1$. Because every smoother above uses an unpenalized intercept, the term for approximation of the target vanishes exactly.  The stability rate is therefore
governed only by the learner complexity defined in
Subsection~\ref{subsec:perm-stab-learner-setup}, leading to the following explicit stability rates for common smoothers.

\begin{corollary}[Explicit GCM stability rates for common smoothers]
\label{cor:gcm-explicit-smoother-rates}
Assume $\E[\bm X^2]<\infty$. Then the following results hold.
\begin{itemize}
  \item \textbf{Learner complexity results.} We have $\tilde E_{n,x}=O_\PP(\mathcal C_{\mathrm{OLS}}/n_\star)$ for OLS, $\tilde E_{n,x}=O_\PP(\mathcal C_{\mathrm{ridge}}(\lambda_n)/n_\star)$ for linear ridge regression, and $\tilde E_{n,x}=O_\PP(\mathcal C_{\mathrm{KRR}}(\lambda_n)/n_\star)$ for kernel ridge regression.
  \item \textbf{Permutation consistency.} GCM permutation consistency holds whenever the corresponding variance complexity $\mathcal C_{\mathrm{OLS}}$, $\mathcal C_{\mathrm{ridge}}(\lambda_n)$, or $\mathcal C_{\mathrm{KRR}}(\lambda_n)$ is $o_\PP(n_\star)$.
  \item \textbf{Product of permutation errors.} The product condition in Assumption~\ref{assumption:perm} holds if, in addition, $E_{n,y}$ times the same complexity is $o_\PP(1)$.
\end{itemize}
\end{corollary}

\begin{theorem}[Lasso verifies GCM permutation stability]
\label{thm:lasso-gcm}
Use the lasso setup above with $u_i=X_{\Pi(i)}$. Define
\begin{align*}
  \Delta_n^{\mathrm{GCM}}\equiv\Delta_n(A)=\left\|\frac1{n_\star}\sum_{i=1}^{n_\star}\{\phi(Z_i)-\bar\phi\}\{X_{\Pi(i)}-\bar X\}\right\|_\infty.
\end{align*}
If
\begin{equation}\label{eq:lasso-gcm-score-domination}
  \PP\!\left(\Delta_n^{\mathrm{GCM}}\le\lambda_n/2\right)\longrightarrow1,
\end{equation}
then
\begin{equation}\label{eq:lasso-gcm-zero-stability}
  \PP(\tilde E_{n,x}=0)\longrightarrow1.
\end{equation}
Consequently, $\tilde E_{n,x}=o_\PP(1)$ and
$n_\star E_{n,y}\tilde E_{n,x}=o_\PP(1)$, so both parts of Assumption~\ref{assumption:perm} hold.
Suppose also that $d=d_n\to\infty$ and, for
some fixed $c_G>0$, with probability tending to one,
\begin{equation}\label{eq:lasso-gcm-score-tail}
  \PP\!\left(\Delta_n^{\mathrm{GCM}}>t\mid\cF_n\right)\le 2d\exp(-c_Gn_\star t^2),\qquad t>0.
\end{equation}
Then there exists a constant $C>0$ so that $\lambda_n=C\sqrt{\frac{\log d}{n_\star}}$ is sufficient for~\eqref{eq:lasso-gcm-score-domination}.
\end{theorem}

\begin{remark}[A heavy tail scenario]
  If only the weaker second-moment assumptions $\E[\bm X^2]<\infty$ and $v_{\phi,\max}=O_\PP(1)$ are imposed, Lemma~\ref{lem:lasso-perm-score} instead gives the fallback bound $\Delta_n^{\mathrm{GCM}}=O_\PP(\sqrt{d/n_\star})$, so $\sqrt{d/n_\star}=o(\lambda_n)$ remains sufficient without exponential tails.
\end{remark}

\begin{remark}[A primitive GCM score condition]\label{rmk:gcm-lasso-score-primitive}
Suppose that, with probability tending to one, the centered observations
$X_k-\bar X$ and centered features $\phi_{c,\ell}(Z_i)$ are uniformly
bounded in absolute value by fixed positive constants $K_X$ and $K_\phi$,
respectively. The stronger part of Lemma~\ref{lem:lasso-perm-score} then
gives the exponential score bound~\eqref{eq:lasso-gcm-score-tail} with
tail constant $(32K_X^2K_\phi^2)^{-1}$. Consequently, when $d\to\infty$,
choosing the penalty as $C\sqrt{(\log d)/n_\star}$, with
$C>8\sqrt{2}\,K_XK_\phi$, ensures the score-domination
condition~\eqref{eq:lasso-gcm-score-domination}.
\end{remark}

\subsubsection{PCM}
\label{subsec:perm-stab-results-pcm}

For the PCM, write the finite-sample and population permutation targets as $\mathbf m=(B_{i\cdot})_{i\in\cD_1}$ and $\mathbf b=(b^{\mathrm{perm}}(Z_i))_{i\in\cD_1}$. For finite-dimensional learners such as OLS, linear ridge regression and lasso, suppose
\begin{equation}\label{eq:pcm-target-linear-decomp}
  b^{\mathrm{perm}}(z)=\alpha_n+\phi(z)^\top\boldsymbol\gamma_n+r_n(z),\qquad P_Z\text{-a.s.},
\end{equation}
Write $\mathbf r_n=(r_n(Z_i))_{i\in\cD_1}$.
For KRR, suppose instead
\[
b^{\mathrm{perm}}(z)=\alpha_{n,\mathrm{KRR}}+h_n(z)+r_{n,\mathrm{KRR}}(z), \qquad h_n\in\mathcal H_k.
\]
Now we present the explicit PCM stability rates for common smoothers.

\begin{corollary}[Explicit PCM stability rates for common smoothers]\label{cor:pcm-explicit-smoother-rates}
Assume the conditions of Proposition~\ref{prop:pcm-product-law-bridge} and $\sigma_{f,n,\mathrm{perm}}^2>0$. Under the target decompositions above, the following results hold.
\begin{itemize}
  \item \textbf{Learner complexity results.} $\tilde E_{n,f}^{\mathrm{PCM}}=O_\PP\!\left(\mathcal C_{\mathrm{OLS}}\sigma_{f,n,\mathrm{perm}}^2+\sum_i r_n(Z_i)^2\right)/n_\star$ for OLS, $\tilde E_{n,f}^{\mathrm{PCM}}=O_\PP\!\left(\mathcal C_{\mathrm{ridge}}(\lambda_n)\sigma_{f,n,\mathrm{perm}}^2+\sum_i r_n(Z_i)^2+\lambda_n\|\boldsymbol\gamma_n\|_2^2\right)/n_\star$ for linear ridge regression, and $\tilde E_{n,f}^{\mathrm{PCM}}=O_\PP\!\left(\mathcal C_{\mathrm{KRR}}(\lambda_n)\sigma_{f,n,\mathrm{perm}}^2+\sum_i r_{n,\mathrm{KRR}}(Z_i)^2+\lambda_n\|h_n\|_{\mathcal H_k}^2\right)/n_\star$ for kernel ridge regression.
  \item \textbf{Permutation consistency.} The first part of Assumption~\ref{assumption:pcm-perm} follows whenever the corresponding right-hand side is $o_\PP(\sigma_{f,n,\mathrm{perm}}^2)$.
  \item \textbf{Product of permutation errors.} The condition on the product of the errors follows if the same right-hand side multiplied by $n_\star E_{n,y}^{\mathrm{PCM}}$ is $o_\PP(\sigma_{f,n,\mathrm{perm}}^2)$.
\end{itemize}
\end{corollary}

Now we present the explicit PCM permutation stability result for the lasso.
\begin{theorem}[Lasso verifies PCM permutation stability]
\label{thm:lasso-vpcm}
Use the lasso setup above with
$u_i=\hat f(X_{\Pi(i)},Z_i)$ on $\cD_1$. Also let $A_{ik}=B_{ik}$ and consider $\Delta_n^{\mathrm{PCM}}\equiv\Delta_n(A)$. Suppose $\sigma_{f,n,\mathrm{perm}}^2>0$, the conditions of Proposition~\ref{prop:pcm-product-law-bridge} hold, and
\eqref{eq:pcm-target-linear-decomp} holds. If
\begin{equation}\label{eq:lasso-pcm-score-domination}
\PP\!\left(\Delta_n^{\mathrm{PCM}}\le\lambda_n/2\right)\to1,
\end{equation}
then
\begin{equation}\label{eq:lasso-pcm-stability-rate}
\frac{\tilde E_{n,f}^{\mathrm{PCM}}}{\sigma_{f,n,\mathrm{perm}}^2}
=O_\PP\!\left(
\frac{\|\mathbf r_n\|_2^2}{n_\star\sigma_{f,n,\mathrm{perm}}^2}
+\frac{\lambda_n\|\boldsymbol\gamma_n\|_1}
{\sigma_{f,n,\mathrm{perm}}^2}
+\frac1{n_\star}
\right).
\end{equation}
Consequently, if the right-hand side is $o_\PP(1)$, the first part of
Assumption~\ref{assumption:pcm-perm} holds; if its product with $n_\star E_{n,y}^{\mathrm{PCM}}$ is also
$o_\PP(1)$, then the second part holds.
\end{theorem}

The preceding theorem does not require sparsity.  When
$S_n\equiv\operatorname{supp}(\boldsymbol\gamma_n)$,
$s_n\equiv|S_n|$, and $d=d_n\to\infty$, define, for $s_n\ge1$, the compatibility constant
\begin{equation}\label{eq:pcm-lasso-compatibility}
\kappa_n^2
\equiv\inf_{\substack{\boldsymbol\delta\ne0\\
\|\boldsymbol\delta_{S_n^c}\|_1\le4\|\boldsymbol\delta_{S_n}\|_1}}
\frac{s_n\|\Phi_c\boldsymbol\delta\|_2^2}
{n_\star\|\boldsymbol\delta_{S_n}\|_1^2}.
\end{equation}
Set $\kappa_n=1$ when $s_n=0$, so the terms $s_n/\kappa_n^2$ below are zero.

\begin{corollary}[Sparse oracle rate for the PCM lasso]
\label{cor:pcm-lasso-sparse}
In the setting of Theorem~\ref{thm:lasso-vpcm}, suppose that, for some $\kappa_0>0$,
$\PP(s_n=0\text{ or }\kappa_n\ge\kappa_0)\to1$. Suppose also that,
for some fixed $c_0>0$, with probability tending to one,
\begin{equation}\label{eq:pcm-lasso-normalized-score-tail}
\PP\!\left(
\Delta_n^{\mathrm{PCM}}>t\,\sigma_{f,n,\mathrm{perm}}
\mid\cF_n\right)
\le2d\exp(-c_0n_\star t^2),
\qquad t>0.
\end{equation}
Choose
\begin{equation}\label{eq:pcm-lasso-sparse-penalty}
\lambda_n
=C\,\sigma_{f,n,\mathrm{perm}}\sqrt{\frac{\log d}{n_\star}},
\qquad C>\frac{2}{\sqrt{c_0}}.
\end{equation}
Then
\begin{equation}\label{eq:pcm-lasso-sparse-rate}
\frac{\tilde E_{n,f}^{\mathrm{PCM}}}{\sigma_{f,n,\mathrm{perm}}^2}
=O_\PP\!\left(
\frac{\|\mathbf r_n\|_2^2}{n_\star\sigma_{f,n,\mathrm{perm}}^2}
+\frac{s_n\log d}{n_\star\kappa_n^2}
+\frac1{n_\star}
\right)
=O_\PP\!\left(
\frac{\|\mathbf r_n\|_2^2}{n_\star\sigma_{f,n,\mathrm{perm}}^2}
+\frac{s_n\log d}{n_\star}
+\frac1{n_\star}
\right).
\end{equation}
Hence the first part of Assumption~\ref{assumption:pcm-perm} follows whenever
\[
\frac{\|\mathbf r_n\|_2^2}{n_\star\sigma_{f,n,\mathrm{perm}}^2}
+\frac{s_n\log d}{n_\star}=o_\PP(1),
\]
and the condition on the product of the errors of Assumption~\ref{assumption:pcm-perm} follows if the left-hand side,
plus $n_\star^{-1}$, multiplied by $n_\star E_{n,y}^{\mathrm{PCM}}$ is $o_\PP(1)$.
In particular, if $\mathbf r_n=0$, both parts of Assumption~\ref{assumption:pcm-perm} follow provided
\[
\frac{s_n\log d}{n_\star}=o_\PP(1),\qquad
E_{n,y}^{\mathrm{PCM}}(s_n\log d+1)=o_\PP(1).
\]
\end{corollary}

\begin{remark}[A primitive score condition]\label{rmk:pcm-lasso-score-primitive}
The normalized score bound~\eqref{eq:pcm-lasso-normalized-score-tail} follows
from the stronger part of Lemma~\ref{lem:lasso-perm-score}.  For example, it
holds with $c_0=(32K_A^2K_\phi^2)^{-1}$ if, with probability tending to one,
\[
\max_{i,k}|B_{ik}-B_{i\cdot}|
\le K_A\sigma_{f,n,\mathrm{perm}},
\qquad
\max_{i,\ell}|\phi_{c,\ell}(Z_i)|\le K_\phi
\]
for fixed constants $K_A,K_\phi$.
\end{remark}

\subsection{Proofs for linear smoothers}
\label{subsec:perm-stab}

\subsubsection{General bound for linear smoothers}

Write $\mathbf m=\E[\mathbf u\mid\cF_n]$,
$\Omega_\Pi=\cov(\mathbf u\mid\cF_n)$, and
$\tilde E_n=n_\star^{-1}\|S\mathbf u-\mathbf m\|_2^2$.

\begin{lemma}[Bias--variance principle for permutation refits]\label{lem:perm-stab-linear-smoother}
For any $\cF_n$-measurable matrix $S$ representing a linear smoother,
\begin{align}
  \E[\tilde E_n\mid\cF_n]
  &\label{eq:ls-bv-identity}
  =\frac1{n_\star}\Tr(S\Omega_\Pi S^\top)+\frac1{n_\star}\|(I-S)\mathbf m\|_2^2\\
  &\label{eq:ls-cond-moments}
  \le\frac{\|\Omega_\Pi\|_{\mathrm{op}}\|S\|_F^2}{n_\star}+\frac1{n_\star}\|(I-S)\mathbf m\|_2^2.
\end{align}
Consequently,
\begin{equation*}
\tilde E_n
=O_\PP\!\left(
\frac{\|\Omega_\Pi\|_{\mathrm{op}}\|S\|_F^2
+\|(I-S)\mathbf m\|_2^2}{n_\star}
\right).
\end{equation*}
\end{lemma}

\begin{theorem}[Permutation stability of linear smoothers for the GCM]
\label{thm:ls-gcm}
Suppose the permuted $X$-on-$Z$ regression is a linear smoother with
$\cF_n$-measurable hat matrix $S$. If $S\bm1=\bm1$ and $\E[\bm X^2]<\infty$, then
\begin{align*}
  \tilde E_{n,x}=O_\PP\!\left(\frac{\|S\|_F^2}{n_\star}\right)
  \quad\text{and}\quad
  n_\star E_{n,y}\tilde E_{n,x}=O_\PP\!\left(E_{n,y}\|S\|_F^2\right).
\end{align*}
Consequently, $\|S\|_F^2=o_\PP(n_\star)$ implies
$\tilde E_{n,x}=o_\PP(1)$. If, in addition,
$E_{n,y}\|S\|_F^2=o_\PP(1)$, then
$n_\star E_{n,y}\tilde E_{n,x}=o_\PP(1)$, so both parts of
Assumption~\ref{assumption:perm} hold.
\end{theorem}

\begin{theorem}[Permutation stability of linear smoothers for the PCM]\label{thm:ls-pcm}
Suppose the permuted projection regression on $\cD_1$ is a linear smoother with $\cF_n$-measurable hat matrix $S$, assume $\PP(\sigma_{f,n,\mathrm{perm}}^2>0)\to1$, and suppose the normalized bound on the row variances
\[
\frac{\displaystyle\max_i n_\star^{-1}\sum_{k\in\cD_1}(B_{ik}-B_{i\cdot})^2}{\sigma_{f,n,\mathrm{perm}}^2}=O_\PP(1)
\]
holds. Then, on $\{\sigma_{f,n,\mathrm{perm}}^2>0\}$,
\begin{align}
\frac{\tilde E_{n,f}^{\mathrm{PCM}}}{\sigma_{f,n,\mathrm{perm}}^2}
&=O_\PP\!\left(
\frac{\|S\|_F^2}{n_\star}
+
\frac{\|(I-S)\mathbf m\|_2^2}
{n_\star\sigma_{f,n,\mathrm{perm}}^2}
\right),\label{eq:pcm-ls-relative}\\
\frac{n_\star E_{n,y}^{\mathrm{PCM}}\tilde E_{n,f}^{\mathrm{PCM}}}{\sigma_{f,n,\mathrm{perm}}^2}
&=O_\PP\!\left(
E_{n,y}^{\mathrm{PCM}}\|S\|_F^2
+
\frac{E_{n,y}^{\mathrm{PCM}}\|(I-S)\mathbf m\|_2^2}
{\sigma_{f,n,\mathrm{perm}}^2}
\right).\label{eq:pcm-ls-relative-product}
\end{align}
Consequently, $\|S\|_F^2=o_\PP(n_\star)$ and
\[
\frac{\|(I-S)\mathbf m\|_2^2}
{n_\star\sigma_{f,n,\mathrm{perm}}^2}=o_\PP(1)
\]
imply the first part of Assumption~\ref{assumption:pcm-perm}. If, in addition,
\[
E_{n,y}^{\mathrm{PCM}}\|S\|_F^2=o_\PP(1),
\qquad
\frac{E_{n,y}^{\mathrm{PCM}}\|(I-S)\mathbf m\|_2^2}
{\sigma_{f,n,\mathrm{perm}}^2}=o_\PP(1),
\]
then the second part also holds.
\end{theorem}

\begin{proof}[Proof of Lemma~\ref{lem:perm-stab-linear-smoother}]
The decomposition
\[
S\mathbf u-\mathbf m
=S(\mathbf u-\mathbf m)-(I-S)\mathbf m
\]
has conditional cross term zero, since
$\E[\mathbf u-\mathbf m\mid\cF_n]=0$. Thus
\begin{align*}
\E[\tilde E_n\mid\cF_n]
&=\frac1{n_\star}
  \E[\|S(\mathbf u-\mathbf m)\|_2^2\mid\cF_n]
  +\frac{\|(I-S)\mathbf m\|_2^2}{n_\star}\\
&=\frac{\Tr(S\Omega_\Pi S^\top)}{n_\star}
  +\frac{\|(I-S)\mathbf m\|_2^2}{n_\star}.
\end{align*}
Because $0\preceq\Omega_\Pi\preceq\|\Omega_\Pi\|_{\mathrm{op}}I$,
\[
\Tr(S\Omega_\Pi S^\top)
\le\|\Omega_\Pi\|_{\mathrm{op}}\Tr(SS^\top)
=\|\Omega_\Pi\|_{\mathrm{op}}\|S\|_F^2.
\]
This proves~\eqref{eq:ls-bv-identity}--\eqref{eq:ls-cond-moments}.
Conditional Markov gives the stated stochastic order: for any nonnegative
$\cF_n$-measurable bound $b_n$ on $\E[\tilde E_n\mid\cF_n]$,
$\PP(\tilde E_n>Mb_n,b_n>0)\le M^{-1}$, while
$\tilde E_n=0$ a.s.\ on $\{b_n=0\}$.
\end{proof}

\begin{lemma}[Covariance of a permutation response generated from an array]
\label{lem:perm-array-cov}
Let $n_\star\ge2$ and let $(A_{ik})_{1\le i,k\le n_\star}$ be an $\cF_n$-measurable real array and let
$\Pi$ be uniform on the symmetric group, independently of $\cF_n$.  Define
\[
u_i\equiv A_{i,\Pi(i)},
\qquad
A_{i\cdot}\equiv\frac1{n_\star}\sum_{k=1}^{n_\star}A_{ik},
\qquad
v_{A,\max}\equiv\max_i\frac1{n_\star}\sum_{k=1}^{n_\star}(A_{ik}-A_{i\cdot})^2.
\]
Then $\mathbf m=(A_{i\cdot})_{i=1}^{n_\star}$ and
\begin{equation}\label{eq:array-cov-op}
\|\Omega_\Pi\|_{\mathrm{op}}
\le \frac{n_\star}{n_\star-1}v_{A,\max}.
\end{equation}
For GCM, $A_{ik}=X_k$ and $v_{A,\max}=O_\PP(1)$ if $\E[\bm X^2]<\infty$.
For PCM, $A_{ik}=B_{ik}$; Proposition~\ref{prop:pcm-product-law-bridge} gives $v_{A,\max}/\sigma_{f,n,\mathrm{perm}}^2=O_\PP(1)$ under its stated envelope condition.
\end{lemma}

\begin{proof}
For every $q\in\R^{n_\star}$, Lemma~\ref{lem:perm-array-moments}, applied to
$h_{ik}=q_i(A_{ik}-A_{i\cdot})$, gives
\begin{align*}
q^\top\Omega_\Pi q
&=\var\!\left(\sum_i q_iA_{i,\Pi(i)}\middle|\cF_n\right)\\
&\le\frac1{n_\star-1}\sum_iq_i^2\sum_k(A_{ik}-A_{i\cdot})^2\\
&\le\frac{n_\star v_{A,\max}}{n_\star-1}\sum_iq_i^2
=\frac{n_\star}{n_\star-1}v_{A,\max}\|q\|_2^2.
\end{align*}
Taking the supremum over $\|q\|_2=1$ proves~\eqref{eq:array-cov-op}.
For GCM, $v_{A,\max}=n_\star^{-1}\sum_k(X_k-\bar X)^2=O_\PP(1)$.
For PCM, use~\eqref{eq:pcm-envelope-row}.
\end{proof}

\begin{proof}[Proof of Theorem~\ref{thm:ls-gcm}]
  For the GCM,
  \(\mathbf m=\bar X\mathbf1\).  Therefore
  \[
  \frac1{n_\star}\|(I-S)\mathbf m\|_2^2
  =\bar X^2\frac1{n_\star}\|(I-S)\mathbf1\|_2^2=0
  \]
  because \(S\mathbf1=\mathbf1\).  Lemma~\ref{lem:perm-array-cov} and
  \(\E[\bm X^2]<\infty\) give
  \(\|\Omega_\Pi\|_{\mathrm{op}}=O_\PP(1)\).  Lemma~\ref{lem:perm-stab-linear-smoother} now yields
  \[
  \tilde E_{n,x}
  =O_\PP\!\left(\frac{\|S\|_F^2}{n_\star}\right).
  \]
  Moreover,
  \[
  n_\star E_{n,y}\tilde E_{n,x}
  =O_\PP(E_{n,y}\|S\|_F^2).
  \]
  The two stated sufficient conditions yield the corresponding
  $o_\PP(1)$ conclusions.
  \end{proof}

\begin{proof}[Proof of Theorem~\ref{thm:ls-pcm}]
For $u_i=B_{i,\Pi(i)}$, the conditional target is $m_i=B_{i\cdot}$.
Lemma~\ref{lem:perm-array-cov} and the assumed bound on the row variances give
\[
\frac{\|\Omega_\Pi\|_{\mathrm{op}}}{\sigma_{f,n,\mathrm{perm}}^2}
\le\frac{n_\star}{n_\star-1}
\frac{\max_i n_\star^{-1}\sum_k(B_{ik}-B_{i\cdot})^2}
{\sigma_{f,n,\mathrm{perm}}^2}=O_\PP(1).
\]
The conditional bound~\eqref{eq:ls-cond-moments} therefore becomes
\begin{align*}
\E\!\left[
\frac{\tilde E_{n,f}^{\mathrm{PCM}}}{\sigma_{f,n,\mathrm{perm}}^2}
\middle|\cF_n\right]
&\le
\frac{\|\Omega_\Pi\|_{\mathrm{op}}}{\sigma_{f,n,\mathrm{perm}}^2}
\frac{\|S\|_F^2}{n_\star}
+\frac{\|(I-S)\mathbf m\|_2^2}{n_\star\sigma_{f,n,\mathrm{perm}}^2}.
\end{align*}
Conditional Markov, as in Lemma~\ref{lem:perm-stab-linear-smoother},
gives~\eqref{eq:pcm-ls-relative}.
Multiplying by the $\cF_n$-measurable quantity $n_\star E_{n,y}^{\mathrm{PCM}}$
gives~\eqref{eq:pcm-ls-relative-product} and the stated sufficient conditions.
\end{proof}

\subsubsection{OLS, ridge, and KRR}
\label{subsec:perm-stab-smoother-verification}

\begin{lemma}[Properties of the three linear smoothers]
\label{lem:smoother-properties}
For each of the OLS, ridge, and KRR matrices in
Subsection~\ref{subsec:perm-stab-learner-setup},
\[
S\mathbf1=\mathbf1,\qquad
\|S\|_{\mathrm{op}}=1,\qquad
\|I-S\|_{\mathrm{op}}\le1.
\]
Their variance complexities are the expressions given there.
\end{lemma}
\begin{proof}
\noindent\textit{OLS.}\quad
By the assumption of full column rank on $\widetilde\Phi$,
\[
S_{\mathrm{OLS}}^\top=S_{\mathrm{OLS}},\qquad
S_{\mathrm{OLS}}^2=S_{\mathrm{OLS}},\qquad
S_{\mathrm{OLS}}\widetilde\Phi=\widetilde\Phi.
\]
Thus $S_{\mathrm{OLS}}$ is the orthogonal projection onto
$\operatorname{col}(\widetilde\Phi)$.  Since the first column of
$\widetilde\Phi$ is $\mathbf1$, it reproduces constants, and
\[
S_{\mathrm{OLS}}\mathbf1=\mathbf1,\qquad
\|S_{\mathrm{OLS}}\|_F^2
=\Tr(S_{\mathrm{OLS}}^\top S_{\mathrm{OLS}})
=\Tr(S_{\mathrm{OLS}})=d+1.
\]

\smallskip
\noindent\textit{Ridge.}\quad
Eliminating the intercept gives the normal equations and fitted vector
\begin{align*}
(\Phi_c^\top\Phi_c+\lambda_n I_d)
\hat{\boldsymbol\gamma}^{(\Pi)}
&=\Phi_c^\top\mathbf u,\\
\hat{\boldsymbol\mu}^{(\Pi)}
&=P_1\mathbf u+\Phi_c\hat{\boldsymbol\gamma}^{(\Pi)}\\
&=\left\{P_1+\Phi_c(\Phi_c^\top\Phi_c+\lambda_n I_d)^{-1}
\Phi_c^\top\right\}\mathbf u.
\end{align*}
Because $\Phi_c^\top\mathbf1=0$, this matrix is symmetric and
$S_{\mathrm{ridge}}(\lambda_n)\mathbf1=\mathbf1$.
Choose orthonormal eigenvectors $v_j$ of $\Phi_c^\top\Phi_c$.
For $\rho_j>0$, put $q_j=\Phi_cv_j/\sqrt{\rho_j}$.  These vectors are
orthonormal and orthogonal to $\mathbf1$, and
\[
S_{\mathrm{ridge}}(\lambda_n)q_j
=\frac{\Phi_c(\Phi_c^\top\Phi_c+\lambda_n I_d)^{-1}
\Phi_c^\top\Phi_cv_j}{\sqrt{\rho_j}}
=\frac{\rho_j}{\rho_j+\lambda_n}q_j.
\]
On the orthogonal complement of
$\operatorname{span}(\mathbf1)+\operatorname{col}(\Phi_c)$, the smoother
is zero. Including the zero $\rho_j$ in the sum gives~\eqref{eq:ridge-frob}:
\[
\|S_{\mathrm{ridge}}(\lambda_n)\|_F^2
=1+\sum_{j=1}^d\left(\frac{\rho_j}{\rho_j+\lambda_n}\right)^2
\le d+1.
\]

\smallskip
\noindent\textit{KRR.}\quad
The centered kernel matrix satisfies
\[
K_c\mathbf1=0,\qquad HK_c=K_cH=K_c.
\]
Consequently $S_{\mathrm{KRR}}(\lambda_n)\mathbf1=\mathbf1$.
Choose an orthonormal eigenbasis for $K_c$ containing
$n_\star^{-1/2}\mathbf1$.  For any other vector $q_j$ in this basis,
$Hq_j=q_j$ and $K_cq_j=\nu_jq_j$, so
\[
S_{\mathrm{KRR}}(\lambda_n)q_j
=P_1q_j+K_c(K_c+\lambda_n I_{n_\star})^{-1}Hq_j
=\frac{\nu_j}{\nu_j+\lambda_n}q_j.
\]
This gives a real orthonormal eigenbasis for the smoother, proving
symmetry, and gives
\[
\|S_{\mathrm{KRR}}(\lambda_n)\|_F^2
=1+\sum_{j=1}^{n_\star}
\left(\frac{\nu_j}{\nu_j+\lambda_n}\right)^2,
\]
where the summand for the constant direction is zero.
This is~\eqref{eq:eff-dim}.

\smallskip
In all three cases, the eigenvalues of $S$ lie in $[0,1]$, and $1$ is
an eigenvalue.  Symmetry therefore gives
\[
\|S\|_{\mathrm{op}}=1,\qquad
\|I-S\|_{\mathrm{op}}\le1.
\]
\end{proof}

\begin{proof}[Proof of Corollary~\ref{cor:gcm-explicit-smoother-rates}]
Apply Theorem~\ref{thm:ls-gcm} using the reproduction of constants
and the complexities established in Lemma~\ref{lem:smoother-properties}.
\end{proof}

\subsubsection{PCM target approximation and explicit rates}
\label{subsec:perm-stab-pcm-tools}

We use~\eqref{eq:pcm-target-empirical-population} to transfer the decompositions of the population target to the exact permutation target.

\begin{proposition}[Population approximation of the PCM permutation target]\label{prop:pcm-target-approx}
Suppose $\sigma_{f,n,\mathrm{perm}}^2>0$ and the conditions of
Proposition~\ref{prop:pcm-product-law-bridge} hold.  Use the target decompositions and measurability conditions in
Subsection~\ref{subsec:perm-stab-results-pcm}.  Then
\begin{align}
\frac{\|(I-S_{\mathrm{OLS}})\mathbf m\|_2^2}
{n_\star\sigma_{f,n,\mathrm{perm}}^2}
&\le
2\frac{\sum_{i\in\cD_1}r_n(Z_i)^2}
{n_\star\sigma_{f,n,\mathrm{perm}}^2}
+O_\PP(n_\star^{-1}),\label{eq:pcm-target-ols-bound}\\
\frac{\|(I-S_{\mathrm{ridge}}(\lambda_n))\mathbf m\|_2^2}
{n_\star\sigma_{f,n,\mathrm{perm}}^2}
&\le
2\frac{\sum_{i\in\cD_1}r_n(Z_i)^2
+\lambda_n\|\boldsymbol\gamma_n\|_2^2}
{n_\star\sigma_{f,n,\mathrm{perm}}^2}
+O_\PP(n_\star^{-1}).\label{eq:pcm-target-ridge-bound}
\end{align}
For KRR,
\begin{equation}\label{eq:pcm-target-krr-bound}
\frac{\|(I-S_{\mathrm{KRR}}(\lambda_n))\mathbf m\|_2^2}
{n_\star\sigma_{f,n,\mathrm{perm}}^2}
\le
2\frac{\sum_{i\in\cD_1}r_{n,\mathrm{KRR}}(Z_i)^2
+\lambda_n\|h_n\|_{\mathcal H_k}^2}
{n_\star\sigma_{f,n,\mathrm{perm}}^2}
+O_\PP(n_\star^{-1}).
\end{equation}
\end{proposition}

\begin{proof}
The exact and population target vectors satisfy
\[
\|\mathbf m-\mathbf b\|_2^2
=\sum_{i\in\cD_1}\{B_{i\cdot}-b^{\mathrm{perm}}(Z_i)\}^2.
\]
Thus Proposition~\ref{prop:pcm-product-law-bridge}, specifically
\eqref{eq:pcm-target-empirical-population}, gives
\begin{equation}\label{eq:pcm-target-transfer-error}
\frac{\|\mathbf m-\mathbf b\|_2^2}
{n_\star\sigma_{f,n,\mathrm{perm}}^2}=O_\PP(n_\star^{-1}).
\end{equation}
For any of the three smoothers, Lemma~\ref{lem:smoother-properties}
gives $\|I-S\|_{\mathrm{op}}\le1$.  Consequently,
\begin{align*}
\|(I-S)\mathbf m\|_2^2
&=\|(I-S)\mathbf b+(I-S)(\mathbf m-\mathbf b)\|_2^2\\
&\le2\|(I-S)\mathbf b\|_2^2
+2\|(I-S)(\mathbf m-\mathbf b)\|_2^2\\
&\le2\|(I-S)\mathbf b\|_2^2+2\|\mathbf m-\mathbf b\|_2^2.
\end{align*}
Dividing by $n_\star\sigma_{f,n,\mathrm{perm}}^2$ yields the common bound
\begin{equation}\label{eq:pcm-target-contraction}
\frac{\|(I-S)\mathbf m\|_2^2}{n_\star\sigma_{f,n,\mathrm{perm}}^2}
\le
2\frac{\|(I-S)\mathbf b\|_2^2}{n_\star\sigma_{f,n,\mathrm{perm}}^2}
+O_\PP(n_\star^{-1}).
\end{equation}
It remains to bound the residual for the population target for each smoother.

\smallskip
\noindent\textit{OLS.}\quad
By~\eqref{eq:pcm-target-linear-decomp},
$\mathbf b=\alpha_n\mathbf1+\Phi\boldsymbol\gamma_n+\mathbf r_n$.
The first two terms lie in the range of the OLS projection, so
\begin{align*}
(I-S_{\mathrm{OLS}})\mathbf b
&=(I-S_{\mathrm{OLS}})\mathbf r_n,\\
\|(I-S_{\mathrm{OLS}})\mathbf b\|_2^2
&\le\|\mathbf r_n\|_2^2
=\sum_{i\in\cD_1}r_n(Z_i)^2.
\end{align*}
Substitution into~\eqref{eq:pcm-target-contraction} proves~\eqref{eq:pcm-target-ols-bound}.

\smallskip
\noindent\textit{Ridge.}\quad
Let $(\hat\alpha_b,\hat{\boldsymbol\gamma}_b)$ minimize the ridge
criterion with response $\mathbf b$.  Its fitted vector is
$\hat\alpha_b\mathbf1+\Phi\hat{\boldsymbol\gamma}_b
=S_{\mathrm{ridge}}(\lambda_n)\mathbf b$.
Nonnegativity of the penalty and optimality give
\begin{align*}
\|(I-S_{\mathrm{ridge}}(\lambda_n))\mathbf b\|_2^2
&\le\|\mathbf b-\hat\alpha_b\mathbf1
-\Phi\hat{\boldsymbol\gamma}_b\|_2^2
+\lambda_n\|\hat{\boldsymbol\gamma}_b\|_2^2\\
&\le\|\mathbf b-\alpha_n\mathbf1
-\Phi\boldsymbol\gamma_n\|_2^2
+\lambda_n\|\boldsymbol\gamma_n\|_2^2\\
&=\sum_{i\in\cD_1}r_n(Z_i)^2
+\lambda_n\|\boldsymbol\gamma_n\|_2^2.
\end{align*}
Substitution into~\eqref{eq:pcm-target-contraction} proves~\eqref{eq:pcm-target-ridge-bound}.

\smallskip
\noindent\textit{KRR.}\quad
Let $(\hat\alpha_b,\hat h_b)$ minimize the KRR criterion with response
$\mathbf b$, and write $\hat{\mathbf h}_b=(\hat h_b(Z_i))_{i\in\cD_1}$.
Then $\hat\alpha_b\mathbf1+\hat{\mathbf h}_b
=S_{\mathrm{KRR}}(\lambda_n)\mathbf b$.
Optimality and nonnegativity of the penalty give
\begin{align*}
\|(I-S_{\mathrm{KRR}}(\lambda_n))\mathbf b\|_2^2
&\le\|\mathbf b-\hat\alpha_b\mathbf1-\hat{\mathbf h}_b\|_2^2
+\lambda_n\|\hat h_b\|_{\mathcal H_k}^2\\
&\le\sum_{i\in\cD_1}
\{b^{\mathrm{perm}}(Z_i)-\alpha_{n,\mathrm{KRR}}-h_n(Z_i)\}^2
+\lambda_n\|h_n\|_{\mathcal H_k}^2\\
&=\sum_{i\in\cD_1}r_{n,\mathrm{KRR}}(Z_i)^2
+\lambda_n\|h_n\|_{\mathcal H_k}^2.
\end{align*}
Substitution into~\eqref{eq:pcm-target-contraction} proves~\eqref{eq:pcm-target-krr-bound}.
\end{proof}

\begin{proof}[Proof of Corollary~\ref{cor:pcm-explicit-smoother-rates}]
Substitute the three bounds of Proposition~\ref{prop:pcm-target-approx}
and the complexities in Lemma~\ref{lem:smoother-properties} into
Theorem~\ref{thm:ls-pcm}. Each $\mathcal C(S)\ge1$, so the additional
$O_\PP(n_\star^{-1})$ term is absorbed by $\mathcal C(S)/n_\star$.
Multiplication by $n_\star E_{n,y}^{\mathrm{PCM}}$ gives the claims concerning products of errors.
\end{proof}

\subsection{Proofs for lasso}
\label{subsec:lasso-perm-stab}

\subsubsection{A generic bound for the permutation score}
\label{subsec:lasso-perm-score-bound}

\begin{lemma}[Permutation score bound for lasso]
\label{lem:lasso-perm-score}
Let $(A_{ik})_{1\le i,k\le n_\star}$ be the array in
Lemma~\ref{lem:perm-array-cov}, with $v_{A,\max}$ as defined there, and use
$\Delta_n(A),v_{\phi,\max}$ from~\eqref{eq:lasso-score-generic}. Then
\begin{equation}\label{eq:lasso-delta-bound}
\E[\Delta_n(A)^2\mid\cF_n]
\le
\frac{d\,v_{A,\max}v_{\phi,\max}}{n_\star-1}.
\end{equation}
Consequently, if $v_{A,\max}=O_\PP(1)$ and
$v_{\phi,\max}=O_\PP(1)$, then
$\Delta_n(A)=O_\PP(\sqrt{d/n_\star})$.

If, in addition, for some positive $\cF_n$-measurable bounds $M_{A,n}$ and
$M_{\phi,n}$,
\[
\max_{i,k}|A_{ik}-A_{i\cdot}|\le M_{A,n},
\qquad
\max_{i,\ell}|\phi_{c,\ell}(Z_i)|\le M_{\phi,n},
\]
then, conditionally on $\cF_n$,
\begin{equation}\label{eq:lasso-delta-subgauss}
\PP\{\Delta_n(A)\ge t\mid\cF_n\}
\le
2d\exp\!\left\{-\frac{n_\star t^2}
{32M_{A,n}^2M_{\phi,n}^2}\right\},
\qquad t>0.
\end{equation}
\end{lemma}

\begin{proof}
For each $\ell\le d$, set
$T_\ell=n_\star^{-1}\sum_i\phi_{c,\ell}(Z_i)(A_{i,\Pi(i)}-A_{i\cdot})$.
Lemma~\ref{lem:perm-array-moments} gives $\E[T_\ell\mid\cF_n]=0$ and
\begin{align*}
\E[T_\ell^2\mid\cF_n]
&\le\frac{\sum_i\phi_{c,\ell}(Z_i)^2\sum_k(A_{ik}-A_{i\cdot})^2}
{n_\star^2(n_\star-1)}\\
&\le\frac{v_{A,\max}}{n_\star-1}
\left\{\frac1{n_\star}\sum_i\phi_{c,\ell}(Z_i)^2\right\}
\le\frac{v_{A,\max}v_{\phi,\max}}{n_\star-1}.
\end{align*}
Since $\Delta_n(A)^2=\max_\ell T_\ell^2\le\sum_\ell T_\ell^2$,
summing proves~\eqref{eq:lasso-delta-bound}; conditional Markov then gives
the stated $O_\PP(\sqrt{d/n_\star})$ rate.

It remains to prove \eqref{eq:lasso-delta-subgauss}.  Fix a coordinate
$\ell$ and view $T_\ell=g_\ell(\Pi)$ as a function of the uniform
permutation.  Expose $\Pi(1),\ldots,\Pi(n_\star)$ sequentially and let
\[
\mathcal M_r
\equiv\E[T_\ell\mid\cF_n,\Pi(1),\ldots,\Pi(r)],
\qquad r=0,\ldots,n_\star,
\]
be the corresponding Doob martingale.  Conditional on the first $r-1$
images, take two distinct possible values $a$ and $b$ for $\Pi(r)$.  Couple any
completion $\pi$ with $\pi(r)=a$ to a completion $\pi'$ with
$\pi'(r)=b$ by swapping the images of rows $r$ and $s=\pi^{-1}(b)>r$.
Only these two rows change, so
\begin{align*}
g_\ell(\pi)-g_\ell(\pi')
&=\frac1{n_\star}\left\{
\phi_{c,\ell}(Z_r)(A_{ra}-A_{rb})
+\phi_{c,\ell}(Z_s)(A_{sb}-A_{sa})\right\},\\
|g_\ell(\pi)-g_\ell(\pi')|
&\le\frac{M_{\phi,n}}{n_\star}
\left\{|A_{ra}-A_{r\cdot}|+|A_{rb}-A_{r\cdot}|
+|A_{sb}-A_{s\cdot}|+|A_{sa}-A_{s\cdot}|\right\}\\
&\le\frac{4M_{A,n}M_{\phi,n}}{n_\star}.
\end{align*}
The swap is a bijection between the two sets of completions. Averaging shows that the
conditional expectations corresponding to any two admissible values of
$\Pi(r)$ differ by at most the same amount.  Since $\mathcal M_{r-1}$ is a
convex combination of those conditional expectations,
\[
|\mathcal M_r-\mathcal M_{r-1}|
\le\frac{4M_{A,n}M_{\phi,n}}{n_\star}
\qquad\text{a.s. conditionally on }\cF_n.
\]
Moreover, $\mathcal M_0=\E[T_\ell\mid\cF_n]=0$ and
$\mathcal M_{n_\star}=T_\ell$.  Azuma--Hoeffding therefore yields
\begin{align*}
\PP(|T_\ell|\ge t\mid\cF_n)
&\le
2\exp\!\left\{-\frac{t^2}
{2\sum_{r=1}^{n_\star}(4M_{A,n}M_{\phi,n}/n_\star)^2}\right\}\\
&=
2\exp\!\left\{-\frac{n_\star t^2}
{32M_{A,n}^2M_{\phi,n}^2}\right\}.
\end{align*}
Finally, $\{\Delta_n(A)\ge t\}\subseteq\bigcup_{\ell=1}^d
\{|T_\ell|\ge t\}$, so a union bound proves
\eqref{eq:lasso-delta-subgauss}.
\end{proof}

\subsubsection{Proof for the GCM}
\label{subsec:lasso-perm-stab-gcm}

\begin{proof}[Proof of Theorem~\ref{thm:lasso-gcm}]
Write $X^{(\Pi)}=(X_{\Pi(i)})_{i=1}^{n_\star}$.
A permutation preserves $\bar X$, so
$HX^{(\Pi)}=X^{(\Pi)}-\bar X\mathbf1$.
Specializing~\eqref{eq:centered-objective} to lasso and comparing its
minimizer with zero gives
\[
\frac{\|HX^{(\Pi)}-\Phi_c\hat{\boldsymbol\gamma}^{(\Pi)}\|_2^2}{2n_\star}
+\lambda_n\|\hat{\boldsymbol\gamma}^{(\Pi)}\|_1
\le\frac{\|HX^{(\Pi)}\|_2^2}{2n_\star}.
\]
Expanding the square and cancelling $\|HX^{(\Pi)}\|_2^2/(2n_\star)$,
\begin{align*}
\frac{\|\Phi_c\hat{\boldsymbol\gamma}^{(\Pi)}\|_2^2}{2n_\star}
+\lambda_n\|\hat{\boldsymbol\gamma}^{(\Pi)}\|_1
&\le\frac{(HX^{(\Pi)})^\top\Phi_c\hat{\boldsymbol\gamma}^{(\Pi)}}{n_\star}\\
&\le\left\|\frac{\Phi_c^\top(X^{(\Pi)}-\bar X\mathbf1)}{n_\star}\right\|_\infty
\|\hat{\boldsymbol\gamma}^{(\Pi)}\|_1\\
&=\Delta_n^{\mathrm{GCM}}\|\hat{\boldsymbol\gamma}^{(\Pi)}\|_1,
\end{align*}
where the second inequality is H\"older's inequality. Thus, on
$\{\Delta_n^{\mathrm{GCM}}\le\lambda_n/2\}$,
\[
0\le
\frac{\|\Phi_c\hat{\boldsymbol\gamma}^{(\Pi)}\|_2^2}{2n_\star}
+\frac{\lambda_n}{2}\|\hat{\boldsymbol\gamma}^{(\Pi)}\|_1
\le0.
\]
Since $\lambda_n>0$, this implies $\hat{\boldsymbol\gamma}^{(\Pi)}=0$.
The fitted vector is therefore $\bar X\mathbf1$, exactly the permutation
target, and $\tilde E_{n,x}=0$ on this event. This proves
\eqref{eq:lasso-gcm-zero-stability}; moreover, for every $\varepsilon>0$,
\[
\PP(n_\star E_{n,y}\tilde E_{n,x}>\varepsilon)
\le\PP(\Delta_n^{\mathrm{GCM}}>\lambda_n/2)\longrightarrow0.
\]

Under the second-moment conditions, Lemma~\ref{lem:lasso-perm-score}
with $A_{ik}=X_k$ gives
$\Delta_n^{\mathrm{GCM}}=O_\PP(\sqrt{d/n_\star})$ because
$v_{A,\max}=n_\star^{-1}\sum_k(X_k-\bar X)^2=O_\PP(1)$.
Hence $\sqrt{d/n_\star}=o(\lambda_n)$ implies score domination.
For the exponential bound, let $\mathcal G_n$ be the event on which
\eqref{eq:lasso-gcm-score-tail} holds. Since $\PP(\mathcal G_n)\to1$,
$d\to\infty$, and $C>2/\sqrt{c_G}$, substituting
$\lambda_n=C\sqrt{\log d/n_\star}$ gives
\[
\PP(\Delta_n^{\mathrm{GCM}}>\lambda_n/2)
\le\PP(\mathcal G_n^c)+2d^{\,1-c_GC^2/4}\longrightarrow0.\qedhere
\]
\end{proof}

\subsubsection{Proofs for the PCM}
\label{subsec:lasso-perm-stab-pcm}

\begin{proof}[Proof of Theorem~\ref{thm:lasso-vpcm}]
\noindent\textit{Centered fit.}\quad
Write $\boldsymbol\varepsilon^{(\Pi)}=\mathbf u-\mathbf m$.
The population approximation and its transfer to the permutation target give
\begin{equation}\label{eq:pcm-lasso-population-decomp}
\mathbf m
=\mathbf b+(\mathbf m-\mathbf b)
=\alpha_n\mathbf1+\Phi\boldsymbol\gamma_n+\mathbf r_n
+(\mathbf m-\mathbf b),
\end{equation}
and~\eqref{eq:pcm-target-transfer-error} gives
\[
\frac{\|\mathbf m-\mathbf b\|_2^2}
{n_\star\sigma_{f,n,\mathrm{perm}}^2}=O_\PP(n_\star^{-1}).
\]
Specializing~\eqref{eq:centered-objective} to lasso, its minimizing
property yields
\begin{align*}
&\frac{\|H\mathbf u-\Phi_c\hat{\boldsymbol\gamma}^{(\Pi)}\|_2^2}{2n_\star}
+\lambda_n\|\hat{\boldsymbol\gamma}^{(\Pi)}\|_1\\
&\qquad\le
\frac{\|H\mathbf u-\Phi_c\boldsymbol\gamma_n\|_2^2}{2n_\star}
+\lambda_n\|\boldsymbol\gamma_n\|_1.
\end{align*}
For either coefficient vector $\boldsymbol\gamma$, since $H^2=H$ and
$H\Phi_c=\Phi_c$,
\begin{align*}
\|H\mathbf u-\Phi_c\boldsymbol\gamma\|_2^2
&=\|H\boldsymbol\varepsilon^{(\Pi)}\|_2^2
+\|H\mathbf m-\Phi_c\boldsymbol\gamma\|_2^2\\
&\quad+2(\boldsymbol\varepsilon^{(\Pi)})^\top H\mathbf m
-2(\boldsymbol\varepsilon^{(\Pi)})^\top\Phi_c\boldsymbol\gamma.
\end{align*}
The terms $\|H\boldsymbol\varepsilon^{(\Pi)}\|_2^2$ and
$2(\boldsymbol\varepsilon^{(\Pi)})^\top H\mathbf m$ cancel in the comparison.
On $\{\Delta_n^{\mathrm{PCM}}\le\lambda_n/2\}$, H\"older's inequality gives
\begin{align}
&\frac{\|H\mathbf m-\Phi_c\hat{\boldsymbol\gamma}^{(\Pi)}\|_2^2}{2n_\star}
+\lambda_n\|\hat{\boldsymbol\gamma}^{(\Pi)}\|_1\notag\\
&\quad\le\frac{\|H\mathbf m-\Phi_c\boldsymbol\gamma_n\|_2^2}{2n_\star}
+\lambda_n\|\boldsymbol\gamma_n\|_1
+\frac{(\boldsymbol\varepsilon^{(\Pi)})^\top\Phi_c
(\hat{\boldsymbol\gamma}^{(\Pi)}-\boldsymbol\gamma_n)}{n_\star}\notag\\
&\quad\le\frac{\|H\mathbf m-\Phi_c\boldsymbol\gamma_n\|_2^2}{2n_\star}
+\lambda_n\|\boldsymbol\gamma_n\|_1
+\frac{\lambda_n}{2}\|\hat{\boldsymbol\gamma}^{(\Pi)}-\boldsymbol\gamma_n\|_1.
\label{eq:pcm-lasso-basic-inequality}
\end{align}
Here
$\|\Phi_c^\top\boldsymbol\varepsilon^{(\Pi)}/n_\star\|_\infty
=\Delta_n^{\mathrm{PCM}}$.
Using
\begin{align*}
\|\hat{\boldsymbol\gamma}^{(\Pi)}-\boldsymbol\gamma_n\|_1
&\le\|\hat{\boldsymbol\gamma}^{(\Pi)}\|_1+\|\boldsymbol\gamma_n\|_1,\\
H\mathbf m-\Phi_c\boldsymbol\gamma_n
&=H\mathbf r_n+H(\mathbf m-\mathbf b),
\end{align*}
we obtain
\begin{align}
\frac{\|H\mathbf m-\Phi_c\hat{\boldsymbol\gamma}^{(\Pi)}\|_2^2}{n_\star}
+\lambda_n\|\hat{\boldsymbol\gamma}^{(\Pi)}\|_1
&\le\frac{\|H\mathbf m-\Phi_c\boldsymbol\gamma_n\|_2^2}{n_\star}
+3\lambda_n\|\boldsymbol\gamma_n\|_1\notag\\
&\le\frac{2\|\mathbf r_n\|_2^2+2\|\mathbf m-\mathbf b\|_2^2}{n_\star}
+3\lambda_n\|\boldsymbol\gamma_n\|_1.
\label{eq:pcm-lasso-centered-fit-bound}
\end{align}
The last inequality uses $\|H\|_{\mathrm{op}}\le1$ and
$\|a+b\|_2^2\le2\|a\|_2^2+2\|b\|_2^2$.

\noindent\textit{Intercept and total error.}\quad
With $\bar m=n_\star^{-1}\mathbf1^\top\mathbf m$, the error of the fitted
vector is
\[
\bar u\mathbf1+\Phi_c\hat{\boldsymbol\gamma}^{(\Pi)}-\mathbf m
=(\bar u-\bar m)\mathbf1
+\Phi_c\hat{\boldsymbol\gamma}^{(\Pi)}-H\mathbf m.
\]
The two terms on the right are orthogonal because
$\mathbf1^\top\Phi_c=0$ and $\mathbf1^\top H=0$. Hence
\begin{equation}\label{eq:pcm-lasso-orthogonal-decomp}
\tilde E_{n,f}^{\mathrm{PCM}}
=(\bar u-\bar m)^2
+\frac{\|\Phi_c\hat{\boldsymbol\gamma}^{(\Pi)}-H\mathbf m\|_2^2}{n_\star}.
\end{equation}
By Lemma~\ref{lem:perm-array-cov} and~\eqref{eq:pcm-envelope-row}, with $A=B$,
\begin{align*}
\E\!\left[\frac{(\bar u-\bar m)^2}
{\sigma_{f,n,\mathrm{perm}}^2}\middle|\cF_n\right]
&=\frac{\mathbf1^\top\Omega_\Pi\mathbf1}{n_\star^2\sigma_{f,n,\mathrm{perm}}^2}\\
&\le\frac{\|\Omega_\Pi\|_{\mathrm{op}}}{n_\star\sigma_{f,n,\mathrm{perm}}^2}
\le\frac{v_{A,\max}}{(n_\star-1)\sigma_{f,n,\mathrm{perm}}^2}
=O_\PP(n_\star^{-1}).
\end{align*}
Conditional Markov gives
\begin{equation}\label{eq:pcm-lasso-intercept-rate}
\frac{(\bar u-\bar m)^2}{\sigma_{f,n,\mathrm{perm}}^2}
=O_\PP(n_\star^{-1}).
\end{equation}
Combining~\eqref{eq:pcm-lasso-centered-fit-bound} and
\eqref{eq:pcm-lasso-orthogonal-decomp}, on the event where the score bound holds,
\begin{align*}
\frac{\tilde E_{n,f}^{\mathrm{PCM}}}{\sigma_{f,n,\mathrm{perm}}^2}
&\le\frac{(\bar u-\bar m)^2}{\sigma_{f,n,\mathrm{perm}}^2}
+\frac{2\|\mathbf r_n\|_2^2+2\|\mathbf m-\mathbf b\|_2^2}
{n_\star\sigma_{f,n,\mathrm{perm}}^2}\\
&\quad+\frac{3\lambda_n\|\boldsymbol\gamma_n\|_1}{\sigma_{f,n,\mathrm{perm}}^2}.
\end{align*}
This event has probability tending to one; using
\eqref{eq:pcm-target-transfer-error} and~\eqref{eq:pcm-lasso-intercept-rate}
therefore proves~\eqref{eq:lasso-pcm-stability-rate}.
\end{proof}

\begin{proof}[Proof of Corollary~\ref{cor:pcm-lasso-sparse}]
\noindent\textit{Score domination.}\quad
Let $\mathcal G_n$ denote the event on which the conditional tail bound
\eqref{eq:pcm-lasso-normalized-score-tail} holds, so
$\PP(\mathcal G_n)\to1$. On $\mathcal G_n$, substitution of
$t=(C/2)\sqrt{\log d/n_\star}$ gives
\[
\PP\!\left(\Delta_n^{\mathrm{PCM}}>\lambda_n/2\mid\cF_n\right)
\le2d\exp\!\left(-\frac{c_0C^2}{4}\log d\right)
=2d^{\,1-c_0C^2/4}.
\]
Since $C>2/\sqrt{c_0}$ and $d\to\infty$, the tower property yields
\[
\PP\!\left(\Delta_n^{\mathrm{PCM}}>\lambda_n/2\right)
\le\PP(\mathcal G_n^c)+2d^{\,1-c_0C^2/4}\longrightarrow0,
\]
which proves~\eqref{eq:lasso-pcm-score-domination}.

\noindent\textit{Sparse prediction bound.}\quad
Let
\begin{align*}
\boldsymbol\delta
&\equiv\hat{\boldsymbol\gamma}^{(\Pi)}-\boldsymbol\gamma_n,\\
\mathcal E_n&\equiv\frac1{n_\star}\|H\mathbf m-
\Phi_c\hat{\boldsymbol\gamma}^{(\Pi)}\|_2^2,\\
R_n&\equiv\frac1{n_\star}\|H\mathbf r_n+H(\mathbf m-\mathbf b)\|_2^2.
\end{align*}
Equation~\eqref{eq:pcm-lasso-population-decomp} gives
\[
\mathcal E_n=\frac1{n_\star}\|H\mathbf r_n+H(\mathbf m-\mathbf b)-
\Phi_c\boldsymbol\delta\|_2^2.
\]
Also, by~\eqref{eq:pcm-target-transfer-error},
\begin{align}
R_n
&\le\frac{2\|\mathbf r_n\|_2^2+2\|\mathbf m-\mathbf b\|_2^2}{n_\star},\notag\\
\frac{R_n}{\sigma_{f,n,\mathrm{perm}}^2}
&=O_\PP\!\left(
\frac{\|\mathbf r_n\|_2^2}
{n_\star\sigma_{f,n,\mathrm{perm}}^2}+\frac1{n_\star}
\right).
\label{eq:pcm-lasso-Rn-transfer}
\end{align}
On $\{\Delta_n^{\mathrm{PCM}}\le\lambda_n/2\}$, apply the basic inequality
\eqref{eq:pcm-lasso-basic-inequality} and the support decomposition
\begin{align*}
\|\boldsymbol\gamma_n+\boldsymbol\delta\|_1
&=\|\boldsymbol\gamma_{n,S_n}+\boldsymbol\delta_{S_n}\|_1
+\|\boldsymbol\delta_{S_n^c}\|_1\\
&\ge\|\boldsymbol\gamma_n\|_1-\|\boldsymbol\delta_{S_n}\|_1
+\|\boldsymbol\delta_{S_n^c}\|_1.
\end{align*}
Multiplying by two and collecting the two parts of
$\|\boldsymbol\delta\|_1$ yields
\begin{equation}\label{eq:pcm-lasso-cone-basic}
\mathcal E_n+\lambda_n\|\boldsymbol\delta_{S_n^c}\|_1
\le R_n+3\lambda_n\|\boldsymbol\delta_{S_n}\|_1.
\end{equation}
If $s_n=0$, this gives $\mathcal E_n\le R_n$; if
$\boldsymbol\delta=0$, then $\mathcal E_n=R_n$ by definition.
For $s_n\ge1$ and $\boldsymbol\delta\ne0$, work on
$\{\kappa_n\ge\kappa_0\}$ and distinguish two cases.
If $\lambda_n\|\boldsymbol\delta_{S_n}\|_1<R_n$, then
\eqref{eq:pcm-lasso-cone-basic} gives $\mathcal E_n\le4R_n$.
Otherwise,
\[
\lambda_n\|\boldsymbol\delta_{S_n^c}\|_1
\le R_n+3\lambda_n\|\boldsymbol\delta_{S_n}\|_1
\le4\lambda_n\|\boldsymbol\delta_{S_n}\|_1,
\]
so $\boldsymbol\delta$ belongs to the cone in
\eqref{eq:pcm-lasso-compatibility}. Compatibility and the triangle inequality
give
\[
\|\boldsymbol\delta_{S_n}\|_1
\le\frac{\sqrt{s_n}}{\kappa_n}
\frac{\|\Phi_c\boldsymbol\delta\|_2}{\sqrt{n_\star}}
\le\frac{\sqrt{s_n}}{\kappa_n}
(\sqrt{R_n}+\sqrt{\mathcal E_n}).
\]
Substituting into~\eqref{eq:pcm-lasso-cone-basic},
\[
\mathcal E_n
\le R_n+\frac{3\sqrt{s_n}\lambda_n}{\kappa_n}
(\sqrt{\mathcal E_n}+\sqrt{R_n}).
\]
The two applications of $xy\le(x^2+y^2)/2$ are
\begin{align*}
\frac{3\sqrt{s_n}\lambda_n}{\kappa_n}\sqrt{\mathcal E_n}
&\le\frac{\mathcal E_n}{2}+\frac{9s_n\lambda_n^2}{2\kappa_n^2},\\
\frac{3\sqrt{s_n}\lambda_n}{\kappa_n}\sqrt{R_n}
&\le\frac{R_n}{2}+\frac{9s_n\lambda_n^2}{2\kappa_n^2}.
\end{align*}
Consequently,
\[
\mathcal E_n\le\frac32R_n+\frac12\mathcal E_n
+\frac{9s_n\lambda_n^2}{\kappa_n^2},
\qquad
\mathcal E_n\le3R_n+\frac{18s_n\lambda_n^2}{\kappa_n^2}.
\]
Combining all cases gives, with probability tending to one,
\begin{equation}\label{eq:pcm-lasso-sparse-centered-bound}
\mathcal E_n\le4R_n+\frac{18s_n\lambda_n^2}{\kappa_n^2}.
\end{equation}
Using~\eqref{eq:pcm-lasso-Rn-transfer}, the exact decomposition
\eqref{eq:pcm-lasso-orthogonal-decomp}, and
\eqref{eq:pcm-lasso-intercept-rate},
\[
\frac{\tilde E_{n,f}^{\mathrm{PCM}}}
{\sigma_{f,n,\mathrm{perm}}^2}
=O_\PP\!\left(
\frac{\|\mathbf r_n\|_2^2}
{n_\star\sigma_{f,n,\mathrm{perm}}^2}
+\frac{s_n\lambda_n^2}
{\kappa_n^2\sigma_{f,n,\mathrm{perm}}^2}
+\frac1{n_\star}
\right).
\]
Substitution of
$\lambda_n^2=C^2\sigma_{f,n,\mathrm{perm}}^2\log d/n_\star$
proves the first equality in~\eqref{eq:pcm-lasso-sparse-rate};
$\PP(s_n=0\text{ or }\kappa_n\ge\kappa_0)\to1$ gives the second.
\end{proof}

\section{Local power under permutation calibration}\label{app:power}\label{sec:power}

Fix $\alpha\in(0,1)$. Throughout this appendix, we consider right-tailed tests
calibrated by the ideal conditional permutation quantile.

\subsection{A common lemma for transferring critical values}
\label{subsec:power-transfer}

\begin{lemma}[Transfer of critical values under contiguity]
\label{lem:power-contiguity-transfer}
Let $P_0^{(n)}$ and $P_n^{(n)}$ be sequences of laws on the same sample space and
suppose $P_n^{(n)}$ is contiguous with respect to $P_0^{(n)}$.  Let $c_n$ be a
possibly data-dependent critical value and let $c\in\mathbb R$.  If
\[
  c_n\xrightarrow{P_0^{(n)}}c,
\]
then also $c_n\xrightarrow{P_n^{(n)}}c$.  If, in addition,
$T_n\convd T$ under $P_n^{(n)}$ and $\PP(T=c)=0$, then
\begin{equation}\label{eq:power-transfer}
  P_n^{(n)}(T_n>c_n)\longrightarrow \PP(T>c).
\end{equation}
\end{lemma}

\begin{proof}
For every $\varepsilon>0$, contiguity gives
\[
P_0^{(n)}(|c_n-c|>\varepsilon)\to0
\quad\Longrightarrow\quad
P_n^{(n)}(|c_n-c|>\varepsilon)\to0.
\]
Hence, under $P_n^{(n)}$, Slutsky's theorem yields
$T_n-c_n\convd T-c$. Since $\PP(T-c=0)=0$, the Portmanteau theorem gives
\[
P_n^{(n)}(T_n>c_n)
=P_n^{(n)}(T_n-c_n>0)\longrightarrow\PP(T-c>0).
\]
\end{proof}

\subsection{Local power of the GLM score with permutation calibration}
\label{subsec:glm-power}

Fix a null parameter $(0,\beta_0,\vartheta_0)$ in the GLM of Section~\ref{sec:glm-warmup}.
Let $P_0$ denote its law for a single observation, $P_0^{(n)}=P_0^{\otimes n}$,
and $\E_0$ expectation under $P_0$. Write
$\mu_0(\bm Z)\equiv g^{-1}(\bm Z^\top\beta_0)$, $v_0\equiv V_{\vartheta_0}(\mu_0(\bm Z))$,
$g_0'\equiv g'(\mu_0(\bm Z))$, $\xi_0\equiv(\bm Y-\mu_0(\bm Z))/(v_0g_0')$, and
$W_0\equiv1/\{v_0(g_0')^2\}$.
Let
\begin{equation}\label{eq:glm-power-projection}
M_0\equiv\E_0[W_0\bm Z\bm Z^\top],\qquad
r_0\equiv\E_0[W_0\bm Z\bm X],\qquad
\theta_0\equiv M_0^{-1}r_0,\qquad
\delta_0\equiv \bm X-\theta_0^\top \bm Z,
\end{equation}
and define the profile information
\begin{equation}\label{eq:glm-power-info}
I_{\mathrm{GLM}}
\equiv\E_0[W_0\delta_0^2]
=\E_0[W_0\bm X^2]-r_0^\top M_0^{-1}r_0.
\end{equation}
The second equality follows from
$\E_0[W_0\delta_0\bm Z]=0$.

\begin{assumption}[Local GLM alternatives for power]
\label{assumption:glm-power}
Assume that, in a neighbourhood of $(0,\beta_0)$, the true conditional
law of $\bm Y\mid \bm X,\bm Z$ has density
$p\bigl(\,\cdot\,;g^{-1}(\bm X\gamma+\bm Z^\top\beta),\vartheta_0\bigr)$, while the marginal law
of $(\bm X,\bm Z)$ is fixed.  Suppose the model is differentiable in quadratic mean at
$(0,\beta_0)$. Under $P_0$, assume Assumptions~\ref{assumption:glm-reg},
\ref{assumption:reg-outcome}, and \ref{assumption:reg-glm} hold, with
$\vartheta_\star=\vartheta_0$ in Assumption~\ref{assumption:glm-reg}(i).
For fixed $h_\gamma\in\mathbb R$ and
$k\in\mathbb R^p$, consider
\begin{equation}\label{eq:glm-local-path}
  \gamma_n=\frac{h_\gamma}{\sqrt n},
  \qquad
  \beta_n=\beta_0+\frac{k}{\sqrt n}.
\end{equation}
Write $h=(h_\gamma,k)$ and denote the corresponding $n$-sample law by
$P_{n,h}^{(n)}$.
\end{assumption}

Here $h_\gamma$ specifies the direction and magnitude of the departure of
$\gamma_n$ from zero on the $n^{-1/2}$ scale, while $k$ specifies the local
change in the nuisance coefficient $\beta_n$.

\begin{theorem}[GLM score: local power is unchanged by permutation calibration]
\label{thm:glm-power}
Under Assumption~\ref{assumption:glm-power},
\begin{equation}\label{eq:glm-local-limit}
T_n^{\mathrm{GLM}}
\convd N\!\left(h_\gamma\sqrt{I_{\mathrm{GLM}}},1\right)
\qquad\text{under }P_{n,h}^{(n)}.
\end{equation}
Moreover, if
\[
q_{n,\alpha}^{\mathrm{GLM},\pi}
\equiv\mathbb Q_{1-\alpha}
\bigl(T_n^{\mathrm{GLM},(\Pi)}\mid\cF_n\bigr),
\]
then
\begin{equation}\label{eq:glm-perm-critical-local}
q_{n,\alpha}^{\mathrm{GLM},\pi}
=z_{1-\alpha}+o_{P_{n,h}^{(n)}}(1),
\end{equation}
and therefore
\begin{equation}\label{eq:glm-power-formula}
P_{n,h}^{(n)}
\!\left(T_n^{\mathrm{GLM}}>q_{n,\alpha}^{\mathrm{GLM},\pi}\right)
\longrightarrow
1-\Phi\!\left(z_{1-\alpha}-h_\gamma\sqrt{I_{\mathrm{GLM}}}\right).
\end{equation}
The same expression is the local power of the usual score test
$\indc{T_n^{\mathrm{GLM}}>z_{1-\alpha}}$ calibrated using a normal distribution.
The limiting rejection probability equals $\alpha$ when $h_\gamma=0$ and
is strictly increasing in $h_\gamma$.
\end{theorem}

\begin{proof}
At $(0,\beta_0)$ the score in direction $(h_\gamma,k)$ is
$\dot\ell_h=(h_\gamma \bm X+k^\top \bm Z)\xi_0$, with
$\E_0(\xi_0\mid \bm X,\bm Z)=0$ and $\E_0(\xi_0^2\mid \bm X,\bm Z)=W_0$.
By differentiability in quadratic mean, the standard LAN theorem
\citep[Theorem~7.2]{vanderVaart1998} gives
\[
  \log\frac{dP_{n,h}^{(n)}}{dP_0^{(n)}}
  =\frac1{\sqrt n}\sum_{i=1}^n\dot\ell_{h,i}
  -\frac12\E_0(\dot\ell_h^2)+o_{P_0^{(n)}}(1),
\]
so $P_{n,h}^{(n)}\triangleleft P_0^{(n)}$ by Le Cam's first lemma \citep[Lemma~6.4]{vanderVaart1998}.
Equation~\eqref{eq:glm-emp-numerator-limit} in the proof of
Proposition~\ref{prop:glm-unperm-emp}, evaluated at the null parameter,
gives
\[
\frac1{\sqrt n}\sum_{i=1}^n
\hat\nu_i^{\mathrm{GLM}}\hat\epsilon_i
=
\frac1{\sqrt n}\sum_{i=1}^n
\delta_{0,i}\xi_{0,i}
+o_{P_0^{(n)}}(1).
\]
For the studentizer, equation~\eqref{eq:glm-emp-studentizer-limit} and
$\E_0(\xi_0^2\mid\bm X,\bm Z)=W_0$ give
\[
\widehat V_n^{\mathrm{GLM}}
\xrightarrow{P_0^{(n)}}
\E_0[\delta_0^2\xi_0^2]
=\E_0\!\left[\delta_0^2\E_0(\xi_0^2\mid\bm X,\bm Z)\right]
=\E_0[\delta_0^2W_0]
=I_{\mathrm{GLM}}.
\]
Hence
\begin{equation}\label{eq:glm-power-oracle-equiv}
T_n^{\mathrm{GLM}}
=
\frac{1}{\sqrt{nI_{\mathrm{GLM}}}}
\sum_{i=1}^n\delta_{0,i}\xi_{0,i}
+o_{P_0^{(n)}}(1).
\end{equation}
Contiguity transfers the remainder in \eqref{eq:glm-power-oracle-equiv} to
$o_{P_{n,h}^{(n)}}(1)$.

Since
$\E_0(W_0\bm Z\delta_0)=0$ and $\bm X=\delta_0+\theta_0^\top \bm Z$,
\begin{align*}
  \operatorname{Cov}_0(\delta_0\xi_0,\dot\ell_h)
&=\E_0\{W_0\delta_0(h_\gamma \bm X+k^\top \bm Z)\}\\
&=h_\gamma\E_0(W_0\delta_0^2)
=h_\gamma I_{\mathrm{GLM}}.
\end{align*}
Both summands are centered and square integrable under $P_0$:
$\E_0[\delta_0^2\xi_0^2]=I_{\mathrm{GLM}}<\infty$, and
$\E_0[\dot\ell_h^2]<\infty$ by differentiability in quadratic mean.
Thus the multivariate CLT gives, under $P_0^{(n)}$,
\[
\frac1{\sqrt n}\sum_{i=1}^n
\begin{pmatrix}
\delta_{0,i}\xi_{0,i}/\sqrt{I_{\mathrm{GLM}}}\\
\dot\ell_{h,i}
\end{pmatrix}
\convd
N_2\!\left(
\begin{pmatrix}0\\0\end{pmatrix},
\begin{pmatrix}
1 & h_\gamma\sqrt{I_{\mathrm{GLM}}}\\
h_\gamma\sqrt{I_{\mathrm{GLM}}} & \E_0[\dot\ell_h^2]
\end{pmatrix}
\right).
\]
Together with the likelihood-ratio expansion, Le Cam's third lemma
\citep[Example~6.7]{vanderVaart1998} gives
\[
\frac1{\sqrt{nI_{\mathrm{GLM}}}}
\sum_{i=1}^n\delta_{0,i}\xi_{0,i}
\convd N(h_\gamma\sqrt{I_{\mathrm{GLM}}},1)
\quad\text{under }P_{n,h}^{(n)}.
\]
Equation~\eqref{eq:glm-power-oracle-equiv} and the transferred remainder
then yield~\eqref{eq:glm-local-limit}.

Finally, Theorem~\ref{thm:glm-perm-emp} gives
$q_{n,\alpha}^{\mathrm{GLM},\pi}\to z_{1-\alpha}$ under $P_0^{(n)}$.
Lemma~\ref{lem:power-contiguity-transfer} transfers this convergence to
$P_{n,h}^{(n)}$ and, together with the preceding continuous Gaussian limit,
yields \eqref{eq:glm-perm-critical-local}--\eqref{eq:glm-power-formula}.
\end{proof}

\subsection{Local power of the GCM with permutation calibration}
\label{subsec:gcm-power}

\begin{assumption}[GPLM local alternatives for GCM power]
\label{assumption:gcm-power}
Let $P_{\beta,g}$ be the law of a single observation under which $(\bm X,\bm Z)$ has
fixed distribution $P_{XZ}$ and $\bm Y\mid\bm X,\bm Z$ has density
\begin{equation}\label{eq:gcm-gplm-model}
f_\eta(y)=\exp\{y\eta-\psi(\eta)\},
\qquad \eta=\beta\bm X+g(\bm Z),
\end{equation}
with respect to a fixed measure $\lambda$, where $f_\eta$ is a regular
one-parameter canonical exponential family with open natural parameter
space $J$. Write $P_0=P_{0,g_0}$, $P_0^{(n)}=P_0^{\otimes n}$, and
$\E_0$ for expectation under $P_0$.
For fixed $h=(h_\beta,h_g)$ with $h_\beta\in\mathbb R$ and
$h_g\in L^2(P_Z)$, consider
\begin{equation}\label{eq:gcm-local-alt}
\beta_n=\frac{h_\beta}{\sqrt n},\qquad
 g_n=g_0+\frac{h_g}{\sqrt n},\qquad
 P_{n,h}^{(n)}=P_{\beta_n,g_n}^{\otimes n}.
\end{equation}
Assume $g_0(\bm Z)+t\{h_\beta\bm X+h_g(\bm Z)\}\in J$ almost surely
for every $t$ in a neighbourhood of zero. Under $P_0$, assume
Assumptions~\ref{assumption:unperm}, \ref{assumption:perm}, \ref{assumption:reg-outcome}, and
\ref{assumption:reg-gcm} hold. Assume also either
\begin{enumerate}[(i)]
\item $\psi''$ is a positive constant on $J$; or
\item the support of $(\bm X,\bm Z)$ is compact, $g_0$ and $h_g$ are
continuous on the support of $\bm Z$, and, for some $\delta>0$,
$\psi$ is twice continuously differentiable on an open interval in $J$ containing
\[
\{g_0(z)+t\{h_\beta x+h_g(z)\}:
(x,z)\in\operatorname{supp}(\bm X,\bm Z),\ |t|\le\delta\}.
\]
\end{enumerate}
\end{assumption}

Throughout this subsection, $\mu_x$, $\mu_y$, and
$\sigma_{\mathrm{GCM}}^2$ are evaluated under $P_0$. In particular,
$\mu_x(z)=\E_0[\bm X\mid\bm Z=z]$ and $\mu_y(z)=\psi'(g_0(z))$.
Since $\bm X\indep\bm Y\mid\bm Z$ under $P_0$,
\begin{equation}\label{eq:gcm-s2}
\sigma_{\mathrm{GCM}}^2
=\E_0[\var_0(\bm X\mid\bm Z)\var_0(\bm Y\mid\bm Z)]>0,
\end{equation}
where positivity is Assumption~\ref{assumption:reg-gcm}(ii).

\begin{theorem}[GCM: local power is unchanged by permutation calibration]
\label{thm:gcm-power}
Under Assumption~\ref{assumption:gcm-power},
\begin{equation}\label{eq:gcm-local-limit}
T_n^{\mathrm{GCM}}\convd N(h_\beta\sigma_{\mathrm{GCM}},1)
\qquad\text{under }P_{n,h}^{(n)}.
\end{equation}
If
\[
q_{n,\alpha}^{\mathrm{GCM},\pi}
\equiv\mathbb Q_{1-\alpha}
\bigl(T_n^{\mathrm{GCM},(\Pi)}\mid\cF_n\bigr),
\]
then
\begin{equation}\label{eq:gcm-perm-critical-local}
q_{n,\alpha}^{\mathrm{GCM},\pi}
=z_{1-\alpha}+o_{P_{n,h}^{(n)}}(1),
\end{equation}
and
\begin{equation}\label{eq:gcm-power-formula}
P_{n,h}^{(n)}\!\left(T_n^{\mathrm{GCM}}>q_{n,\alpha}^{\mathrm{GCM},\pi}\right)
\longrightarrow
1-\Phi\!\left(z_{1-\alpha}-h_\beta\sigma_{\mathrm{GCM}}\right).
\end{equation}
The GCM calibrated using a normal distribution with cutoff $z_{1-\alpha}$ has the same limit.
\end{theorem}

\begin{proof}
Define
\begin{equation}\label{eq:gcm-oracle-power-summand}
U\equiv(\bm X-\mu_x(\bm Z))(\bm Y-\mu_y(\bm Z)).
\end{equation}
Conditional independence under $P_0$ gives $\E_0[U]=0$ and
\begin{equation}\label{eq:gcm-oracle-power-var}
\E_0[U^2]
=\E_0[\var_0(\bm X\mid\bm Z)\var_0(\bm Y\mid\bm Z)]
=\sigma_{\mathrm{GCM}}^2.
\end{equation}
Equations~\eqref{eq:gcm-observed-reductions} and
\eqref{eq:gcm-observed-denominator} in the proof of Theorem~\ref{thm:gcm-valid} give
\[
\frac1{\sqrt n}\sum_{i=1}^n\hat R_i^{\mathrm{GCM}}
=\frac1{\sqrt n}\sum_{i=1}^nU_i+o_{P_0^{(n)}}(1),\qquad
\widehat V_n^{\mathrm{GCM}}=\sigma_{\mathrm{GCM}}^2+o_{P_0^{(n)}}(1).
\]
Since $\sigma_{\mathrm{GCM}}>0$ and
$n^{-1/2}\sum_iU_i=O_{P_0^{(n)}}(1)$, it follows that
\begin{equation}\label{eq:gcm-power-oracle-equiv}
T_n^{\mathrm{GCM}}
=\frac1{\sigma_{\mathrm{GCM}}\sqrt n}\sum_{i=1}^nU_i
+o_{P_0^{(n)}}(1).
\end{equation}

Put $\Delta(\bm X,\bm Z)=h_\beta\bm X+h_g(\bm Z)$ and consider
$P_t=P_{t h_\beta,g_0+t h_g}$.
Its density $p_t$ with respect to $P_{XZ}\otimes\lambda$ satisfies
\[
\partial_t\sqrt{p_t(x,y,z)}
=\tfrac12\Delta(x,z)
\{y-\psi'(g_0(z)+t\Delta(x,z))\}\sqrt{p_t(x,y,z)}.
\]
Thus its score at zero is
$\dot\ell_h=\Delta(\bm X,\bm Z)\{\bm Y-\mu_y(\bm Z)\}$, and
\[
4\int(\partial_t\sqrt{p_t})^2\,d(P_{XZ}\otimes\lambda)
=\E_0[\Delta^2\psi''\{g_0(\bm Z)+t\Delta\}]
\equiv I(t).
\]
Under Assumption~\ref{assumption:gcm-power}(i), writing $v=\psi''>0$,
\[
I(t)=v\E_0[\Delta^2]
\le2v\{h_\beta^2\E_0[\bm X^2]+\E_0[h_g(\bm Z)^2]\}<\infty,
\]
by Assumption~\ref{assumption:reg-gcm}(i) and $h_g\in L^2(P_Z)$.
Hence $I(t)$ is constant. Under (ii), $\Delta$ is bounded and the
specified set of natural parameters is compact. Continuity of $\psi''$
therefore gives
\[
\sup_{|t|\le\delta}
\left|\Delta^2\psi''\{g_0(\bm Z)+t\Delta\}\right|
\le\|\Delta\|_\infty^2
\sup_{\substack{(x,z)\in\operatorname{supp}(\bm X,\bm Z)\\|t|\le\delta}}
\left|\psi''\{g_0(z)+t\Delta(x,z)\}\right|<\infty.
\]
Dominated convergence gives $I(t)\to I(0)$ as $t\to0$.
The square-root density is continuously differentiable pointwise, so
Lemma~7.6 of \citet{vanderVaart1998} gives differentiability in quadratic
mean at zero. Theorem~7.2 therein yields
\[
\log\frac{dP_{n,h}^{(n)}}{dP_0^{(n)}}
=\frac1{\sqrt n}\sum_{i=1}^n\dot\ell_{h,i}
-\frac12 I(0)+o_{P_0^{(n)}}(1).
\]
Since $\E_0[\dot\ell_h]=0$ and $\E_0[\dot\ell_h^2]=I(0)<\infty$,
the CLT and Le Cam's first lemma
\citep[Lemma~6.4]{vanderVaart1998} give
$P_{n,h}^{(n)}\triangleleft P_0^{(n)}$.

Using conditional independence and $\E_0[\bm X-\mu_x(\bm Z)\mid\bm Z]=0$,
\begin{align*}
\E_0[U\dot\ell_h]
&=\E_0\!\left[\var_0(\bm Y\mid\bm Z)
\E_0\{(\bm X-\mu_x(\bm Z))(h_\beta\bm X+h_g(\bm Z))\mid\bm Z\}\right]\\
&=h_\beta\E_0[\var_0(\bm Y\mid\bm Z)\var_0(\bm X\mid\bm Z)]
=h_\beta\sigma_{\mathrm{GCM}}^2.
\end{align*}
The centered summands have finite second moments, so under $P_0^{(n)}$,
\[
\frac1{\sqrt n}\sum_{i=1}^n
\begin{pmatrix}U_i/\sigma_{\mathrm{GCM}}\\\dot\ell_{h,i}\end{pmatrix}
\convd N_2\!\left(
\begin{pmatrix}0\\0\end{pmatrix},
\begin{pmatrix}
1 & h_\beta\sigma_{\mathrm{GCM}}\\
h_\beta\sigma_{\mathrm{GCM}} & I(0)
\end{pmatrix}\right).
\]
Together with the likelihood-ratio expansion, Le Cam's third lemma
\citep[Example~6.7]{vanderVaart1998} gives
\[
\frac1{\sigma_{\mathrm{GCM}}\sqrt n}\sum_{i=1}^nU_i
\convd N(h_\beta\sigma_{\mathrm{GCM}},1)
\quad\text{under }P_{n,h}^{(n)}.
\]
Contiguity transfers the remainder in~\eqref{eq:gcm-power-oracle-equiv},
proving~\eqref{eq:gcm-local-limit}. Finally, Theorem~\ref{thm:gcm-valid} and
Lemma~\ref{lem:quantile} give
$q_{n,\alpha}^{\mathrm{GCM},\pi}\to z_{1-\alpha}$ under $P_0^{(n)}$.
Lemma~\ref{lem:power-contiguity-transfer} transfers the cutoff to
$P_{n,h}^{(n)}$ and yields
\eqref{eq:gcm-perm-critical-local}--\eqref{eq:gcm-power-formula}.
\end{proof}

\subsection{Local power of the PCM with permutation calibration}
\label{subsec:pcm-power}

In Algorithm~\ref{alg:vpcm}, use OLS with an intercept in $Z$ and follow the rules for fitting and
splitting the sample in Appendix~\ref{sec:practical}.
Allow deterministic fold sizes $n_1,n_2\to\infty$, $n_1+n_2=n$,
replacing $n/2$ by $n_1$ in both statistics.

\begin{assumption}[Local linear alternatives for OLS PCM power]
\label{assumption:pcm-power}
Let $(\bm X,\bm Z,\zeta)$ have a fixed joint law, with $p$ fixed,
$\E[\zeta\mid\bm X,\bm Z]=0$, and $\zeta\indep\bm X\mid\bm Z$.
For fixed $h\in\mathbb R$ and $\gamma_0\in\mathbb R^p$, consider
\begin{equation}\label{eq:pcm-linear-local-model}
\bm Y=\beta_n\bm X+\gamma_0^\top\bm Z+\zeta,
\qquad \beta_n=h/\sqrt n.
\end{equation}
Write $P_{n,h}^{(n)}$ for the resulting $n$-sample law and $P_0^{(n)}$
for the law with $\beta=0$. Assume $P_{n,h}^{(n)}\triangleleft P_0^{(n)}$,
\[
\E[|\bm X|^4+\|\bm Z\|^4+|\zeta|^4]<\infty,
\qquad
\E[WW^\top]\succ0,\qquad W=(\bm X,\bm Z^\top)^\top.
\]
Define
\[
\eta_X=\{\E[\bm Z\bm Z^\top]\}^{-1}\E[\bm Z\bm X],\qquad
\xi=\bm X-\eta_X^\top\bm Z,\qquad
\sigma_\xi^2=\E[\xi^2],
\]
and assume $\sigma_{\xi\zeta}^2\equiv\E[\xi^2\zeta^2]>0$.
\end{assumption}

Here $h$ specifies the departure of $\beta_n$ from zero on the
$n^{-1/2}$ scale. Write
\[
\sigma_\beta^2=\frac{\sigma_{\xi\zeta}^2}{\sigma_\xi^4},
\]
and define
\begin{equation}\label{eq:pcm-abn}
a_n=\frac{\sqrt{n_2}\,\beta_n}{\sigma_\beta},\qquad
b_n=\frac{\sqrt{n_1}\,\beta_n\sigma_\xi^2}{\sigma_{\xi\zeta}}.
\end{equation}

\begin{theorem}[PCM with OLS: local power is unchanged by permutation calibration]
\label{thm:pcm-power}
Fix $\alpha\in(0,1)$. Under Assumption~\ref{assumption:pcm-power}, suppose $a_n\to a$ and
$b_n\to b$. Then, under $P_{n,h}^{(n)}$,
\begin{equation}\label{eq:pcm-original-local-limit}
T_n^{\mathrm{PCM}}\convd T_\infty^{\mathrm{PCM}},\qquad
T_\infty^{\mathrm{PCM}}\sim\Phi(a)N(b,1)+\Phi(-a)N(-b,1).
\end{equation}
Moreover, the conditional permutation quantile
\[
q_{n,\alpha}^{\mathrm{PCM},\pi}
\equiv\mathbb Q_{1-\alpha}
\bigl(T_n^{\mathrm{PCM},(\Pi)}\mid\cF_n\bigr)
\]
satisfies
\begin{equation}\label{eq:pcm-perm-critical-local-power}
q_{n,\alpha}^{\mathrm{PCM},\pi}
=z_{1-\alpha}+o_{P_{n,h}^{(n)}}(1),
\end{equation}
and
\begin{equation}\label{eq:pcm-power-formula}
P_{n,h}^{(n)}\!\left(T_n^{\mathrm{PCM}}>q_{n,\alpha}^{\mathrm{PCM},\pi}\right)
\longrightarrow
\Phi(a)\Phi(z_\alpha+b)+\Phi(-a)\Phi(z_\alpha-b).
\end{equation}
The statistic calibrated using a normal distribution with cutoff $z_{1-\alpha}$ has the same
local power, as in \citet[Proposition~S37]{lundborg2022projected}.
For $n_1=n_2=n/2$,
\begin{equation}\label{eq:pcm-ab-balanced}
a=b=\frac{h\sigma_\xi^2}{\sqrt2\,\sigma_{\xi\zeta}}.
\end{equation}
\end{theorem}

\begin{proof}
\noindent\textit{OLS factorization and sampling limit.}\quad
Probability orders in this part refer to $P_{n,h}^{(n)}$.
Write $\hat m_p(x,z)=\hat\beta x+\hat\gamma^\top z$ and
$\hat m_q(z)=\hat\chi^\top z$. Let $Z_1,\mathbf X_1$ be the
design matrix for the inference fold and exposure vector, and $H_1$ the
orthogonal projection onto the column space of $Z_1$. Then
\[
\widehat{\boldsymbol\nu}^{\mathrm{PCM}}
=(I-H_1)\{\hat\beta\mathbf X_1+Z_1(\hat\gamma-\hat\chi)\}
=\hat\beta(I-H_1)\mathbf X_1.
\]
For the OLS coefficients for the inference fold $\hat\eta_X,\hat\eta_Y$, set
$R_i=(X_i-Z_i^\top\hat\eta_X)(Y_i-Z_i^\top\hat\eta_Y)$.
Thus
\begin{equation}\label{eq:pcm-power-sign-factorization}
\hat R_i^{\mathrm{PCM}}=\hat\beta R_i,\qquad
T_n^{\mathrm{PCM}}=\operatorname{sgn}(\hat\beta)Q_n,\qquad
Q_n=\frac{n_1^{-1/2}\sum_{i\in\cD_1}R_i}{\sqrt{\mathcal S_{n_1}(R)}}.
\end{equation}
Here $\mathcal S_{n_1}(R)=n_1^{-1}\sum_iR_i^2-(n_1^{-1}\sum_iR_i)^2$;
set $Q_n=0$ when this variance vanishes and $\operatorname{sgn}(0)=0$.
The variables $\hat\beta$ and $Q_n$ use independent folds.

The population coefficient of $Y$ on $Z$ is
$\eta_{Y,n}=\gamma_0+\beta_n\eta_X$, with residual
$\epsilon_{n,i}=\zeta_i+\beta_n\xi_i$.
Put $d_X=\hat\eta_X-\eta_X$ and $d_Y=\hat\eta_Y-\eta_{Y,n}$.
Positive definiteness and the law of large numbers make the fold design
matrices invertible with probability tending to one. On this event,
\[
\begin{pmatrix}d_X&d_Y\end{pmatrix}
=\left(\frac1{n_1}\sum_{i\in\cD_1}Z_iZ_i^\top\right)^{-1}
\frac1{n_1}\sum_{i\in\cD_1}Z_i
\begin{pmatrix}\xi_i&\epsilon_{n,i}\end{pmatrix}
=O_\PP(n_1^{-1/2}),
\]
since $Z_i\xi_i$ and $Z_i\epsilon_{n,i}$ are centered with bounded
second moments. In the following calculations for the inference fold, sums are
over $\cD_1$. The normal equations give
\[
\frac1{\sqrt{n_1}}\sum_i(R_i-\xi_i\epsilon_{n,i})
=-\frac{d_X^\top Z_1^\top Z_1d_Y}{\sqrt{n_1}}
=O_\PP(n_1^{-1/2}).
\]
Moreover,
$R_i-\xi_i\zeta_i=-\xi_iZ_i^\top d_Y-\epsilon_{n,i}Z_i^\top d_X
+(Z_i^\top d_X)(Z_i^\top d_Y)+\beta_n\xi_i^2$, so
\begin{align*}
\frac1{n_1}\sum_i(R_i-\xi_i\zeta_i)^2
\le{}&4\|d_Y\|^2\frac1{n_1}\sum_i\xi_i^2\|Z_i\|^2
+4\|d_X\|^2\frac1{n_1}\sum_i\epsilon_{n,i}^2\|Z_i\|^2\\
&+4\|d_X\|^2\|d_Y\|^2\frac1{n_1}\sum_i\|Z_i\|^4
+\frac{4\beta_n^2}{n_1}\sum_i\xi_i^4\\
&=O_\PP(n_1^{-1}+\beta_n^2)=o_\PP(1).
\end{align*}
Each average on the right is $O_\PP(1)$ by the fourth-moment assumption
and Cauchy--Schwarz. The oracle scores $\xi_i\zeta_i$ are i.i.d.\ with
mean zero and variance $\sigma_{\xi\zeta}^2$. Their CLT and law of large
numbers, together with Lemma~\ref{lem:studentizer-transfer}, give the
first line below. The second follows from the sum bound above and
$\beta_n n_1^{-1/2}\sum_i(\xi_i^2-\sigma_\xi^2)=O_\PP(|\beta_n|)=o_\PP(1)$:
\begin{equation}\label{eq:pcm-power-ols-expansion}
\begin{aligned}
\mathcal S_{n_1}(R)&=\sigma_{\xi\zeta}^2+o_\PP(1),\\
Q_n&=\frac1{\sigma_{\xi\zeta}\sqrt{n_1}}\sum_i\xi_i\zeta_i+b_n+o_\PP(1).
\end{aligned}
\end{equation}
Here $b_n=O(1)$ justifies replacing the studentizer by $\sigma_{\xi\zeta}$.

For the training coefficient, with $W_i=(X_i,Z_i^\top)^\top$, the
block inverse formula for $\E[WW^\top]$ yields
\begin{align*}
\sqrt{n_2}(\hat\beta-\beta_n)
&=\left[
\left(\frac1{n_2}\sum_{i\in\cD_2}W_iW_i^\top\right)^{-1}
\frac1{\sqrt{n_2}}\sum_{i\in\cD_2}W_i\zeta_i
\right]_1\\
&=\frac1{\sigma_\xi^2\sqrt{n_2}}\sum_{i\in\cD_2}\xi_i\zeta_i+o_\PP(1),
\end{align*}
where $[\cdot]_1$ denotes the first coordinate. Hence the foldwise CLTs
and independence give
\[
\begin{pmatrix}\sqrt{n_2}\hat\beta/\sigma_\beta\\Q_n\end{pmatrix}
\convd N_2\!\left(\begin{pmatrix}a\\b\end{pmatrix},I_2\right).
\]
The first limiting coordinate has no atom at zero, so
$\PP(\hat\beta=0)\to0$. Applying the continuous mapping theorem to
\eqref{eq:pcm-power-sign-factorization} proves
\eqref{eq:pcm-original-local-limit}.

\noindent\textit{Permutation distribution under the base null.}\quad
Here probability orders refer to $P_0^{(n)}$ and the independent
permutation; all sums and maxima are over $\cD_1$. Set
$c_i=X_i-\bar X_1$, with $\bar X_1=n_1^{-1}\sum_iX_i$, and
$e_i=Y_i-Z_i^\top\hat\eta_Y$.
The intercept and OLS orthogonality give
$\sum_i c_i=\sum_i e_i=0$, $H_1e=0$, and
$e_i=\zeta_i-Z_i^\top d_Y$ with $d_Y=O_\PP(n_1^{-1/2})$.
Since $n_1^{-1}\sum_i(Z_i^\top d_Y)^2=O_\PP(n_1^{-1})$, the law of
large numbers gives
\[
\frac1{n_1}\sum_i c_i^2\convp\var(\bm X)>0,\qquad
\frac1{n_1}\sum_i e_i^2\convp\E[\zeta^2]>0.
\]
For any i.i.d.\ $V_i$ with $\E|V_i|^4<\infty$,
\[
\PP\!\left(\max_i|V_i|^4>\varepsilon n_1\right)
\le\varepsilon^{-1}\E[|V_1|^4\indc{|V_1|^4>\varepsilon n_1}]
\longrightarrow0,\qquad \varepsilon>0.
\]
Apply this to $X_i,\zeta_i,\|Z_i\|$. The bound
\[
c_i^4+e_i^4
\le8\{|X_i|^4+|\bar X_1|^4+|\zeta_i|^4+\|d_Y\|^4\|Z_i\|^4\}
\]
and the law of large numbers imply
$n_1^{-1}\sum_i(c_i^4+e_i^4)=O_\PP(1)$ and
$n_1^{-1}\max_i(c_i^4+e_i^4)=o_\PP(1)$.
Define
\[
\sigma_{p,n}^2=\left(\frac1{n_1}\sum_i e_i^2\right)
\left(\frac1{n_1}\sum_i c_i^2\right)
\convp\var(\bm X)\E[\zeta^2]>0.
\]
The array $C_{n,ik}=e_ic_k$ has zero row and column sums, and
\[
\frac{\max_{i,k}C_{n,ik}^2}{n_1\sigma_{p,n}^2}
=\frac{(\max_i e_i^4/n_1)^{1/2}(\max_k c_k^4/n_1)^{1/2}}
{\sigma_{p,n}^2}=o_\PP(1).
\]
Its permutation sum has conditional variance
$n_1^2\sigma_{p,n}^2/(n_1-1)$ by Lemma~\ref{lem:comb-clt}(b).
Lemma~\ref{lem:comb-clt-prime} therefore gives
\begin{equation}\label{eq:pcm-power-perm-oracle}
\frac{\sum_i e_ic_{\Pi(i)}}{\sqrt{n_1}\,\sigma_{p,n}}
\ \Big|\ \cF_n\ \convdp\ N(0,1).
\end{equation}
For the second moment, Lemma~\ref{lem:hoeffding} gives
\begin{align*}
\E_\Pi\!\left[\frac1{n_1}\sum_i e_i^2c_{\Pi(i)}^2\middle|\cF_n\right]
&=\sigma_{p,n}^2,\\
\var_\Pi\!\left(\frac1{n_1}\sum_i e_i^2c_{\Pi(i)}^2\middle|\cF_n\right)
&\le\frac{(n_1^{-1}\sum_i e_i^4)(n_1^{-1}\sum_i c_i^4)}{n_1-1}
=O_\PP(n_1^{-1}).
\end{align*}
Lemma~\ref{lem:cond-markov}(b) thus gives
$n_1^{-1}\sum_i e_i^2c_{\Pi(i)}^2-\sigma_{p,n}^2=o_\PP(1)$.

The OLS refit gives
$\hat R_i^{\mathrm{PCM},(\Pi)}=\hat\beta R_i^{(\Pi)}$, where
$R_i^{(\Pi)}=e_i\{c_{\Pi(i)}-(H_1c_\Pi)_i\}$.
Since $H_1e=0$, $\sum_iR_i^{(\Pi)}=\sum_i e_ic_{\Pi(i)}$ exactly.
Moreover, uniform permutation of the centered vector $c$ gives
$\E_\Pi[c_\Pi c_\Pi^\top\mid\cF_n]
=\|c\|^2(I-\mathbf1\mathbf1^\top/n_1)/(n_1-1)$.
Using $H_1^2=H_1$, $H_1\mathbf1=\mathbf1$, and
$\operatorname{rank}(H_1)\le p$,
\begin{align*}
\E_\Pi\!\left[\frac1{n_1}\sum_i
\{R_i^{(\Pi)}-e_ic_{\Pi(i)}\}^2\middle|\cF_n\right]
&\le\frac{\max_i e_i^2}{n_1}\E_\Pi[\|H_1c_\Pi\|^2\mid\cF_n]\\
&=\frac{\max_i e_i^2}{n_1}\,
\frac{\|c\|^2}{n_1-1}\{\operatorname{rank}(H_1)-1\}=o_\PP(1).
\end{align*}
Lemma~\ref{lem:cond-markov}(a) gives the mean-square error $o_\PP(1)$.
The normalized oracle sum is $O_\PP(1)$ by
\eqref{eq:pcm-power-perm-oracle} and Lemma~\ref{lem:cond-to-marg}.
Lemma~\ref{lem:studentizer-transfer} therefore yields
$\mathcal S_{n_1}(R^{(\Pi)})/\sigma_{p,n}^2\convp1$.
By Lemma~\ref{lem:slutsky},
\[
\frac{\sum_iR_i^{(\Pi)}}{\sqrt{n_1\mathcal S_{n_1}(R^{(\Pi)})}}
\ \Big|\ \cF_n\ \convdp\ N(0,1).
\]
Multiplication by the $\cF_n$-measurable sign of $\hat\beta$ preserves
this limit by normal symmetry and $P_0^{(n)}(\hat\beta=0)\to0$.
Lemma~\ref{lem:quantile} gives
$q_{n,\alpha}^{\mathrm{PCM},\pi}\xrightarrow{P_0^{(n)}}z_{1-\alpha}$.

Finally, contiguity and Lemma~\ref{lem:power-contiguity-transfer}, applied
to the continuous mixture in~\eqref{eq:pcm-original-local-limit}, give
\eqref{eq:pcm-perm-critical-local-power} and~\eqref{eq:pcm-power-formula}.
The deterministic cutoff $z_{1-\alpha}$ has the same limit by
\eqref{eq:pcm-original-local-limit}; the expression for equally sized folds follows
from~\eqref{eq:pcm-abn} and the definition of $\sigma_\beta$.
\end{proof}

\end{document}